\pdfoutput=1

\documentclass[11pt]{article}

\usepackage[preprint]{acl}

\usepackage{times}
\usepackage{latexsym}

\usepackage[T1]{fontenc}

\usepackage[utf8]{inputenc}
\usepackage{adjustbox}
\usepackage{pdflscape}
\usepackage{url}
\usepackage{amssymb}
\usepackage{amsmath}
\usepackage{multirow}
\usepackage{graphicx} 
\usepackage{graphics}
\usepackage{multirow}

\usepackage{float}
\usepackage{fancyvrb}
\usepackage{amssymb}
\usepackage{amsmath}
\usepackage{amsfonts}
\usepackage{amssymb}
\usepackage{array}
\usepackage{ragged2e}
\usepackage{tabu}
\usepackage{algorithmicx}
\usepackage{algorithm}
\usepackage{algpseudocode}
\usepackage{booktabs}
\usepackage{adjustbox}
\usepackage{tikz}
\usetikzlibrary{shapes,arrows,positioning,fit,backgrounds,decorations.pathreplacing}

\usepackage{pifont}

\usepackage{tcolorbox}
\newtcolorbox{codebox}{
    colback=green!10,
    colframe=green!50!black,
    boxrule=2pt,
    arc=5pt,
    left=10pt,
    right=10pt,
    top=10pt,
    bottom=10pt
}

\usepackage{microtype}

\title{LoCoBench: A Benchmark for Long-Context Large Language Models \\
in Complex Software Engineering}

\author{Jielin Qiu, Zuxin Liu, Zhiwei Liu, Rithesh Murthy, Jianguo Zhang, Haolin Chen,  \\
\textbf{Shiyu Wang, Ming Zhu, Liangwei Yang, Juntao Tan, Zhepeng Cen, Cheng Qian,} \\
\textbf{Shelby Heinecke, Weiran Yao, Silvio Savarese, Caiming Xiong, Huan Wang}\\
\\
  Salesforce AI Research
  }

\begin{document}
\maketitle

\begin{figure*}[ht]
\centering
\includegraphics[width=\textwidth]{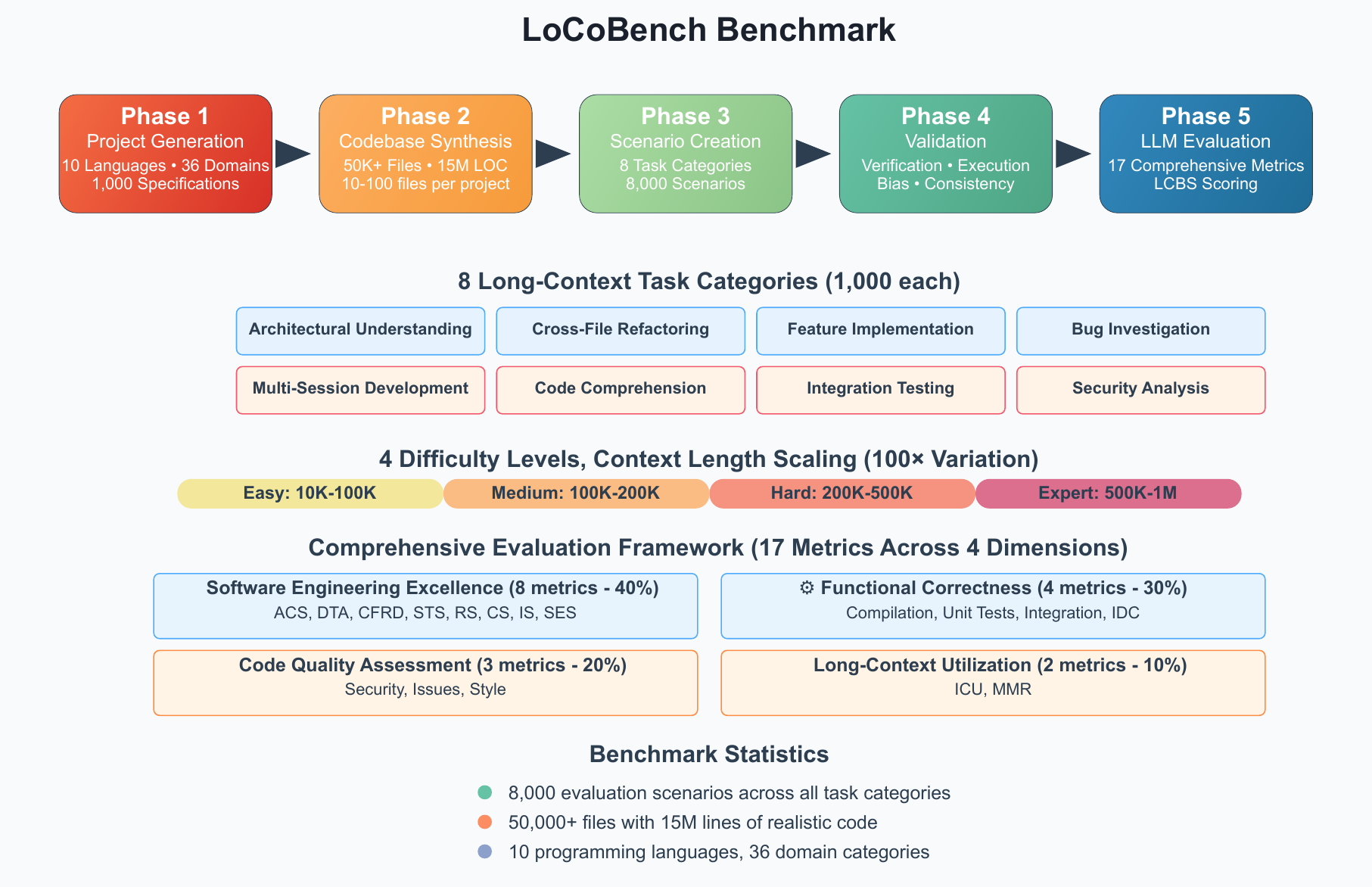}
\caption{LoCoBench Pipeline Architecture. Our systematic 5-phase pipeline transforms high-level specifications into a comprehensive evaluation benchmark. Phase 1 generates 1,000 diverse project specifications across 10 programming languages and 36 domains. Phase 2 creates complete codebases with realistic multi-file architectures, generating over 50K files with 15M lines of code. Phase 3 transforms codebases into 8,000 evaluation scenarios across 8 long-context task categories, with systematic context scaling from 10K to 1M tokens. Phase 4 ensures quality through automated compilation checks, quality metrics validation, and bias detection. Phase 5 evaluates LLMs using 17 comprehensive metrics across 4 evaluation dimensions.}
\label{fig:locobench_pipeline}
\end{figure*}

\begin{abstract}
The emergence of long-context language models with context windows extending to millions of tokens has created new opportunities for sophisticated code understanding and software development evaluation. We propose LoCoBench, a comprehensive benchmark specifically designed to evaluate long-context LLMs in realistic, complex software development scenarios. Unlike existing code evaluation benchmarks that focus on single-function completion or short-context tasks, LoCoBench addresses the critical evaluation gap for long-context capabilities that require understanding entire codebases, reasoning across multiple files, and maintaining architectural consistency across large-scale software systems. Our benchmark provides 8,000 evaluation scenarios systematically generated across 10 programming languages, with context lengths spanning 10K to 1M tokens, a 100× variation that enables precise assessment of long-context performance degradation in realistic software development settings. LoCoBench introduces 8 task categories that capture essential long-context capabilities: architectural understanding, cross-file refactoring, multi-session development, bug investigation, feature implementation, code comprehension, integration testing, and security analysis. Through a 5-phase pipeline, we create diverse, high-quality scenarios that challenge LLMs to reason about complex codebases at unprecedented scale. We introduce a comprehensive evaluation framework with 17 metrics across 4 dimensions including new evaluation metrics: Architectural Coherence Score (ACS), Dependency Traversal Accuracy (DTA), and Multi-Session Memory Retention (MMR), combined in a LoCoBench Score (LCBS). Our evaluation of state-of-the-art long-context models reveals substantial performance gaps, demonstrating that long-context understanding in complex software development represents a significant unsolved challenge that demands more attention.
LoCoBench is released at: \url{https://github.com/SalesforceAIResearch/LoCoBench}.
\end{abstract}

\section{Introduction}

The emergence of long-context language models with context windows extending to millions of tokens has created a new frontier in software development evaluation. As LLMs evolve from simple code completion tools to sophisticated systems capable of reasoning about entire codebases, understanding complex architectural patterns, and handling multi-file development workflows, traditional evaluation frameworks have become fundamentally inadequate.

\textbf{The Long-Context Revolution in Code.} Recent breakthroughs in long-context LLMs with context windows extending to millions of tokens~\cite{reid2024gemini,anthropic2024claude} have unlocked unprecedented opportunities for complex software development tasks. These models can now comprehend entire codebases spanning hundreds of files, understand complex inter-module dependencies, and maintain architectural consistency across large-scale systems. However, recent work reveals that \textit{long-context capabilities remain a critical weakness}: LongCodeBench~\cite{rando2025longcodebench} demonstrates dramatic performance degradation from 29\% to 3\% for Claude 3.5 Sonnet as context length increases, while RULER~\cite{hsieh2024ruler} shows that only half of models claiming 32K+ context sizes can maintain satisfactory performance at that length.

\textbf{The Long-Context Capability Gap.} While existing code evaluation benchmarks have advanced single-function generation~\cite{chen2021evaluating,austin2021program} and repository-level understanding~\cite{jimenez2023swe,liu2023repobench}, they fall short of evaluating the sophisticated \textit{long-context capabilities} required for realistic software development workflows. Complex software development tasks require navigating complex architectural decisions, performing multi-file reasoning, executing coordinated refactoring across dozens of files, and maintaining architectural consistency across large codebases, capabilities that extend far beyond traditional code generation or completion tasks.

\textbf{The Evaluation Challenge.} Current benchmarks exhibit three critical limitations that prevent adequate assessment of long-context software development capabilities:

\textit{Scale Limitations:} Most benchmarks contain fewer than 3K evaluation instances~\cite{jimenez2023swe,hendrycks2021measuring}, providing insufficient coverage for systematic evaluation across languages, complexity levels, and long-context tasks.

\textit{Context Limitations:} Traditional benchmarks operate with short contexts (typically under 10K tokens), failing to test models' ability to understand and operate on realistic enterprise codebase sizes. Even recent long-context benchmarks like $\infty$-Bench~\cite{zhang2024infty} and LongBench~\cite{bai2024longbench} focus primarily on document comprehension rather than complex code understanding.

\textit{Task Scope Limitations:} Existing benchmarks focus on isolated code generation, completion, or bug fixing, neglecting crucial long-context capabilities like architectural understanding, cross-file reasoning, and complex multi-file workflows.

To address these fundamental gaps, we introduce \textbf{LoCoBench}, a comprehensive benchmark specifically designed to evaluate long-context understanding in complex software development scenarios. Our benchmark introduces:

\begin{itemize}
    \item \textbf{Systematic Long-Context Code Evaluation:} LoCoBench provides 8,000 evaluation scenarios with context lengths systematically spanning 10K to 1M tokens, a 100× variation that enables precise assessment of long-context performance degradation in realistic software development settings.
    \vspace{-5pt}
    \item \textbf{Comprehensive Task Categories:} We introduce 8 task categories that capture essential long-context capabilities: architectural understanding, cross-file refactoring, multi-session development, bug investigation, feature implementation, code comprehension, integration testing, and security analysis.
    \vspace{-10pt}
    \item \textbf{New Evaluation Metrics:} We present a comprehensive evaluation framework of 17 metrics across 4 dimensions, including 6 newly proposed metrics specifically designed for long-context capabilities, combined in a unified LoCoBench Score (LCBS).
    \vspace{-5pt}
    \item \textbf{Unprecedented Scale and Diversity:} With 8,000 scenarios across 10 programming languages and 36 domain categories, LoCoBench provides more evaluation instances than the largest existing benchmark while maintaining systematic coverage of difficulty levels and realistic complexity distributions.
\end{itemize}

Our evaluation of state-of-the-art models reveals substantial performance gaps. These findings demonstrate that long-context understanding in complex software development represents a significant unsolved challenge, highlighting the critical need for more benchmarks and models to drive progress in this domain.

\section{Related Work}

\subsection{Code Generation Benchmarks}

Traditional code evaluation benchmarks focus on narrow programming aspects. Function-level benchmarks like HumanEval~\cite{chen2021evaluating} and MBPP~\cite{austin2021program} established foundational evaluation frameworks, with extensions including HumanEval+~\cite{liu2023humaneval+}, MultiPL-E~\cite{cassano2023multiple}, and BigCodeBench~\cite{zhuo2024bigcodebench}. Contest programming benchmarks such as APPS~\cite{hendrycks2021measuring}, LiveCodeBench~\cite{jain2024livecode}, and CodeContests~\cite{li2022competition} test algorithmic problem-solving but do not address software engineering concerns like architectural design or multi-file development.
Recent long-context code benchmarks include LongCodeBench~\cite{rando2025longcodebench}, which demonstrates dramatic performance degradation as context increases.  LongCodeU~\cite{li2025longcodeu} and LongCodeArena~\cite{bogomolov2024long}  focus primarily on code completion rather than comprehensive software development capabilities. Domain-specific benchmarks~\cite{lai2022ds,thakur2023verileval,wang2022cococo,dong2024effibench,du2023classeval} and repository-level evaluation~\cite{liu2023repobench,ding2023crosscodeeval} represent progress toward realistic scenarios but remain limited in scope.

\vspace{-5pt}
\subsection{Software Engineering Benchmarks}

SWE-Bench~\cite{jimenez2023swe} provides real GitHub issues for software engineering evaluation, with recent extensions including SWE-rebench~\cite{swe2025rebench} and LiveSWEBench~\cite{liveswebench2024}. Multi-SWE-Bench \citep{Zan2025MultiSWEbenchAM} extends this approach with high-quality instances across 7 programming languages, curated by expert annotators to address the Python-centric limitations of original SWE-Bench. However, these benchmarks remain limited to bug fixes rather than comprehensive development workflows. DevBench~\cite{li2024devbench} evaluates LLMs across the software development lifecycle but lacks systematic long-context assessment. CodeXGLUE~\cite{lu2021codexglue} addresses code understanding tasks but focuses on existing code analysis rather than development workflows.

\vspace{-5pt}
\subsection{Long-Context Evaluation}

General long-context benchmarks include LongBench~\cite{bai2024longbench}, RULER~\cite{hsieh2024ruler}, $\infty$-Bench~\cite{zhang2024infty}, and others~\cite{yen2024helmet,lee2024loft,an2024longicl,bai2024longalign,dong2024bamboo}. Code-specific long-context evaluation has emerged through LongCodeBench~\cite{rando2025longcodebench}, LongCodeU~\cite{li2025longcodeu}, LongCodeArena~\cite{bogomolov2024long}, and RepoQA~\cite{liu2024repoqa}. However, existing long-context benchmarks primarily focus on natural language tasks or code completion rather than complex multi-file software development capabilities.

\subsection{Limitations and Contributions}

We provided a comprehensive literature discussion in Appendix~\ref{sec:appendix-related-work}. In short, current benchmarks exhibit critical limitations: (1) \textbf{Scale:} Most contain <1,000 instances, insufficient for systematic evaluation; (2) \textbf{Task Scope:} Focus on isolated generation/completion rather than architectural understanding and multi-session development; (3) \textbf{Context Length:} Operate with short contexts (<10K tokens); (4) \textbf{Metrics:} Emphasize functional correctness while ignoring long-context capabilities like architectural coherence and context retention.

LoCoBench addresses these limitations through 8,000 scenarios spanning 10K-1M tokens, comprehensive task categories capturing essential long-context capabilities, and new evaluation metrics designed for complex software development scenarios.

\section{LoCoBench Benchmark}

\subsection{Benchmark Design Principles}

LoCoBench is designed around four core principles that distinguish it from existing code evaluation benchmarks:

\underline{Long-Context Tasks:} Our benchmark focuses on evaluation scenarios that reflect real-world complex software development capabilities, emphasizing tasks that require understanding large codebases, managing complex dependencies, and maintaining consistency across multiple files and development sessions.

\underline{Systematic Scale:} We generate 8,000 evaluation scenarios through a systematic 5-phase pipeline that ensures comprehensive coverage across programming languages, difficulty levels, and task categories while maintaining high quality and diversity.

\underline{Long-Context Focus:} Our scenarios span context lengths from 10K to 1M tokens, systematically testing models' ability to understand and operate on realistic codebase sizes that exceed the scope of traditional benchmarks.

\underline{Comprehensive Metrics:} Beyond traditional functional correctness, we introduce new evaluation metrics that capture long-context capabilities including architectural understanding, cross-file reasoning, and multi-session memory retention.

Figure~\ref{fig:locobench_pipeline} illustrates the complete LoCoBench pipeline, showing the systematic flow from project specifications to validated evaluation scenarios, including data processing, LLM integration, and quality assurance mechanisms.

\subsection{Five-Phase Pipeline}

Our benchmark generation follows a systematic 5-phase pipeline designed to create high-quality, diverse evaluation scenarios at scale:

\textbf{Phase 1: Project Specification Generation}
We generate 1,000 diverse project specifications across 10 programming languages (100 per language). Each specification defines a complete software project with realistic requirements, technical constraints, and architectural patterns. Projects span 36 domain categories including web applications, machine learning systems, data processing pipelines, and system utilities, with complexity levels ranging from simple applications to enterprise-scale systems.

\textbf{Phase 2: Codebase Generation}  
For each project specification, we generate complete, realistic codebases containing 10-100 files per project. This phase creates architecturally coherent codebases that include proper module structure, dependency management, documentation, and realistic code patterns. Generated codebases undergo automated quality validation including compilation checks, complexity analysis, and architectural consistency verification.

\textbf{Phase 3: Evaluation Scenario Creation}
We transform each codebase into 8 evaluation scenarios (1 per task category), resulting in 8,000 total scenarios. Each scenario includes carefully selected file subsets that provide sufficient context while targeting specific long-context capabilities. Context selection employs intelligent algorithms that balance information coverage, difficulty calibration, and realistic development workflows. Our selection algorithm prioritizes files based on dependency graphs, architectural centrality, and task-specific relevance, ensuring scenarios contain the minimum necessary context while maximizing information density and maintaining realistic development patterns.

\textbf{Phase 4: Validation and Quality Assurance}
All generated scenarios undergo comprehensive validation including compilation verification, test execution, complexity scoring, and difficulty calibration. This phase ensures that scenarios are executable, appropriately challenging, and free from generation artifacts that could bias evaluation results. Validation is purely automated using compilation, testing, and metrics, no LLM involvement to prevent bias.

\textbf{Phase 5: LLM Evaluation and Scoring}
We evaluate state-of-the-art models using our comprehensive 17-metric framework across 4 dimensions: Software Engineering Excellence (8 metrics), Functional Correctness (4 metrics), Code Quality Assessment (3 metrics), and Long-Context Utilization (2 metrics). The Software Engineering Excellence dimension includes Architectural Coherence Score (ACS), Dependency Traversal Accuracy (DTA), Cross-File Reasoning Depth (CFRD), System Thinking Score (STS), Robustness Score (RS), Comprehensiveness Score (CS), Innovation Score (IS), and Solution Elegance Score (SES). Functional Correctness comprises Compilation Success, Unit Test Performance, Integration Test Performance, and Incremental Development Capability (IDC). Code Quality Assessment includes Security Analysis Score, Average Issues Found (inverted), and Code Style Adherence. Long-Context Utilization features Information Coverage Utilization (ICU) and Multi-Session Memory Retention (MMR). These metrics are combined into a \textbf{LoCoBench Score (LCBS)} using weighted components: Software Engineering Excellence (40\%), Functional Correctness (30\%), Code Quality Assessment (20\%), and Long-Context Utilization (10\%).

\begin{figure*}[t]
\centering
\begin{minipage}{0.48\textwidth}
    \centering
    \includegraphics[width=\textwidth]{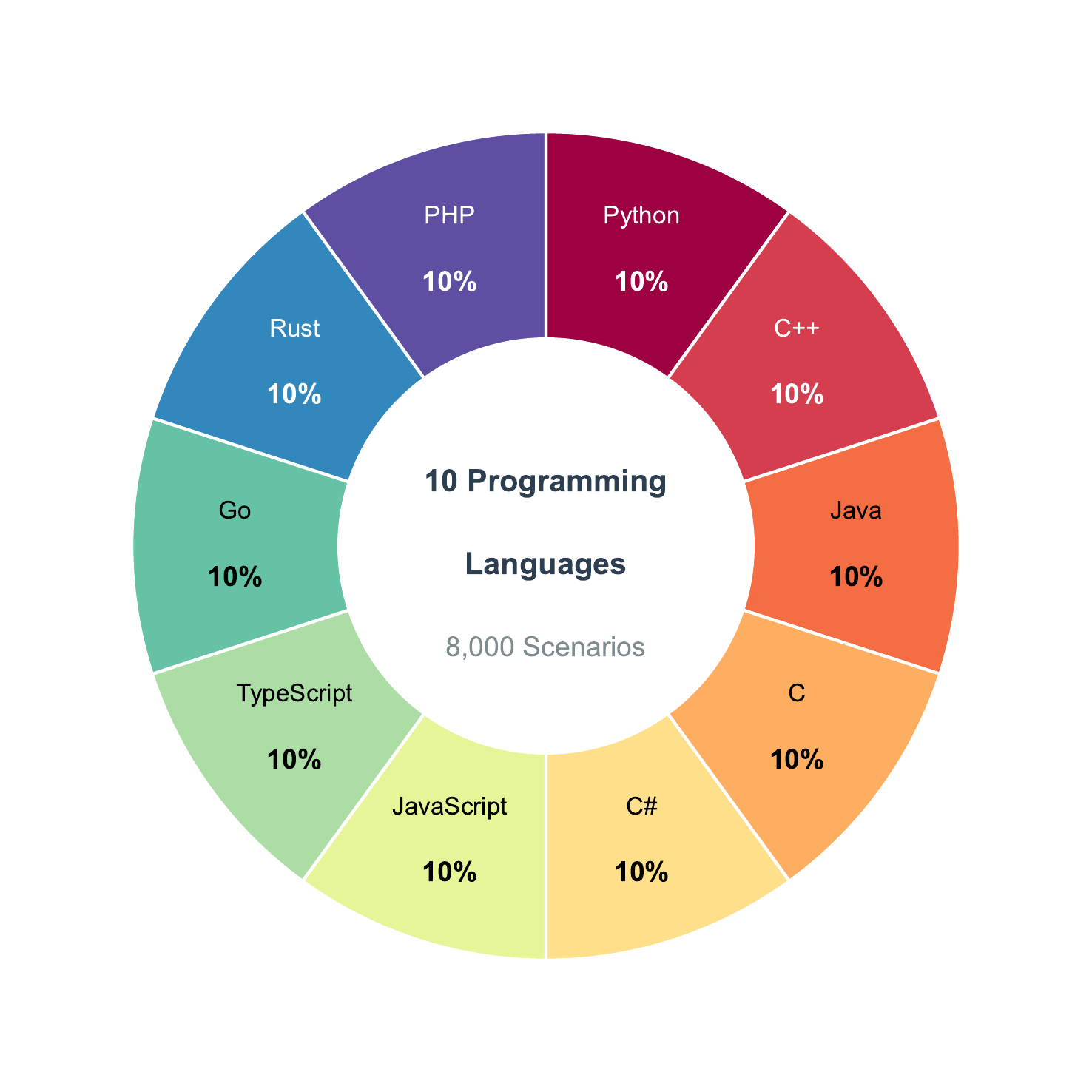}
\end{minipage}
\hfill
\begin{minipage}{0.48\textwidth}
    \centering
    \includegraphics[width=\textwidth]{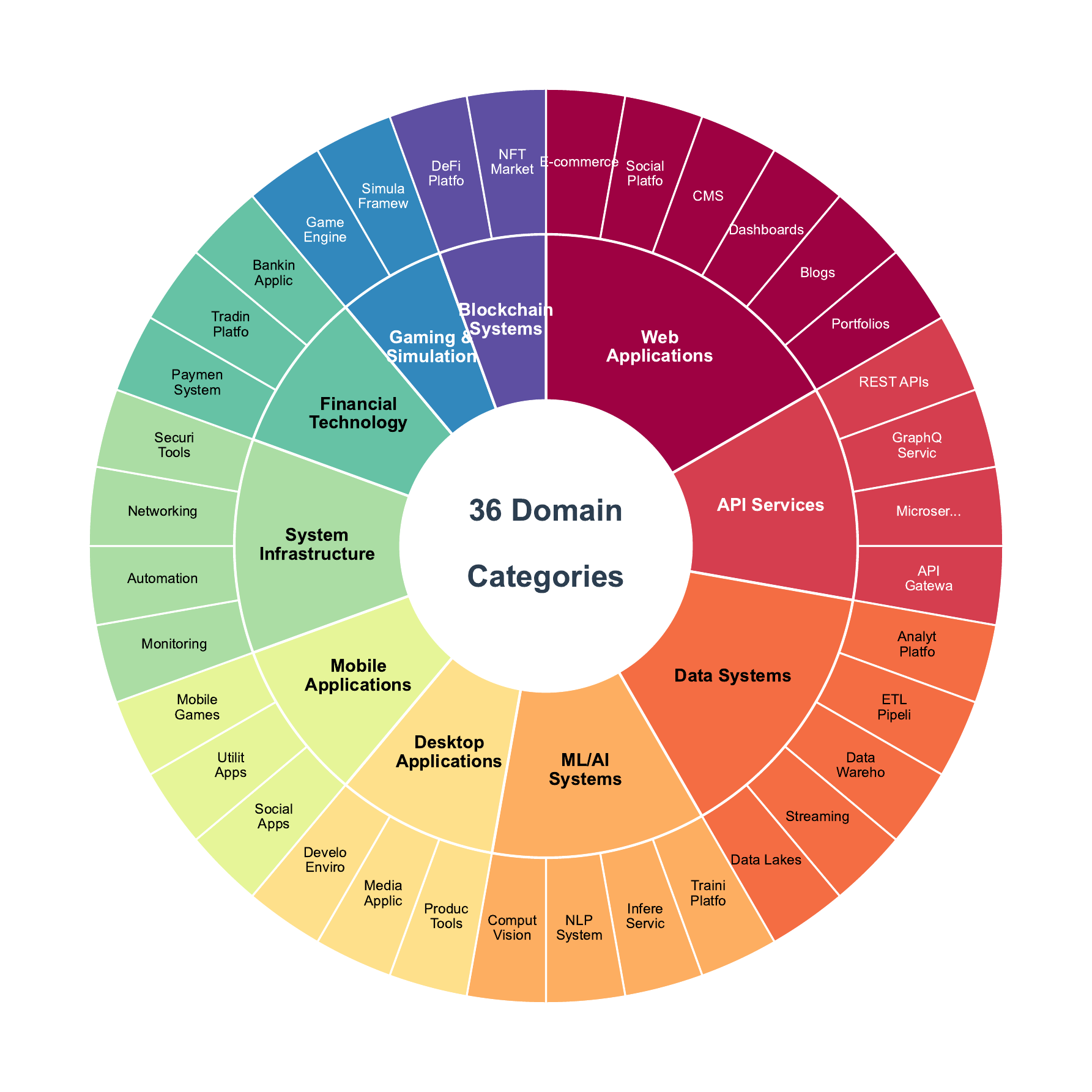}
\end{minipage}
\caption{LoCoBench Coverage Overview. \textbf{Left:} Programming language distribution showing equal representation (10\% each) across 10 languages spanning diverse paradigms from systems programming (C, C++, Rust) to web development (JavaScript, TypeScript, PHP) to enterprise applications (Java, C\#) to modern languages (Go, Python). \textbf{Right:} Hierarchical domain organization with 36 sub-categories grouped into 10 main categories, ensuring comprehensive coverage across web applications, API services, data systems, ML/AI systems, desktop applications, mobile applications, system infrastructure, financial technology, gaming \& simulation, and blockchain systems.}
\vspace{-5pt}
\label{fig:languages_and_categories}
\end{figure*}

\subsection{Task Categories and Long-Context Capabilities}

LoCoBench evaluates eight distinct task categories that capture essential long-context software development capabilities:
\begin{itemize}
    \item \textit{Architectural Understanding:} Scenarios that require LLMs to comprehend complex system designs, identify architectural patterns, and understand component relationships across large codebases.
    \vspace{-5pt}
    \item \textit{Cross-File Refactoring:} Tasks involving code restructuring across multiple files while maintaining functionality and preserving architectural constraints.
    \vspace{-5pt}
    \item \textit{Feature Implementation:} Complex feature development scenarios that require understanding existing code, planning implementation strategies, and integrating new functionality seamlessly.
    \vspace{-5pt}
    \item \textit{Bug Investigation:} Systematic debugging tasks that require analyzing error patterns, tracing execution flows, and identifying root causes across multi-file systems.
    \vspace{-5pt}
    \item \textit{Multi-Session Development:} Scenarios that test long-term memory and context retention across multiple development sessions, simulating realistic project workflows.
    \vspace{-5pt}
    \item \textit{Code Comprehension:} Tasks focused on understanding large, complex codebases and extracting relevant information for development decisions.
    \vspace{-5pt}
    \item \textit{Integration Testing:} Scenarios involving testing component interactions, validating system integration, and ensuring end-to-end functionality.
    \vspace{-5pt}
    \item \textit{Security Analysis:} Tasks requiring identification of security vulnerabilities, assessment of threat vectors, and implementation of security best practices. 
\end{itemize}

\begin{table}[tp]\centering
\caption{8 task categories and details.}
\vspace{-5pt}
\small
\begin{adjustbox}{width=0.95\linewidth}
\begin{tabular}{lp{12cm}}
\toprule
Domains & Details   \\
\midrule
Architectural Understanding & 
Design pattern recognition, dependency analysis, 
System design comprehension, component relationships across large codebases \\
Cross-File Refactoring &
Multi-file restructuring and pattern application, 
Code restructuring across multiple files while maintaining functionality \\
Feature Implementation &
Complex feature development in existing systems,
Understanding existing code, planning implementation strategies, seamless integration \\
Bug Investigation &
Systematic debugging across complex codebases,
Error pattern analysis, execution flow tracing, root cause identification \\
Multi-Session Development &
Context persistence across development sessions,
Long-term memory and incremental building, simulating realistic project workflows \\
Code Comprehension &
Large codebase understanding and explanation,
Information extraction for development decisions, deep codebase analysis \\
Integration Testing &
System-level testing and validation,
Component interaction testing, end-to-end functionality validation \\
Security Analysis &
Security vulnerability assessment,
Threat vector identification, security best practices implementation \\
\bottomrule
\end{tabular}
\end{adjustbox}
\label{table:Domain-Categories}
\vspace{-5pt}
\end{table}

\subsection{Difficulty Calibration and Context Scaling}

Our benchmark systematically varies difficulty across four levels (easy, medium, hard, expert) with corresponding context length ranges:

\begin{itemize}
\item \underline{Easy (10K-100K tokens):} Basic long-context tasks with small to medium codebases.
\vspace{-5pt}
\item \underline{Medium (100K-200K tokens):} Intermediate complexity with larger codebases.
\vspace{-5pt}
\item \underline{Hard (200K-500K tokens):} Advanced scenarios with enterprise-scale codebases.
\vspace{-5pt}
\item \underline{Expert (500K-1M tokens):} Maximum complexity with massive enterprise systems. 
\end{itemize}

This systematic scaling allows precise evaluation of model capabilities as context length increases, providing insights into long-context performance degradation and capabilities.

\subsection{Quality Assurance and Validation}

Every generated scenario undergoes rigorous quality assurance: 
\ding{182}  \textbf{Automated Validation:} All code is validated for compilation, execution, and basic functionality through automated testing pipelines using language-specific compilers (gcc, javac, python, etc.) and testing frameworks.
\ding{183} \textbf{Complexity Metrics:} We employ cyclomatic complexity analysis, dependency depth measurement, and architectural coherence scoring to ensure appropriate difficulty calibration. Scenarios are automatically filtered if complexity metrics fall outside target ranges for their difficulty level.
\ding{184} \textbf{Information Coverage:} Each scenario's information coverage ratio is calculated to ensure sufficient context for task completion while avoiding information redundancy. We target coverage ratios >0.7 for all scenarios.
\ding{185} \textbf{Bias Detection:} Automated analysis identifies and filters scenarios with potential biases, generation artifacts, or unrealistic patterns that could skew evaluation results. This includes detection of repeated code patterns, unrealistic naming conventions, and generation-specific artifacts.

\subsection{Benchmark Statistics and Scale}

LoCoBench represents the largest and most comprehensive evaluation framework for long-context software development to date. Our systematic generation approach produces unprecedented scale and diversity:
\ding{182} \underline{8,000 evaluation scenarios} across 8 task categories.
\ding{183} \underline{1,000 synthetic projects} spanning 36 domain categories. 
\ding{184} \underline{10 programming languages} with balanced coverage.
\ding{185} \underline{Context range} from 10K to 1M tokens (100× variation).
\ding{186} \underline{50,000+ generated files} with realistic code patterns.
\ding{187} \underline{Systematic difficulty distribution} across 4 complexity levels.

\textbf{Language Distribution:} Our benchmark provides balanced coverage across diverse programming paradigms with each language contributing equally (10\%) to our 8,000 scenarios. Languages span from systems programming (C, C++, Rust) to web development (JavaScript, TypeScript, PHP), enterprise applications (Java, C\#), and modern data science/AI frameworks (Python, Go). This equal distribution ensures comprehensive evaluation across different language characteristics while avoiding bias toward any particular programming paradigm.

\begin{table}[htp]\centering
\caption{10 Programming Languages with example usage cases.}
\vspace{-5pt}
\small
\begin{adjustbox}{width=0.7\linewidth}
\begin{tabular}{ll}
\toprule
Programming Language & Usage Cases \\
\midrule
Python  & AI/ML dominance, automation, data science \\
C++ & High-performance, games, embedded systems \\
Java & Enterprise, Android, backend services \\
C & Systems programming, OS development, embedded \\
C\# & Enterprise, Windows, .NET ecosystem \\
JavaScript  & Web development, full-stack \\
TypeScript  & Enterprise web, type safety \\
Go  & Cloud-native, microservices \\
Rust  & Systems, security, memory safety \\
PHP & Web backends, legacy systems \\
\bottomrule
\end{tabular}
\end{adjustbox}
\label{table:Programming-Languages}
\end{table}

\textbf{Domain Coverage:} Projects span 36 distinct domains including web applications (ecommerce, social, dashboard, blog, CMS, portfolio), machine learning systems (training, inference, computer vision, NLP), data processing (analytics, ETL, streaming, warehousing), system utilities (networking, security, monitoring, automation), APIs (REST, GraphQL, microservices, gateway), financial technology (banking, payments, trading), gaming (engine, simulation), blockchain (DeFi, NFT), and mobile applications (utility, social, gaming).

\begin{table}[htp]\centering
\caption{36 domain categories grouped into 10 main domains.}
\vspace{-5pt}
\small
\begin{adjustbox}{width=0.9\linewidth}
\begin{tabular}{llc}
\toprule
Domains & Sub-Domains & Total  \\
\midrule
Web Applications & E-commerce, Social Platforms, CMS, Dashboards, Blogs, Portfolios &6 \\
API Services & REST APIs, GraphQL Services, Microservices, API Gateways &4 \\
Data Systems & Analytics Platforms, ETL Pipelines, Data Warehouses, Streaming, Data Lakes &5 \\
ML/AI Systems & Training Platforms, Inference Services, NLP Systems, Computer Vision &4 \\
Desktop Applications &Productivity Tools, Media Applications, Development Environments &3 \\
Mobile Applications & Social Apps, Utility Apps, Mobile Games &3 \\
System Infrastructure & Monitoring, Automation, Networking, Security Tools &4  \\
Financial Technology & Payment Systems, Trading Platforms, Banking Applications &3 \\
Gaming \& Simulation  &Game Engines, Simulation Frameworks &2 \\
Blockchain Systems & DeFi Platforms, NFT Marketplaces &2 \\
\bottomrule
\end{tabular}
\end{adjustbox}
\label{table:Domain-Categories}
\end{table}

\textbf{Complexity Metrics:} Our generated codebases exhibit realistic complexity distributions with cyclomatic complexity scores ranging from 0.3 to 1.0, file counts between 10-100 per project, and documentation ratios exceeding industry standards. Automated validation ensures all code compiles successfully and maintains architectural coherence.

\begin{table}[htp]\centering
\caption{Additional uniqueness factors.}
\vspace{-5pt}
\small
\begin{adjustbox}{width=0.9\linewidth}
\begin{tabular}{lp{10cm}c}
\toprule
Factor & Details & Total  \\
\midrule
Architecture Patterns & Monolithic, Microservices, Serverless, Event-Driven, Layered, Clean Architecture, Hexagonal, MVC, MVVM, Component-Based &10 \\
Project Themes & Business, Education, Healthcare, Entertainment, Productivity, Social, Utility, Creative & 8 \\
Complexity Levels & Easy (25\%), Medium (25\%), Hard (25\%), Expert (25\%) & 4 \\
\bottomrule
\end{tabular}
\end{adjustbox}
\label{table:Factors}
\end{table}

\begin{figure*}[ht]
\centering
\includegraphics[width=\textwidth]{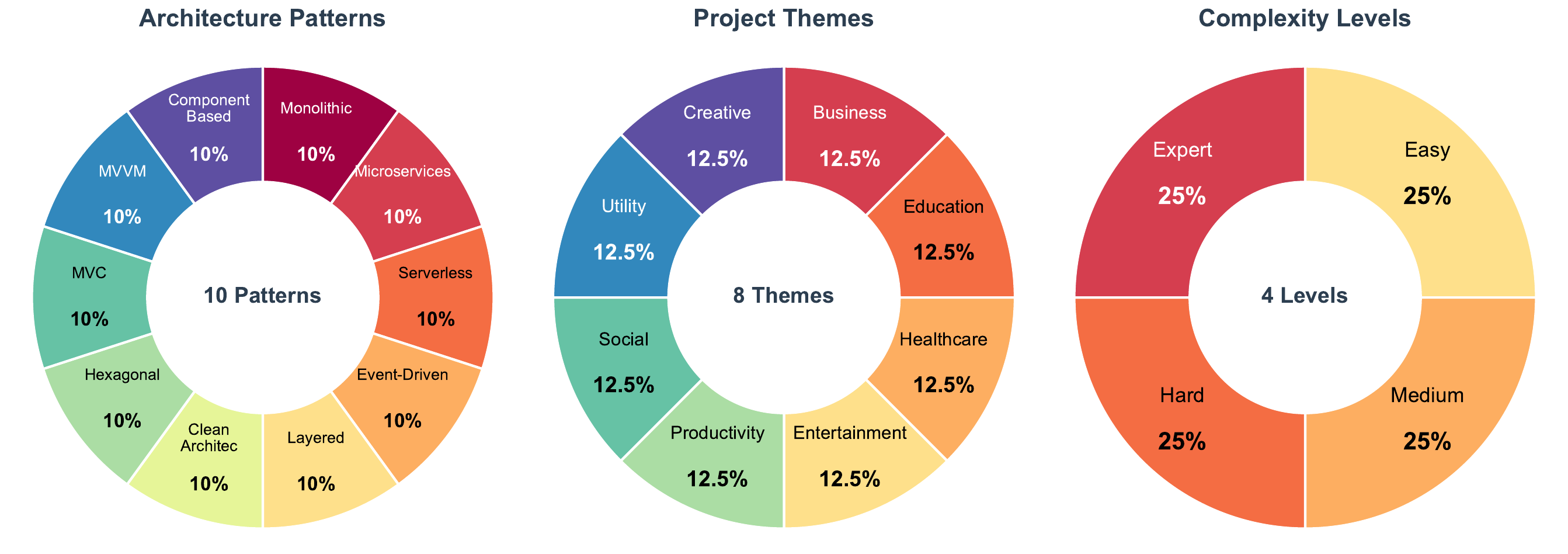}
\caption{Additional uniqueness factors in LoCoBench. Three independent factors provide comprehensive evaluation coverage: \textbf{Left:} 10 architecture patterns including modern paradigms (microservices, serverless, event-driven) and traditional approaches (monolithic, layered, MVC), ensuring evaluation across diverse software architectures. \textbf{Center:} 8 project themes spanning business applications, educational tools, healthcare systems, entertainment platforms, productivity software, social applications, utilities, and creative tools. \textbf{Right:} 4 complexity levels (Easy, Medium, Hard, Expert) with equal 25\% distribution, providing systematic difficulty progression from basic long-context tasks to enterprise-scale challenges.}
\vspace{-10pt}
\label{fig:additional_factors}
\end{figure*}

\begin{figure*}[ht]
\centering
\includegraphics[width=\textwidth]{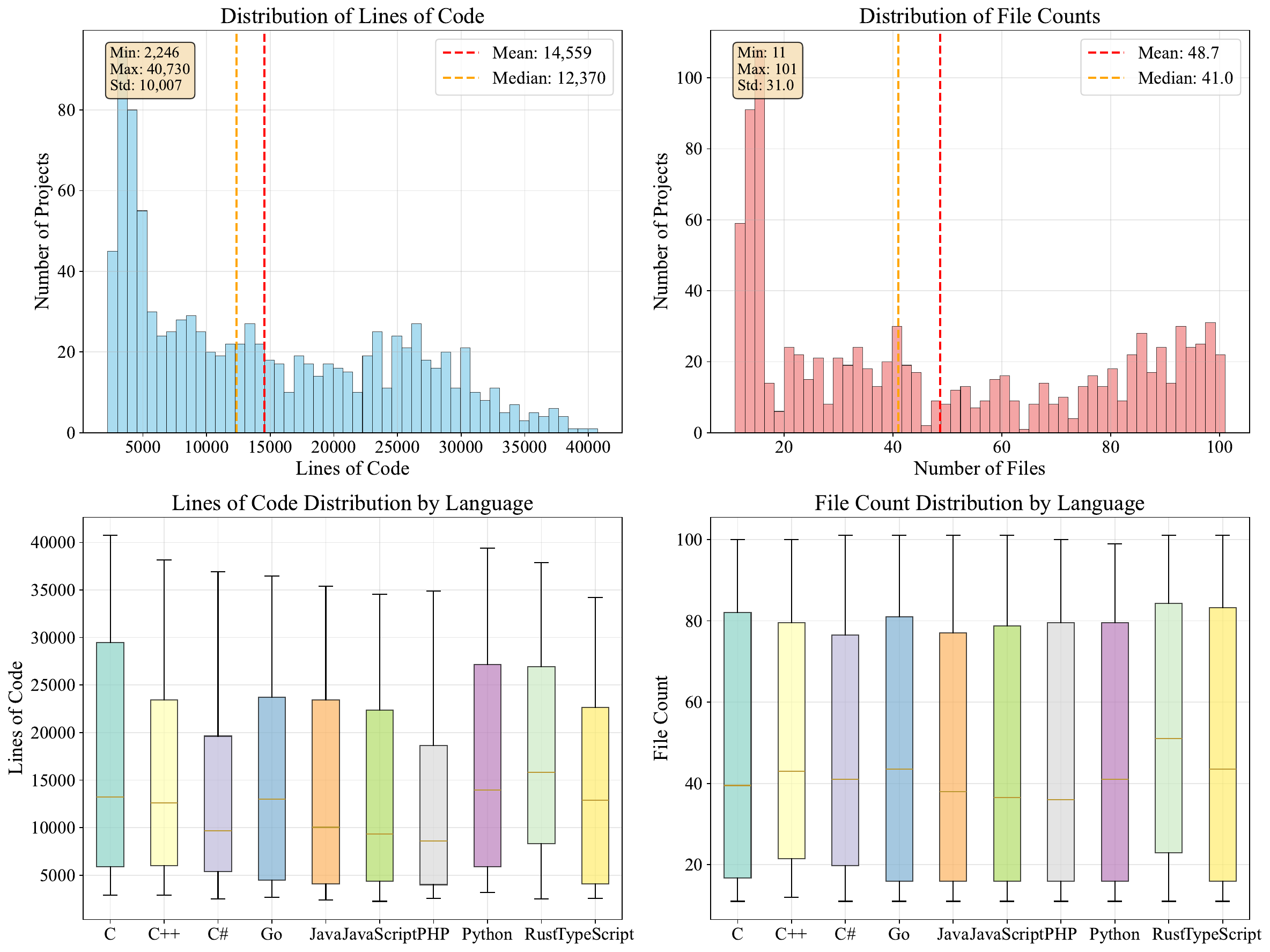}
\caption{LoCoBench's evaluation projects analysis. \textbf{Top row} shows distribution of lines of code (left) and file counts (right) across all evaluation projects. \textbf{Bottom row:} Programming language breakdown displaying lines of code distribution (left) and file count distribution (right) across 10 programming languages.}
\vspace{-5pt}
\label{fig:combined_distributions}
\end{figure*}

\textbf{Line of Code:}  Figure~\ref{fig:combined_distributions} presents the statistical characteristics of LoCoBench's evaluation projects, revealing realistic complexity distributions with a mean of 14,559 lines of code and 48.7 files per project. The right-skewed distributions (top row) mirror real-world software patterns, ranging from compact applications to enterprise-scale systems with over 40,000 lines. The language-specific analysis (bottom row) shows distinct patterns: systems languages (C, Rust) exhibit compact implementations, object-oriented languages (Java, C\#) demonstrate higher complexity with extensive file structures, while web languages (JavaScript, TypeScript, PHP) show intermediate levels. These patterns validate LoCoBench's realistic representation across programming paradigms and complexity levels.

\subsection{Comparison with Existing Benchmarks}

LoCoBench addresses critical limitations in existing code evaluation benchmarks through systematic design choices and comprehensive scope. Table~\ref{tab:benchmark_comparison} provides a comprehensive quantitative comparison highlighting these distinctive features.
While SWE-Bench~\cite{jimenez2023swe} pioneered real-world evaluation using GitHub issues, it remains constrained to Python-only repositories and focuses exclusively on bug-fixing tasks. The benchmark's 2,294 instances provide limited coverage across programming paradigms and development scenarios, failing to capture the diversity of modern software engineering practices.
LongCodeBench~\cite{rando2025longcodebench} introduced long-context evaluation for code but primarily emphasizes code completion and comprehension tasks rather than complex software development workflows. Its focus on single-language evaluation and limited task diversity restricts its ability to assess architectural understanding and multi-file reasoning capabilities essential for enterprise software development.
Despite supporting multiple languages, LongCodeArena~\cite{bogomolov2024long} concentrates on repository-level code completion rather than comprehensive development scenarios. The benchmark lacks systematic evaluation of architectural coherence, cross-file refactoring, and multi-session development workflows that characterize real-world software engineering.
RULER~\cite{hsieh2024ruler} provides valuable long-context evaluation but employs synthetic tasks primarily for natural language processing. Its evaluation paradigm does not capture the unique challenges of software development, including dependency management, architectural consistency, and code quality assessment.

\textbf{LoCoBench's Comprehensive Approach:} Our benchmark uniquely combines: (1) \textit{Multi-language Coverage} across 10 programming languages with equal representation, avoiding language-specific bias; (2) \textit{Complex Task Categories} spanning architectural understanding, cross-file refactoring, and multi-session development that reflect real-world software engineering; (3) \textit{Systematic Context Scaling} from 10K to 1M tokens with 100× variation enabling precise long-context performance analysis; (4) \textit{New Evaluation Metrics} including 6 newly proposed metrics (ACS, DTA, CFRD, ICU, MMR, IDC) specifically designed for long-context capabilities; (5) \textit{Unprecedented Scale} with 8,000 scenarios providing more evaluation instances than the largest existing benchmark while maintaining systematic coverage across difficulty levels and task categories.

\begin{table}[t]
\caption{Comprehensive comparison of LoCoBench with existing benchmarks across important evaluation dimensions. LoCoBench uniquely combines large-scale multi-language evaluation, systematic long-context assessment, complex software engineering tasks, and new metrics specifically designed for long-context capabilities. Columns: Scale - Number of evaluation instances; Languages - Programming language coverage; Context Range - Token length ranges; Task Types - Types of programming tasks; Multi-File - Support for multi-file scenarios; Architecture - Architectural understanding evaluation; New Metrics - New evaluation metrics introduced; Real-World - Real-world applicability. Color-coded symbols:  Green checkmark (\textcolor{green}{\ding{51}}) for full support,  Orange triangle (\textcolor{orange}{$\blacktriangleright$}) for partial support,  Red X (\textcolor{red}{\ding{55}}) for no support.}
\label{tab:benchmark_comparison}
\vspace{-5pt}
\centering
\small
\begin{adjustbox}{width=0.99\linewidth}
\begin{tabular}{lcccccccc}
\toprule
\textbf{Benchmark} & \textbf{Scale} & \textbf{Languages} & \textbf{Context Range} & \textbf{Task Types} & \textbf{Multi-File} & \textbf{Architecture} & \textbf{New Metrics} & \textbf{Real-World} \\
\midrule
HumanEval & 164 & 1 (Python) & Short (<10K) & Algorithm & \textcolor{red}{\ding{55}} & \textcolor{red}{\ding{55}} & \textcolor{red}{\ding{55}} & \textcolor{red}{\ding{55}} \\
SWE-Bench & 2,294 & 1 (Python) & Medium (10-50K) & Bug Fix Only & \textcolor{orange}{$\blacktriangleright$} & \textcolor{red}{\ding{55}} & \textcolor{red}{\ding{55}} & \textcolor{green}{\ding{51}} \\
Multi-SWE-Bench & 1,632 & 7 & Medium (10-50K) & Bug Fix Only & \textcolor{orange}{$\blacktriangleright$} & \textcolor{red}{\ding{55}} & \textcolor{red}{\ding{55}} & \textcolor{green}{\ding{51}} \\
LongCodeBench & 600+ & 1 (Python) & Up to 1M & Completion & \textcolor{orange}{$\blacktriangleright$} & \textcolor{red}{\ding{55}} & \textcolor{red}{\ding{55}} & \textcolor{orange}{$\blacktriangleright$} \\
LongCodeArena & 1,500+ & Multiple & Up to 2M & Completion & \textcolor{green}{\ding{51}} & \textcolor{red}{\ding{55}} & \textcolor{red}{\ding{55}} & \textcolor{orange}{$\blacktriangleright$} \\
DevBench & 200+ & 4 & Short (<10K) & Mixed & \textcolor{orange}{$\blacktriangleright$} & \textcolor{red}{\ding{55}} & \textcolor{red}{\ding{55}} & \textcolor{orange}{$\blacktriangleright$} \\
RULER & 4,000+ & N/A & Up to 128K & NLP Tasks & \textcolor{red}{\ding{55}} & \textcolor{red}{\ding{55}} & \textcolor{red}{\ding{55}} & \textcolor{red}{\ding{55}} \\
\midrule
\textbf{LoCoBench} & \textbf{8,000} & \textbf{10} & \textbf{10K-1M} & \textbf{8 Categories} & \textcolor{green}{\textbf{\ding{51}}} & \textcolor{green}{\textbf{\ding{51}}} & \textbf{6 Metrics} & \textcolor{green}{\textbf{\ding{51}}} \\
\bottomrule
\end{tabular}
\end{adjustbox}
\end{table}

\section{Evaluation Metrics}

\subsection{Metric Overview}

LoCoBench introduces a comprehensive evaluation framework with 17 metrics across 4 dimensions designed to assess capabilities essential for realistic long-context software development scenarios. Our framework combines 6 new metrics specifically designed for long-context LLM evaluation with 11 established metrics adapted from software engineering literature. Table~\ref{tab:metrics_overview} provides a comprehensive overview of all 17 metrics organized by evaluation dimensions.

\begin{table}[ht]
\caption{Complete overview of LoCoBench's 17 evaluation metrics across 4 dimensions. The framework combines 6 new metrics specifically designed for long-context capabilities with 11 established metrics from software engineering literature.}
\label{tab:metrics_overview}
\vspace{-5pt}
\centering
\small
\begin{adjustbox}{width=0.9\linewidth}
\begin{tabular}{llll}
\toprule
\textbf{Dimension} & \textbf{Metric} & \textbf{Abbr.} & \textbf{Source} \\
\midrule
\multirow{8}{*}{\begin{tabular}[c]{@{}l@{}}Software Engineering\\Excellence (8)\end{tabular}} 
& Architectural Coherence Score & ACS & \textcolor{red}{$\bigstar$} \textbf{New} \\
& Dependency Traversal Accuracy & DTA & \textcolor{red}{$\bigstar$} \textbf{New} \\
& Cross-File Reasoning Depth & CFRD & \textcolor{red}{$\bigstar$} \textbf{New} \\
& System Thinking Score & STS & ~\cite{blanchard2016system} \\
& Robustness Score & RS & ~\cite{iso25010} \\
& Comprehensiveness Score & CS & ~\cite{kan2002metrics} \\
& Innovation Score & IS & ~\cite{glass2002research} \\
& Solution Elegance Score & SES & ~\cite{buse2009learning} \\
\midrule
\multirow{4}{*}{\begin{tabular}[c]{@{}l@{}}Functional\\Correctness (4)\end{tabular}} 
& Code Compilation Success & CCS & ~\cite{mccabe1976complexity} \\
& Unit Test Performance & UTP & ~\cite{myers2011art} \\
& Integration Test Performance & ITP & ~\cite{binder1999testing} \\
& Incremental Development Capability & IDC & \textcolor{red}{$\bigstar$} \textbf{New} \\
\midrule
\multirow{3}{*}{\begin{tabular}[c]{@{}l@{}}Code Quality\\Assessment (3)\end{tabular}} 
& Security Analysis Score & SAS & ~\cite{owasp2021top} \\
& Average Issues Found (Inverted) & AIF & ~\cite{campbell2013sonarqube} \\
& Code Style Adherence & CSA & ~\cite{kernighan1999practice} \\
\midrule
\multirow{2}{*}{\begin{tabular}[c]{@{}l@{}}Long-Context\\Utilization (2)\end{tabular}} 
& Information Coverage Utilization & ICU & \textcolor{red}{$\bigstar$} \textbf{New} \\
& Multi-Session Memory Retention & MMR & \textcolor{red}{$\bigstar$} \textbf{New} \\
\bottomrule
\end{tabular}
\end{adjustbox}
\end{table}

\subsection{Software Engineering Excellence (8 metrics)}

This dimension evaluates sophisticated software engineering capabilities essential for complex development scenarios.

\ding{182} \textbf{Architectural Coherence Score (ACS):} We introduce this new metric to evaluate LLMs' ability to maintain system-level design consistency across large codebases. Traditional metrics cannot capture architectural understanding at the scale required for long-context evaluation.

Let $\mathcal{C} = \{c_1, c_2, \ldots, c_n\}$ represent a codebase and $\mathcal{P} = \{p_1, p_2, \ldots, p_m\}$ denote the set of architectural patterns detected in $\mathcal{C}$. For each pattern $p_i \in \mathcal{P}$, we define:

{\small
\begin{equation}
ACS(\mathcal{C}) = \frac{1}{|\mathcal{P}|} \sum_{p \in \mathcal{P}} w(p) \cdot \frac{\alpha(p, \mathcal{C})}{\kappa(p) + \epsilon}
\end{equation}
}

where $w(p) \in [0,1]$ is the criticality weight of pattern $p$, $\alpha(p, \mathcal{C}) \in [0,1]$ measures pattern adherence through SOLID principle compliance and design constraint satisfaction, $\kappa(p) \geq 1$ represents pattern complexity, and $\epsilon > 0$ is a regularization constant preventing division by zero.

\ding{183} \textbf{Dependency Traversal Accuracy (DTA):} This new metric specifically evaluates LLMs' capability to navigate complex inter-module dependencies in long-context scenarios, addressing a key gap in existing evaluation frameworks.

Let $\mathcal{G} = (V, E)$ be the dependency graph where $V$ represents modules and $E$ denotes dependency relationships. For each dependency $d_{ij} \in E$ from module $v_i$ to $v_j$, we define:

{\small
\begin{equation}
DTA(\mathcal{G}) = \frac{1}{|E|} \sum_{d_{ij} \in E} \frac{\mu(d_{ij}) \cdot \gamma(d_{ij}, \mathcal{G})}{\delta(d_{ij}) + 1}
\end{equation}
}

where $\mu(d_{ij}) \in [0,1]$ measures correct usage through import validation and interface compliance, $\gamma(d_{ij}, \mathcal{G}) \in [0,1]$ quantifies contextual awareness of dependency relationships within graph $\mathcal{G}$, and $\delta(d_{ij}) \geq 0$ represents the transitive dependency depth of edge $d_{ij}$.

\ding{184} \textbf{Cross-File Reasoning Depth (CFRD):} We propose this metric to assess LLMs' understanding of multi-file relationships and interactions, a capability crucial for complex software development but not measured by existing benchmarks.

Given a file set $\mathcal{F} = \{f_1, f_2, \ldots, f_n\}$ and the cross-file interaction matrix $\mathbf{R} \in \mathbb{R}^{n \times n}$, we define:

{\small
\begin{equation}
CFRD(\mathcal{F}) = \frac{1}{n(n-1)} \sum_{i=1}^{n} \sum_{\substack{j=1 \\ j \neq i}}^{n} \rho(f_i, f_j) \cdot \iota(f_i, f_j)
\end{equation}
}

where $\rho(f_i, f_j) \in [0,1]$ quantifies reasoning depth between files $f_i$ and $f_j$ through semantic analysis and cross-reference understanding, and $\iota(f_i, f_j) \in [0,1]$ measures interaction complexity based on coupling strength, interface dependencies, and shared abstractions.

\ding{185} {System Thinking Score (STS):} Adapted from systems engineering assessment frameworks~\cite{blanchard2016system}, measuring holistic software system understanding and scalability awareness.

\ding{186} {Robustness Score (RS):} Based on IEEE/ISO 25010 software quality standards~\cite{iso25010}, evaluating code reliability, error handling, and defensive programming practices.

\ding{187} {Comprehensiveness Score (CS):} Derived from software completeness metrics in quality assurance literature~\cite{kan2002metrics}, assessing solution coverage, documentation quality, and requirement fulfillment.

\ding{188} {Innovation Score (IS):} Adapted from creative problem-solving assessment in software engineering research~\cite{glass2002research}, evaluating new approaches, modern practices, and creative solutions.

\ding{189} {Solution Elegance Score (SES):} Based on code aesthetics and design quality metrics~\cite{buse2009learning}, measuring code clarity, theoretical soundness, and adherence to clean code principles.

\subsection{Functional Correctness (4 metrics)}

This dimension assesses the fundamental correctness and executability of generated code.

\ding{182} \textbf{Incremental Development Capability (IDC):} We introduce this metric to evaluate LLMs' ability to build effectively on previous development work across multiple sessions, a crucial capability for long-context software development not addressed by existing metrics.

Let $\mathcal{T} = \{t_1, t_2, \ldots, t_k\}$ represent a sequence of incremental development tasks applied to codebase state transitions $\mathcal{S}_0 \rightarrow \mathcal{S}_1 \rightarrow \cdots \rightarrow \mathcal{S}_k$. For each task $t_i$:

{\small
\begin{equation}
IDC(\mathcal{T}) = \frac{1}{|\mathcal{T}|} \sum_{i=1}^{|\mathcal{T}|} \frac{\xi(t_i, \mathcal{S}_{i-1}) \cdot \sigma(t_i, \mathcal{S}_i)}{\beta(t_i, \mathcal{S}_{i-1}, \mathcal{S}_i) + 1}
\end{equation}
}

where $\xi(t_i, \mathcal{S}_{i-1}) \in [0,1]$ measures extension quality of task $t_i$ relative to previous state $\mathcal{S}_{i-1}$, $\sigma(t_i, \mathcal{S}_i) \in [0,1]$ quantifies integration smoothness in resulting state $\mathcal{S}_i$, and $\beta(t_i, \mathcal{S}_{i-1}, \mathcal{S}_i) \geq 0$ counts breaking changes introduced during the transition.

\ding{183} {Code Compilation Success (CCS):} Binary assessment of syntactic correctness, a fundamental metric established in early software engineering literature~\cite{mccabe1976complexity}.

\ding{184} {Unit Test Performance (UTP):} Individual component testing validation, a standard practice from software testing methodology~\cite{myers2011art}.

\ding{185} {Integration Test Performance (ITP):} System-wide functionality assessment, based on established integration testing frameworks~\cite{binder1999testing}.

\subsection{Code Quality Assessment (3 metrics)}

This dimension evaluates security, maintainability, and adherence to coding standards.

\ding{182} {Security Analysis Score (SAS):} Vulnerability assessment based on OWASP security analysis frameworks~\cite{owasp2021top} and static analysis techniques, evaluating common security issues including SQL injection, XSS, buffer overflows, and insecure cryptographic practices.

\ding{183} {Average Issues Found - Inverted (AIF):} Code quality issue detection derived from static analysis research and modern quality assessment tools~\cite{campbell2013sonarqube}, measuring the absence of code smells, complexity violations, and maintainability issues (lower issue count yields higher score).

\ding{184} {Code Style Adherence (CSA):} Style guide compliance measurement based on coding standards literature~\cite{kernighan1999practice} and automated linting frameworks, evaluating naming conventions, formatting consistency, and language-specific best practices.

\subsection{Long-Context Utilization (2 metrics)}

This dimension specifically evaluates capabilities unique to long-context software development scenarios.

\ding{182} \textbf{Information Coverage Utilization (ICU):} We propose this metric to evaluate how effectively LLMs utilize large context windows, addressing a critical gap in long-context evaluation.

Given context window $\mathcal{W} = \{w_1, w_2, \ldots, w_m\}$ and task-specific information elements $\mathcal{I} = \{i_1, i_2, \ldots, i_n\} \subseteq \mathcal{W}$, we define:

{\small
\begin{equation}
ICU(\mathcal{W}, \mathcal{I}) = \frac{|\mathcal{U}(\mathcal{I})|}{|\mathcal{I}|} \cdot \frac{\sum_{u \in \mathcal{U}(\mathcal{I})} \tau(u)}{\phi(\mathcal{U}(\mathcal{I})) + \epsilon}
\end{equation}
}

where $\mathcal{U}(\mathcal{I}) \subseteq \mathcal{I}$ represents the subset of utilized information elements, $\tau(u) \in [0,1]$ quantifies the task relevance of element $u$, $\phi(\mathcal{U}(\mathcal{I})) \geq 0$ measures redundancy penalty through information overlap, and $\epsilon > 0$ is a regularization constant.

\ding{183} \textbf{Multi-Session Memory Retention (MMR):} This new metric assesses context persistence across extended development sessions, essential for evaluating long-context capabilities in realistic software development workflows.

Consider a sequence of development sessions $\mathcal{S} = \{s_1, s_2, \ldots, s_k\}$ with associated context states $\{\mathcal{C}_1, \mathcal{C}_2, \ldots, \mathcal{C}_k\}$. We define:

{\small
\begin{equation}
MMR(\mathcal{S}) = \frac{1}{|\mathcal{S}|} \sum_{j=1}^{|\mathcal{S}|} \frac{\psi(s_j, \mathcal{C}_{j-1}) \cdot \chi(s_j, \mathcal{C}_j)}{\log(j + 1)}
\end{equation}
}

where $\psi(s_j, \mathcal{C}_{j-1}) \in [0,1]$ measures information retention from previous context state $\mathcal{C}_{j-1}$ to session $s_j$, $\chi(s_j, \mathcal{C}_j) \in [0,1]$ quantifies consistency maintenance in current context state $\mathcal{C}_j$, and the logarithmic decay term $\log(j + 1)$ models expected memory degradation over temporal distance.

\subsection{LoCoBench Score (LCBS)}

We define a unified LoCoBench Score (LCBS) as a weighted aggregate function that maps the 17-dimensional metric space to a scalar evaluation score. Let $\mathcal{M} = \{m_1, m_2, \ldots, m_{17}\}$ represent the complete set of evaluation metrics, partitioned into four evaluation dimensions.

\vspace{5pt}
\textbf{Dimension Partitioning:} The metric space is partitioned as:
{\small
\begin{align}
\mathcal{M}_{SE} &= \{ACS, DTA, CFRD, STS, RS, CS, IS, SES\} \quad |\mathcal{M}_{SE}| = 8 \\
\mathcal{M}_{FC} &= \{CCS, UTP, ITP, IDC\} \quad |\mathcal{M}_{FC}| = 4 \\
\mathcal{M}_{CQ} &= \{SAS, AIF, CSA\} \quad |\mathcal{M}_{CQ}| = 3 \\
\mathcal{M}_{LCU} &= \{ICU, MMR\} \quad |\mathcal{M}_{LCU}| = 2
\end{align}
}

where $\mathcal{M}_{SE} \cup \mathcal{M}_{FC} \cup \mathcal{M}_{CQ} \cup \mathcal{M}_{LCU} = \mathcal{M}$ and the sets are pairwise disjoint.

\vspace{5pt}
\textbf{Normalization Function:} For each metric $m_i \in \mathcal{M}$, we define a normalization function $\mathcal{N}: \mathbb{R} \rightarrow [0,1]$ that maps raw metric values to the unit interval:

{\small
\begin{equation}
\mathcal{N}(m_i) = \frac{m_i - \min(m_i)}{\max(m_i) - \min(m_i)}
\end{equation}
}

\textbf{Dimension Aggregation:} Each dimension score is computed as the arithmetic mean of its constituent normalized metrics:

{\small
\begin{align}
SE &= \frac{1}{|\mathcal{M}_{SE}|} \sum_{m \in \mathcal{M}_{SE}} \mathcal{N}(m) = \frac{1}{8} \sum_{i=1}^{8} \mathcal{N}(m_i^{SE}) \\
FC &= \frac{1}{|\mathcal{M}_{FC}|} \sum_{m \in \mathcal{M}_{FC}} \mathcal{N}(m) = \frac{1}{4} \sum_{i=1}^{4} \mathcal{N}(m_i^{FC}) \\
CQ &= \frac{1}{|\mathcal{M}_{CQ}|} \sum_{m \in \mathcal{M}_{CQ}} \mathcal{N}(m) = \frac{1}{3} \sum_{i=1}^{3} \mathcal{N}(m_i^{CQ}) \\
LCU &= \frac{1}{|\mathcal{M}_{LCU}|} \sum_{m \in \mathcal{M}_{LCU}} \mathcal{N}(m) = \frac{1}{2} \sum_{i=1}^{2} \mathcal{N}(m_i^{LCU})
\end{align}
}

\textbf{Weight Vector:} We define the weight vector $\mathbf{w} = [w_{SE}, w_{FC}, w_{CQ}, w_{LCU}]^T$ where:
{\small
\begin{equation}
\mathbf{w} = [0.4, 0.3, 0.2, 0.1]^T \quad \text{such that} \quad \sum_{i} w_i = 1
\end{equation}
}

The weights are empirically determined to reflect the relative importance of each dimension in long-context software development scenarios, with software engineering excellence receiving the highest weight due to its comprehensive nature.

\vspace{5pt}
\textbf{Final Score:} The LoCoBench Score is defined as a weighted linear combination scaled to the interval [0,5]:

{\small
\begin{equation}
LCBS = 5 \cdot \mathbf{w}^T \cdot [SE, FC, CQ, LCU]^T = 5 \cdot (w_{SE} \cdot SE + w_{FC} \cdot FC + w_{CQ} \cdot CQ + w_{LCU} \cdot LCU)
\end{equation}
}

Substituting the weight values:
{\small
\begin{equation}
LCBS = 5 \cdot (0.4 \cdot SE + 0.3 \cdot FC + 0.2 \cdot CQ + 0.1 \cdot LCU)
\end{equation}
}

\section{Experiments, Results and Discussions}

\subsection{Evaluation Infrastructure and Process}

LoCoBench provides a comprehensive evaluation infrastructure designed for reliable, scalable assessment:
\ding{182} \textbf{Model Integration:} Our framework supports evaluation of any long-context LLM through standardized APIs, including OpenAI GPT, Google Gemini, and Anthropic Claude models. Each model is evaluated with consistent hyperparameters to ensure reproducible results.
\ding{183} \textbf{Context Management:} Advanced context windowing techniques handle scenarios exceeding model context limits, with intelligent truncation strategies that preserve essential information while maintaining task solvability.
\ding{184} \textbf{Execution Environment:} Isolated Docker containers provide secure execution environments for code validation, with language-specific toolchains and timeout mechanisms (3600 seconds per evaluation) to prevent infinite loops or resource exhaustion.
\ding{185} \textbf{Error Recovery:} Robust error handling addresses common evaluation challenges including parsing failures, compilation errors, and runtime exceptions, with detailed logging for debugging and analysis.

\subsection{Overall Model Performance Analysis}

\begin{figure}[ht]
\centering
\includegraphics[width=0.6\textwidth]{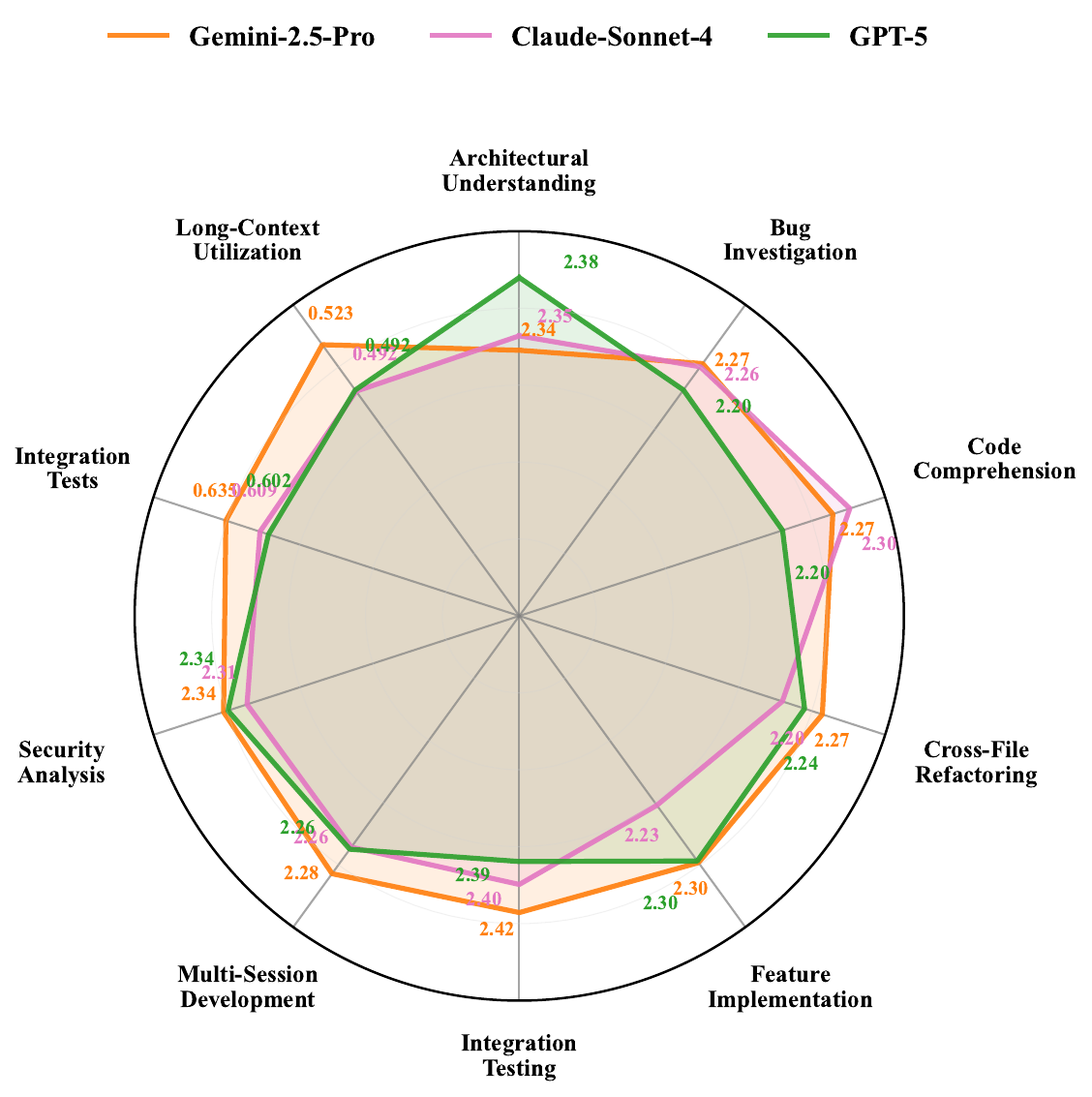}
\caption{Overall performance comparison of GPT-5, Gemini-2.5-Pro, and Claude-Sonnet-4 across 10 LoCoBench dimensions. Gemini-2.5-Pro demonstrates superior performance on many aspects, particularly on cross-file refactoring, long-context utilization, integration tests, and multi-session development capabilities, while GPT-5 excels in architectural understanding. Claude-Sonnet-4 shows balanced performance with particular strength in code comprehension.}
\vspace{-10pt}
\label{fig:overall_performance_radar}
\end{figure}

Figure~\ref{fig:overall_performance_radar} compares the performance of three leading LLMs across 10 evaluation dimensions, showing distinct performance profiles that reflect different architectural strengths and optimization strategies. Gemini-2.5-Pro emerges as the overall leader, demonstrating exceptional performance in cross-file refactoring, long-context utilization, integration tests, and multi-session development. This model shows particular strength in complex software engineering tasks that require deep system-level reasoning and comprehensive testing capabilities. Its superior performance suggests strong capabilities for comprehending large-scale software designs and identifying structural patterns across extensive codebases.

GPT-5 achieves competitive performance, showing remarkable consistency across most evaluation dimensions. Notably, GPT-5 demonstrates the highest performance in architectural understanding, indicating specialized capabilities for recognizing and analyzing complex software design patterns. This strength in architectural comprehension suggests that GPT-5 may be particularly well-suited for tasks involving system design analysis and high-level software architecture evaluation. Claude-Sonnet-4 presents a distinctive performance profile, showing particular excellence in code comprehension, which indicates strong capabilities for understanding and analyzing existing codebases.

Figure~\ref{fig:overall_performance_radar} shows that all three models achieve relatively similar performance levels across many dimensions, with the largest performance gaps occurring in specialized areas such as long-context utilization and specific task categories. This convergence suggests that current state-of-the-art models have reached similar competency levels for basic long-context software development tasks, while differentiation occurs primarily in specialized capabilities requiring domain-specific reasoning patterns. The custom per-axis scaling employed in the visualization effectively highlights these subtle but important performance differences that would be obscured by uniform scaling approaches.

Interestingly, the performance patterns suggest that different models may have been optimized for different aspects of software development workflows. The variation in long-context utilization capabilities across models indicates that handling extended context windows remains a significant technical challenge, with different approaches yielding varying degrees of success. This specialization pattern has important implications for practical deployment, as organizations may benefit from selecting models based on their specific software development needs and the types of long-context tasks they most frequently encounter.
The relatively tight performance clustering among these top-tier models also suggests that the field of long-context code understanding is approaching certain fundamental limitations with current architectures and training methodologies. Future improvements may require new approaches to context management, architectural understanding, and multi-file reasoning rather than incremental refinements to existing techniques.

\subsection{Comprehensive Model Ranking Analysis}

\begin{figure}[ht]
\centering
\includegraphics[width=\textwidth]{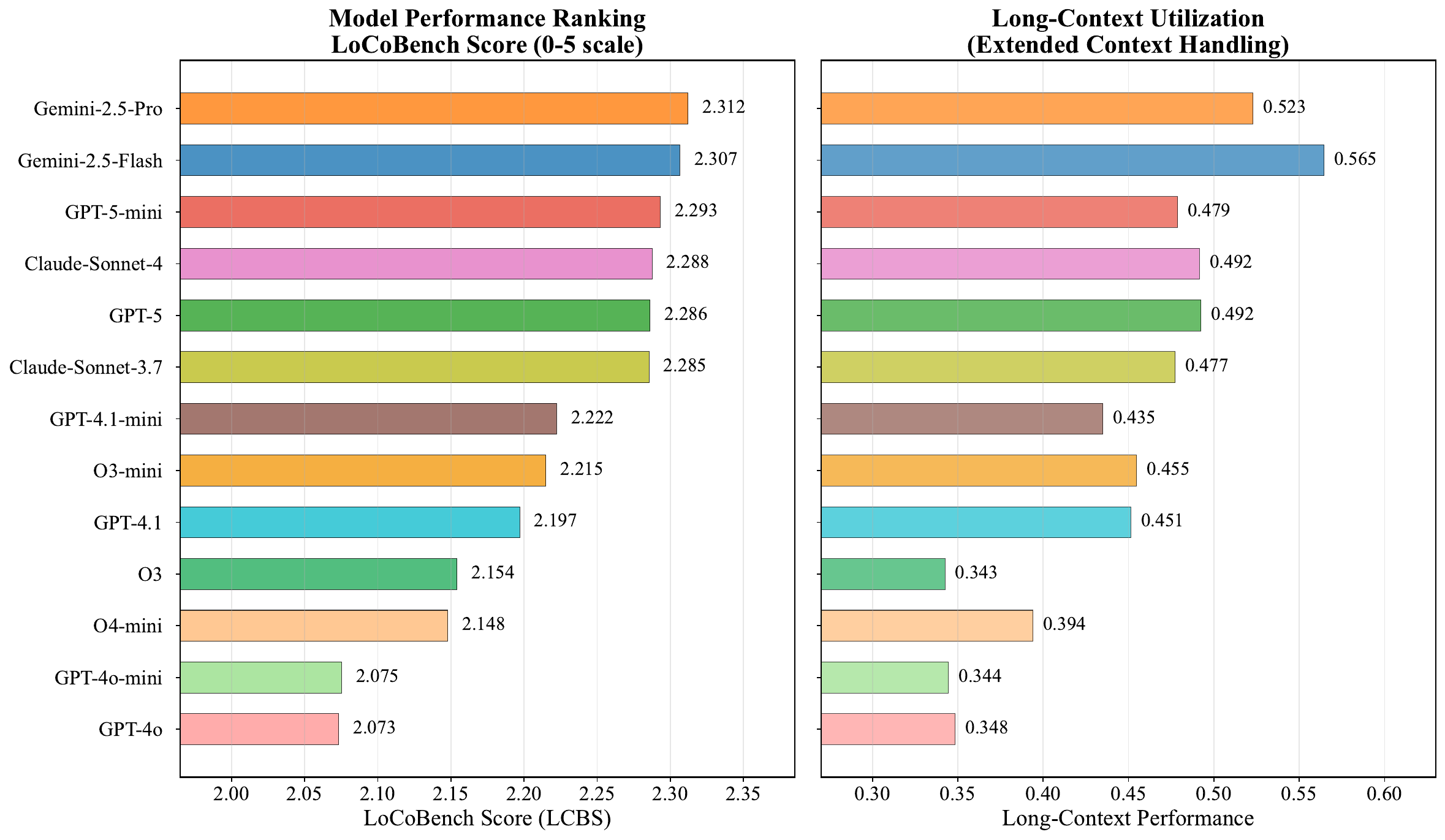}
\caption{Model ranking and long-context utilization comparison. Left chart shows LoCoBench Score (LCBS) rankings. Right chart displays long-context utilization performance.}
\vspace{-5pt}
\label{fig:model_ranking_bar}
\end{figure}

Figure~\ref{fig:model_ranking_bar} presents a comprehensive ranking of all evaluated models across two dimensions: general software engineering competency and specialized long-context processing capabilities. The left chart displays overall LoCoBench Score (LCBS) performance, revealing a clear performance hierarchy with Gemini-2.5-Pro achieving the highest score. The performance distribution shows relatively tight clustering among top-tier models, indicating that leading models have achieved similar competency levels for complex software development tasks. This clustering pattern suggests that the current generation of large language models has reached a plateau in general software engineering capabilities, with incremental improvements rather than dramatic performance leaps.

The right chart focuses specifically on long-context utilization capabilities, revealing markedly different performance patterns compared to overall rankings. Gemini-2.5-Flash demonstrates superior long-context processing abilities, suggesting specialized optimization for extended context handling that may come at the expense of other capabilities. This divergence between overall performance and long-context specialization highlights the distinct challenges posed by extended context scenarios versus general software engineering tasks. The performance gap in long-context utilization is notably larger than in overall scores, indicating that effective context management remains a significant technical challenge requiring specialized architectural solutions.

The dual visualization reveals that model performance varies significantly between comprehensive software engineering evaluation and specialized long-context capabilities. While some models excel in overall software development competency, others show particular strength in processing and utilizing extended context information, suggesting different architectural optimizations and training strategies across model families. This specialization pattern reflects the inherent trade-offs in model design, where optimization for specific capabilities may impact performance in other areas.

The model ranking also demonstrates the importance of evaluating models across multiple dimensions rather than relying on single aggregate scores. Models that appear similar in overall performance may exhibit substantial differences in specific capabilities that are crucial for particular use cases. For organizations deploying these models in production environments, understanding these performance trade-offs is essential for selecting the most appropriate model for their specific long-context software development requirements.

Furthermore, the performance distribution across all evaluated models reveals a clear stratification, with distinct performance tiers emerging. This stratification suggests that while the top-performing models are relatively close in capability, there remain significant gaps between different model generations and architectures. The lower-performing models in the ranking may still be suitable for specific applications or resource-constrained environments, highlighting the importance of comprehensive evaluation frameworks like LoCoBench for understanding the full spectrum of model capabilities.

\subsection{Programming Language Performance Analysis}

\begin{figure}[ht]
\centering
\includegraphics[width=\textwidth]{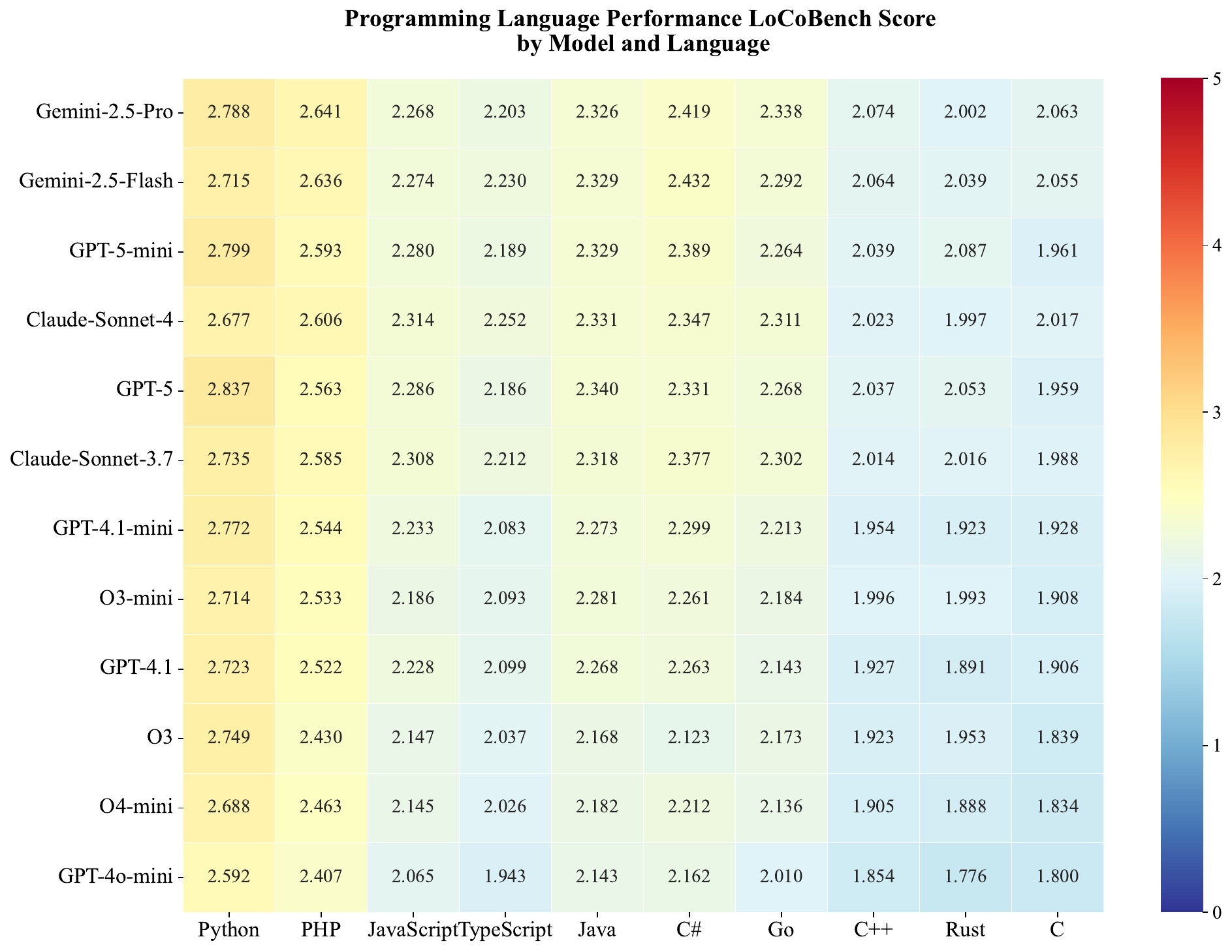}
\caption{Programming language performance heatmap showing model performance across 10 programming languages. Languages are ordered by difficulty from easiest to hardest.}
\label{fig:language_heatmap}
\end{figure}

Figure~\ref{fig:language_heatmap} presents a comprehensive analysis of model performance across 10 programming languages, revealing distinct patterns in language-specific capabilities. The heatmap visualization shows clear performance variations across different programming paradigms, with models demonstrating varying proficiency levels depending on language characteristics and complexity.

The analysis reveals that models generally achieve higher performance on high-level languages such as Python and PHP, while showing more challenging performance patterns on systems programming languages like C and Rust. This performance gradient reflects the inherent complexity differences between languages and the varying amounts of training data typically available for different programming languages. The ordering from easiest to hardest languages demonstrates a consistent difficulty progression that aligns with traditional programming language learning curves and industry adoption patterns.

Language-specific performance patterns indicate that model training and optimization strategies may be influenced by language popularity and representation in training datasets. The consistent performance ordering across most models suggests systematic challenges posed by certain language features, such as memory management in systems languages and complex type systems in modern programming languages. Notably, web development languages like JavaScript and TypeScript show intermediate performance levels, reflecting their moderate complexity and widespread usage in training corpora.

Figure~\ref{fig:language_heatmap} also reveals interesting model-specific strengths and weaknesses across languages. While most models follow similar performance trends, some demonstrate particular proficiency in specific language domains, suggesting that certain architectural choices or training methodologies may be more effective for particular programming paradigms. This language-dependent performance variation has important implications for practical deployment, as organizations working primarily with specific programming languages may benefit from selecting models that demonstrate superior performance in their target language ecosystem.

Furthermore, the performance patterns observed across languages provide insights into the fundamental challenges of long-context code understanding. Systems programming languages, which typically require more precise memory management and lower-level reasoning, consistently pose greater challenges across all evaluated models. This suggests that current long-context LLMs may struggle with the detailed, hardware-aware reasoning required for effective systems programming, highlighting an important area for future model development and training optimization.

\subsection{Task Category Performance and Difficulty Analysis}

\begin{figure}[ht]
\centering
\includegraphics[width=\textwidth]{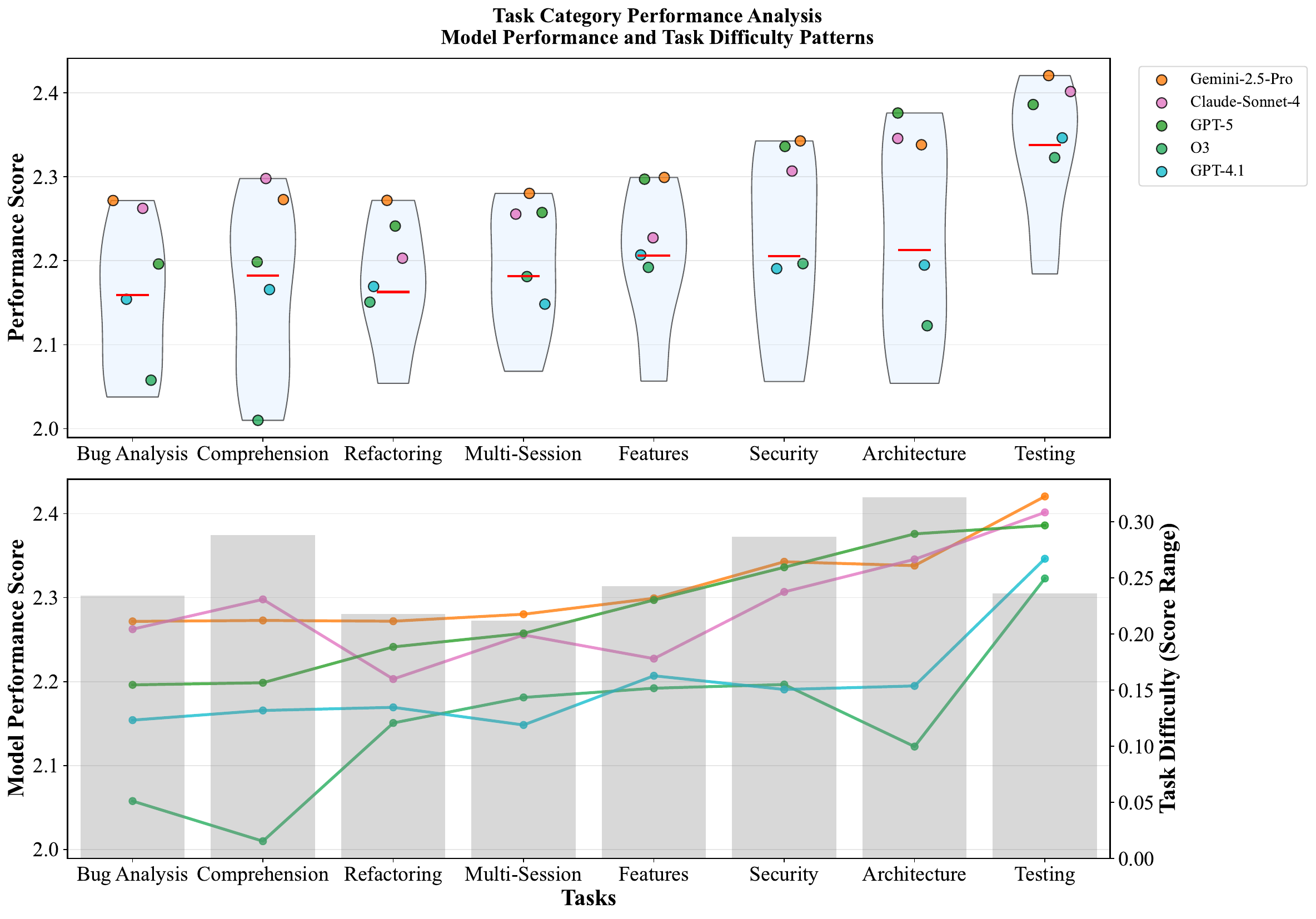}
\caption{Task category performance analysis. Top chart shows performance distribution across all models for each task category, with individual model performance overlaid. Bottom chart displays task difficulty patterns and model performance trends across different software engineering tasks.}
\label{fig:task_category_analysis}
\vspace{-10pt}
\end{figure}

Figure~\ref{fig:task_category_analysis} presents a comprehensive analysis of model performance across eight distinct task categories, revealing both individual model capabilities and inherent task difficulty patterns. The visualization shows overall performance distributions with detailed model-specific analysis, providing insights into both the challenges posed by different software engineering tasks and the varying capabilities of evaluated models.

The top chart shows the complete performance distribution across all evaluated models for each task category, with individual model performances overlaid as scatter points. This figure reveals significant variations in task difficulty, with some categories showing wide performance distributions indicating high variability in model capabilities, while others demonstrate more consistent performance patterns. The plot provide insights into the underlying performance characteristics, with broader distributions indicating tasks where models show more varied success rates, and narrower distributions suggesting more consistent challenge levels across different model architectures.

The bottom chart focuses on task difficulty analysis by displaying the performance range (score variance) for each task category as background bars, while overlaying individual model performance trends as connected line plots. This dual-axis approach effectively illustrates the relationship between inherent task difficulty and model-specific performance patterns. Tasks with larger score ranges indicate greater difficulty variation among models, suggesting that these tasks may be more sensitive to specific architectural optimizations or training methodologies.

The analysis reveals distinct performance patterns across task categories, with integration testing and architectural understanding generally showing higher performance scores, while tasks such as bug investigation and multi-session development present greater challenges for most models. This performance hierarchy reflects the varying complexity of different software engineering activities, with some tasks requiring more sophisticated reasoning capabilities or longer-context understanding than others. The consistent ordering of task difficulty across most models suggests that certain software engineering challenges are fundamentally more difficult for current long-context LLMs, regardless of their specific architectural approaches.

Model-specific performance patterns also emerge from the analysis, with some models demonstrating particular strengths in specific task categories while showing relative weaknesses in others. This specialization pattern indicates that different models may have been optimized for different aspects of software engineering workflows, or that their training data may have contained varying representations of different task types. The performance variations across tasks have important implications for practical deployment, as organizations may benefit from selecting models based on the specific types of software engineering tasks they most frequently encounter.

Figure~\ref{fig:task_category_analysis} shows the importance of considering both absolute performance levels and performance consistency when evaluating models for long-context software development tasks. Tasks that show high performance variance may require more careful model selection and potentially different evaluation strategies, while tasks with consistent performance patterns across models may be more predictable in production environments. This analysis framework provides valuable insights for both model developers seeking to improve specific capabilities and practitioners selecting appropriate models for their software development workflows.

\subsection{Context Length and Difficulty Impact Analysis}

\begin{figure}[ht]
\centering
\includegraphics[width=\textwidth]{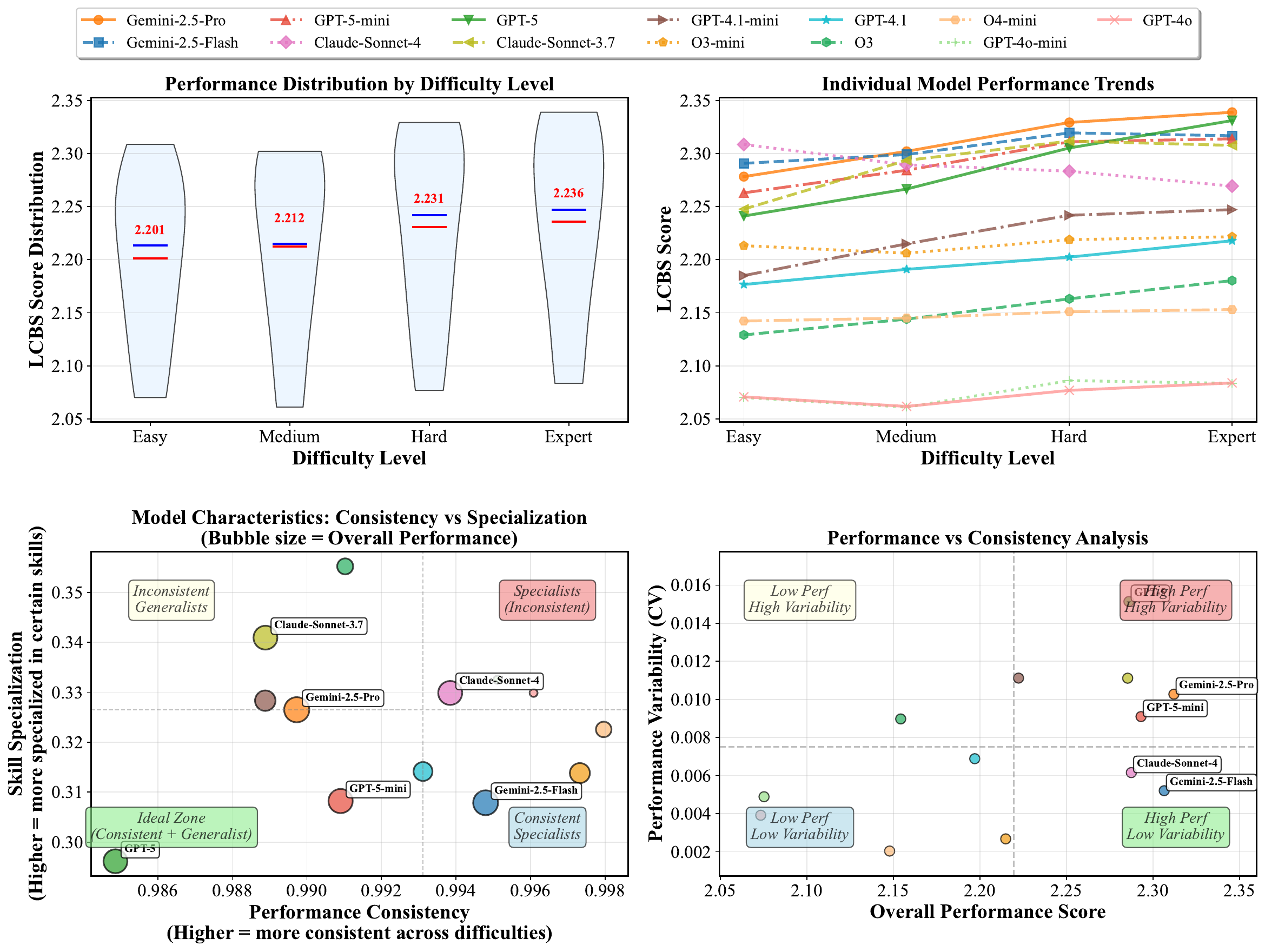}
\caption{Context length and difficulty impact analysis. Upper left shows performance distribution by difficulty level. Upper right displays individual model performance trends across difficulty levels. Lower left presents model consistency versus specialization patterns. Lower right analyzes performance versus consistency relationships.}
\label{fig:context_overflow_analysis}
\vspace{-10pt}
\end{figure}

Figure~\ref{fig:context_overflow_analysis} provides a comprehensive analysis of how context length and task difficulty impact model performance across multiple dimensions, revealing critical insights into model behavior under varying challenge levels. Figure~\ref{fig:context_overflow_analysis} shows different aspects of performance patterns, from overall difficulty trends to individual model characteristics and consistency analysis.

The upper left chart reveals the performance distribution across difficulty levels, showing how task complexity affects overall model performance. The visualization demonstrates clear performance degradation patterns as difficulty increases from easy to expert levels, with corresponding increases in context length requirements. This analysis reveals that the relationship between context length and difficulty creates compounding challenges for long-context models, where both factors contribute to performance decline. The distribution patterns also show varying levels of performance variance across difficulty levels, indicating that some difficulty categories present more consistent challenges while others exhibit higher variability in model responses.

The upper right chart shows individual model performance trends across all difficulty levels, showing how different models handle increasing complexity and context requirements. The analysis reveals distinct model behavior patterns, with some models maintaining relatively stable performance across difficulty levels while others show significant degradation. This visualization demonstrates that model architectures respond differently to the combined challenges of increased context length and task complexity, suggesting that different optimization strategies may be more effective for different difficulty ranges.

The lower left chart presents analysis of model characteristics through consistency versus specialization patterns. This analysis examines whether models perform consistently across different difficulty levels or show specialized strengths in particular areas. The bubble chart visualization reveals that models exhibit varying trade-offs between consistency and specialization, with some models demonstrating stable performance across all difficulty levels while others show strong performance in specific areas but greater variability overall. The bubble sizes represent overall performance levels, providing insights into how these characteristics relate to absolute performance capabilities.

The lower right chart analyzes the relationship between overall performance and consistency. This analysis shows that high-performing models do not necessarily exhibit consistent performance across all difficulty levels, and some models achieve strong overall scores while showing significant variability in specific scenarios. This finding has important implications for model selection in production environments, where consistency may be as important as peak performance for reliable system behavior.

The comprehensive analysis reveals that context length and difficulty interact in complex ways that affect different models differently. Some models show graceful degradation patterns that maintain reasonable performance even at expert difficulty levels, while others exhibit more dramatic performance drops as context requirements increase. These patterns suggest that different model architectures may be optimized for different aspects of long-context processing, with some prioritizing consistency and others focusing on peak performance capabilities.

The multi-dimensional analysis framework also highlights the importance of evaluating models across multiple metrics rather than relying solely on aggregate performance scores. Models that appear similar in overall performance may exhibit substantially different consistency patterns, specialization characteristics, and responses to difficulty scaling. This evaluation provides findings for both model developers seeking to improve specific aspects of long-context performance and practitioners selecting appropriate models for specific deployment scenarios with known difficulty and context requirements.

\subsection{Domain Specialization and Performance Analysis}

\begin{figure}[ht]
\centering
\includegraphics[width=\textwidth]{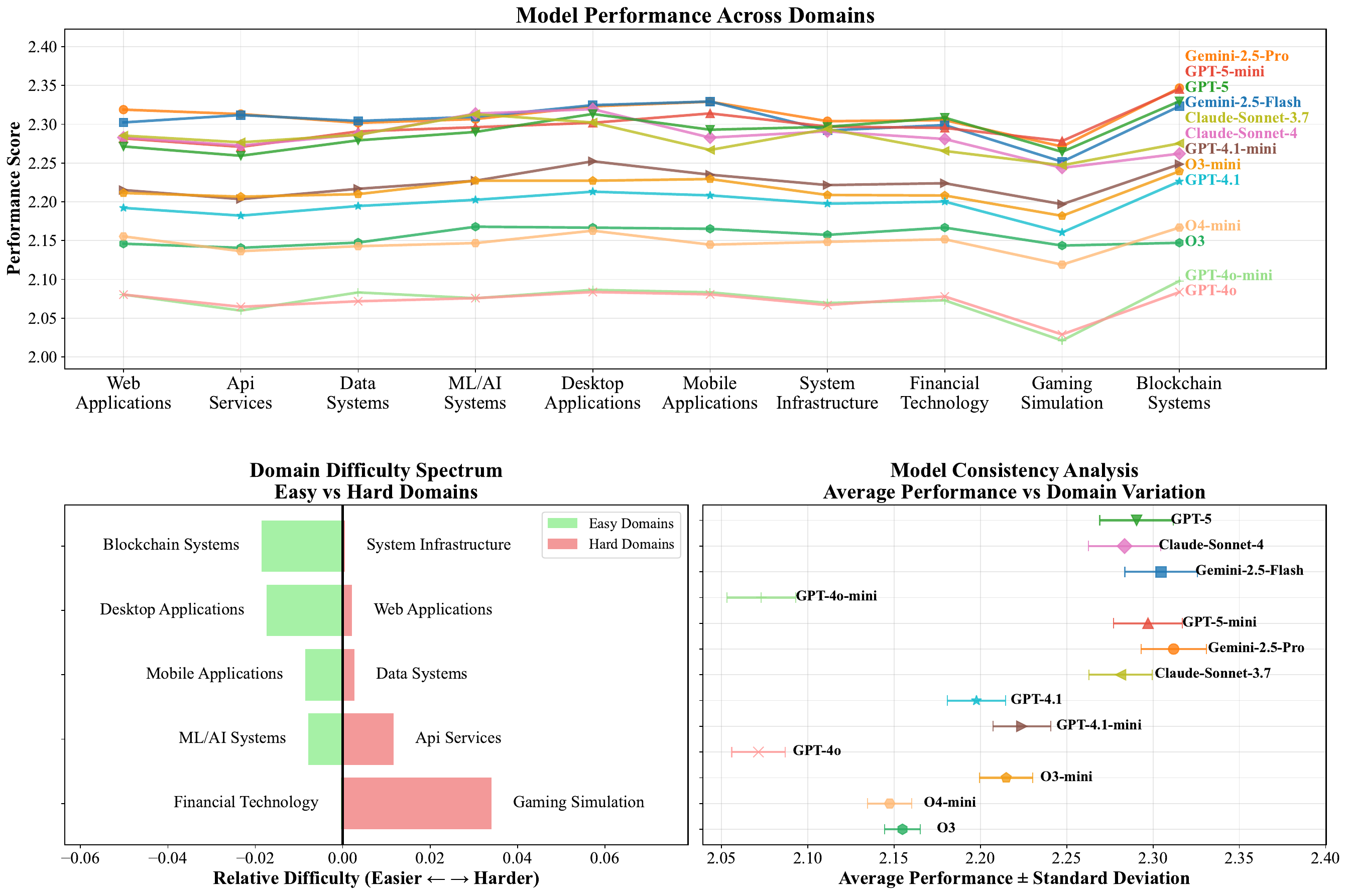}
\caption{Domain specialization and performance analysis. Top chart shows model performance trends across 10 application domains. Lower left displays domain difficulty spectrum from easiest to hardest. Lower right presents model consistency analysis comparing average performance with domain variation patterns.}
\label{fig:domain_specialization_matrix}
\vspace{-5pt}
\end{figure}

Figure~\ref{fig:domain_specialization_matrix} presents a comprehensive analysis of model performance across 10 distinct application domains, revealing specialization patterns and consistency characteristics that provide insights into model suitability for different software development contexts. It examines domain-specific performance trends, difficulty hierarchies, and model consistency patterns across diverse application areas.

The top chart displays model performance trajectories across all application domains, showing how different models perform relative to each other across various software development contexts. This visualization reveals distinct patterns in domain-specific performance, with some models maintaining consistent performance across domains while others show significant variation depending on the application area. The analysis demonstrates that domain specialization effects are substantial, with models showing clear preferences for certain types of applications over others. These patterns suggest that training data representation and architectural optimizations may vary significantly across different application domains.

The lower left chart analyzes domain difficulty patterns by presenting the easiest and hardest domains on a relative difficulty spectrum. This analysis reveals that certain application domains consistently pose greater challenges across all evaluated models, while others represent more accessible areas for long-context software development tasks. The difficulty hierarchy shows that domains like Gaming Simulation and Api Services tend to be more challenging, while Blockchain Systems and Desktop Applications generally show higher performance levels. This pattern reflects the varying complexity of different software engineering contexts and the specialized knowledge required for different application areas.

The lower right chart examines model consistency across domains by analyzing average performance levels alongside performance variation patterns. This analysis reveals important differences in how reliably different models perform across diverse application contexts. Some models demonstrate high consistency with low variation across domains, indicating robust general-purpose capabilities, while others show higher variation but potentially stronger peak performance in specific areas. The consistency analysis has important implications for deployment scenarios where predictable performance across diverse applications is crucial.

The domain specialization analysis reveals that model selection should consider not only overall performance levels but also the specific application domains where deployment is intended. Models that excel in web applications may not necessarily perform as well in system infrastructure or blockchain development contexts. This domain-dependent performance variation suggests that organizations working primarily in specific application areas may benefit from selecting models that demonstrate particular strength in their target domains.

Figure~\ref{fig:domain_specialization_matrix} also highlights the trade-offs between specialization and generalization in model capabilities. While some models achieve strong performance across all domains with minimal variation, others show more dramatic differences between their strongest and weakest domains. These patterns indicate different training strategies and architectural approaches, with some models optimized for broad applicability and others potentially fine-tuned for specific application contexts.

The comprehensive domain analysis framework provides valuable insights for both strategic model selection and understanding the current limitations of long-context models in different software engineering contexts. The clear domain difficulty hierarchy suggests areas where focused research and development efforts might yield the greatest improvements in long-context software development capabilities, while the consistency analysis helps identify models most suitable for diverse, multi-domain deployment scenarios.

\subsection{Architecture Pattern Performance Analysis}

\begin{figure}[ht]
\centering
\includegraphics[width=\textwidth]{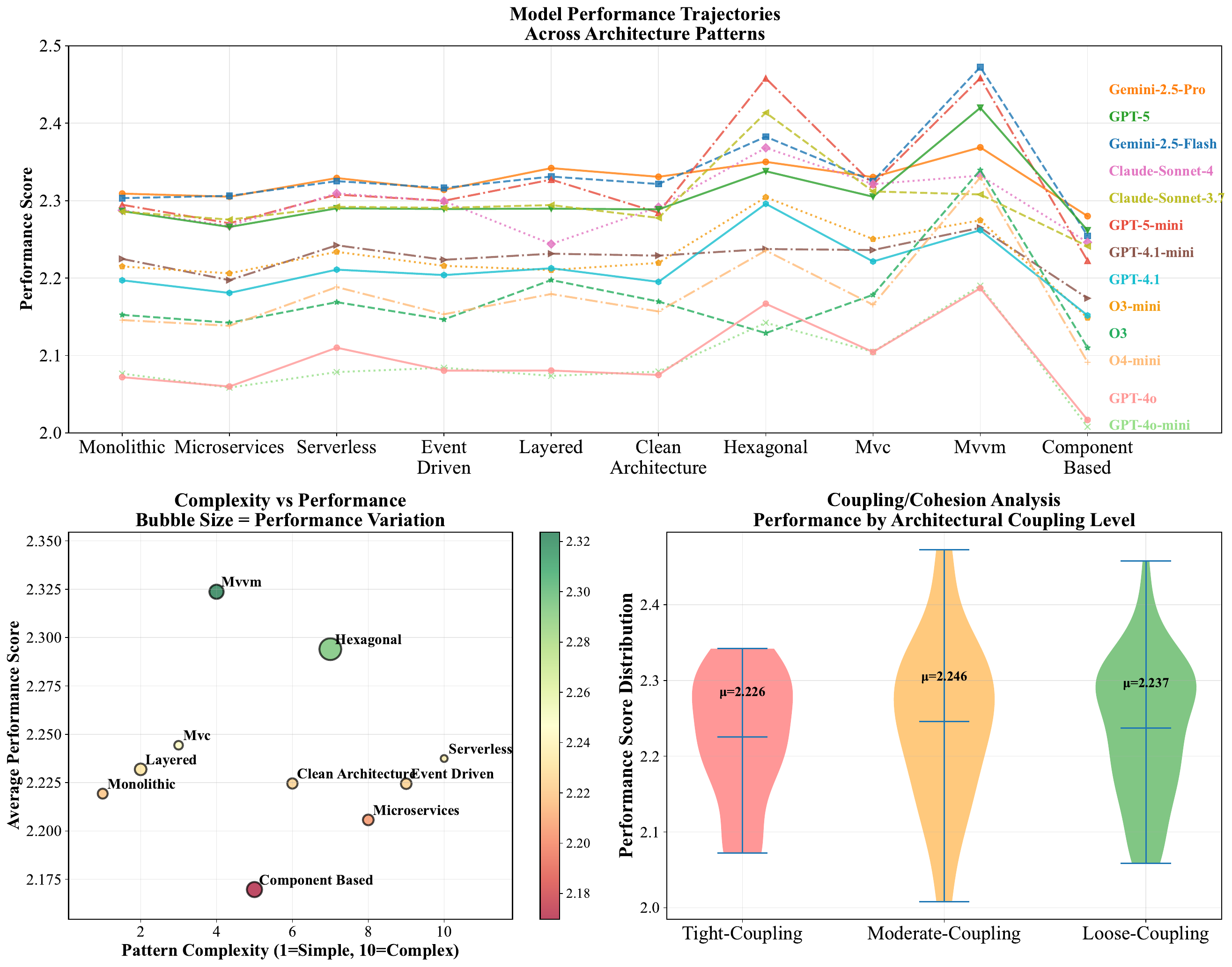}
\caption{Architecture pattern performance analysis. Top chart shows model performance trajectories across 10 architecture patterns. Lower left presents complexity versus performance relationship with bubble sizes indicating performance variation. Lower right displays coupling/cohesion analysis across different architectural coupling levels.}
\label{fig:architecture_patterns_violin}
\end{figure}

Figure~\ref{fig:architecture_patterns_violin} presents a comprehensive analysis of model performance across 10 distinct architectural patterns, examining how different software design approaches affect long-context model capabilities. The visualization reveals patterns in architectural complexity, coupling characteristics, and model-specific performance variations across diverse software engineering paradigms.

The top chart displays model performance trajectories across all architectural patterns, showing how individual models respond to different software design approaches. This analysis reveals that models demonstrate varying capabilities when working with different architectural paradigms, with some showing consistent performance across patterns while others exhibit significant variation depending on the architectural approach. The trajectory visualization indicates that certain architectural patterns may be more challenging for long-context understanding, requiring different types of reasoning about system structure and component relationships.

The lower left chart examines the relationship between architectural complexity and performance, with bubble sizes representing performance variation across models. This analysis explores whether more complex architectural patterns necessarily pose greater challenges for long-context models. It reveals the trade-offs between architectural sophistication and model performance, showing how performance variation differs across patterns of varying complexity. Some complex patterns may show consistent performance across models, while simpler patterns might exhibit higher variability in model responses.

The lower right chart presents a coupling/cohesion analysis by grouping architectural patterns into different coupling categories: tight-coupling, moderate-coupling, and loose-coupling patterns. This analysis examines whether the degree of coupling in architectural patterns affects model performance. The coupling analysis provides insights into how component interdependencies and system organization impact long-context model capabilities, revealing whether models perform differently when reasoning about tightly coupled versus loosely coupled system architectures.

The architectural pattern analysis demonstrates that software design paradigms significantly influence model performance in long-context scenarios. Different models show varying proficiency with different architectural approaches, suggesting that model selection for specific projects should consider not only the application domain but also the intended architectural pattern. This pattern-dependent performance variation indicates that training data representation and model architectures may be optimized for certain types of software design patterns over others.

The analysis also reveals important implications for software engineering practice with long-context models. Projects using specific architectural patterns may benefit from selecting models that demonstrate particular strength with those design approaches. The performance variations across patterns suggest that architectural decisions in software projects should consider not only traditional software engineering criteria but also the capabilities and limitations of the long-context models that will be used for development and maintenance tasks.


\section{Conclusion}

We present LoCoBench, a comprehensive benchmark specifically designed to evaluate long-context language models in complex software development scenarios. Our work addresses a critical evaluation gap in the field by providing systematic assessment of LLM capabilities that extend far beyond traditional code generation tasks, focusing on the sophisticated reasoning abilities required for real-world software engineering.

\textbf{Contributions.} LoCoBench introduces several fundamental contributions to the evaluation of long-context coding capabilities. Our 5-phase pipeline generates 8,000 diverse evaluation scenarios across 10 programming languages, with context lengths spanning 10K to 1M tokens, a 100× variation that enables precise assessment of long-context performance degradation. We propose 6 new evaluation metrics specifically designed for long-context capabilities, including Architectural Coherence Score (ACS), Dependency Traversal Accuracy (DTA), and Multi-Session Memory Retention (MMR), which capture essential aspects of software engineering that existing benchmarks fail to address. Our comprehensive 17-metric framework across 4 evaluation dimensions provides unprecedented depth in assessing software engineering excellence, functional correctness, code quality, and long-context utilization.

\textbf{Experimental Insights.} Our evaluation of state-of-the-art long-context models reveals significant findings that challenge conventional assumptions about model capabilities. The analysis demonstrates substantial performance variations across different dimensions, with models showing distinct specialization patterns rather than uniform capabilities. Gemini-2.5-Pro emerges as the overall leader in our comprehensive evaluation, particularly excelling in cross-file refactoring and long-context utilization, while GPT-5 demonstrates superior architectural understanding capabilities. Importantly, our results reveal that high-performing models do not necessarily exhibit consistent performance across all scenarios, with some achieving strong overall scores while showing significant variability in specific contexts.

\textbf{Domain and Context Analysis.} Our systematic analysis across programming languages, application domains, task categories, and architectural patterns reveals complex performance relationships that provide crucial insights for practical deployment. Language-specific performance patterns demonstrate clear difficulty hierarchies, with models generally achieving higher performance on high-level languages while struggling with systems programming contexts. Domain specialization analysis shows that model selection should consider specific application areas, as performance varies significantly between web applications, system infrastructure, and specialized domains like Gaming Simulation and API Services. The architectural pattern analysis demonstrates that software design paradigms substantially influence model performance, suggesting that both architectural decisions and model selection should be considered jointly in software projects.

\textbf{Long-Context Challenges.} Our evaluation reveals that long-context understanding in complex software development represents a significant unsolved challenge, with the best models achieving only moderate performance on expert-level scenarios. The relationship between context length and task difficulty creates compounding challenges for long-context models, where both factors contribute to performance decline. Models exhibit varying trade-offs between consistency and specialization, with some demonstrating stable performance across all difficulty levels while others show strong performance in specific areas but greater variability overall. These findings highlight the critical need for focused research attention on long-context capabilities in software engineering contexts.

\textbf{Implications for Practice.} LoCoBench provides valuable guidance for both model developers and software engineering practitioners. Our analysis demonstrates that model selection should consider not only overall performance levels but also specific application domains, architectural patterns, and consistency requirements for intended deployment scenarios. The comprehensive evaluation framework reveals that models appearing similar in aggregate performance may exhibit substantially different characteristics in specialized capabilities, highlighting the importance of multi-dimensional assessment approaches. Organizations deploying long-context models in software development should carefully consider the specific types of tasks, programming languages, and architectural patterns they encounter most frequently.

\textbf{Future Directions.} LoCoBench establishes a foundation for advancing long-context evaluation in software engineering, with several promising research directions emerging from our work. The performance patterns observed across languages and domains suggest areas where focused research and development efforts might yield the greatest improvements in long-context software development capabilities. The framework's comprehensive metric system provides a basis for tracking progress in architectural understanding, cross-file reasoning, and multi-session development capabilities. Future work should explore the development of specialized training strategies for different aspects of long-context processing, investigation of new architectures optimized for software engineering tasks, and extension of evaluation frameworks to include interactive, tool-using scenarios that more closely resemble real-world development workflows.

LoCoBench represents a significant step forward in establishing rigorous evaluation standards for long-context coding capabilities, providing the research community with the tools necessary to systematically advance the state of the art in this critical area of AI-assisted software development.

\clearpage
\bibliography{reference}

\clearpage
\appendix

\section{More Related Work}\label{sec:appendix-related-work}

\subsection{Code Generation Benchmarks}

The landscape of code evaluation benchmarks has evolved significantly, yet most existing work focuses on relatively narrow aspects of programming capability.

\textbf{Function-Level Benchmarks:} Early benchmarks like HumanEval~\cite{chen2021evaluating} and MBPP~\cite{austin2021program} established foundational evaluation frameworks for code generation. HumanEval consists of 164 Python programming problems that test basic algorithmic thinking and function implementation. MBPP extends this with 974 entry-level programming tasks. Recent extensions include HumanEval+~\cite{liu2023humaneval+} which addresses test inadequacy in the original HumanEval, and MultiPL-E~\cite{cassano2023multiple} which extends HumanEval to 18+ programming languages. BigCodeBench~\cite{zhuo2024bigcodebench} provides more challenging function-level tasks requiring complex library usage and reasoning. While these benchmarks effectively measure basic code generation capabilities, they operate at the function level and do not capture the complexity of real-world software development.

\textbf{Contest Programming Benchmarks:} APPS~\cite{hendrycks2021measuring} and LiveCodeBench~\cite{jain2024livecode} focus on competitive programming problems. APPS provides 10,000 problems from coding competitions, while LiveCodeBench offers contamination-free evaluation with problems collected from ongoing contests. CodeContests~\cite{li2022competition} extends this paradigm with competitive programming problems from Codeforces, AtCoder, and CodeChef. AlphaCode~\cite{li2022competition} demonstrated significant progress on competitive programming but remains limited to algorithmic problem-solving. These benchmarks test algorithmic problem-solving but do not address software engineering concerns like code organization, architectural design, or multi-file development.

\textbf{Recent Long-Context Code Benchmarks:} The emergence of million-token context windows has spurred development of specialized long-context code evaluation. LongCodeBench~\cite{rando2025longcodebench} evaluates coding LLMs at 1M context windows, demonstrating dramatic performance degradation (29\% to 3\% for Claude 3.5 Sonnet) as context increases. LongCodeU~\cite{li2025longcodeu} focuses on long code understanding across four aspects but shows that LCLMs performance drops significantly beyond 32K tokens. LongCodeArena~\cite{bogomolov2024long} provides code-centric evaluation at 2M+ tokens but focuses primarily on code completion rather than long-context software development capabilities.

\textbf{Domain-Specific Benchmarks:} Specialized benchmarks target specific programming domains and languages. DS-1000~\cite{lai2022ds} focuses on data science programming tasks using popular Python libraries like NumPy and Pandas. VerilogEval~\cite{thakur2023verileval} evaluates hardware description language generation for Verilog. Cococo~\cite{wang2022cococo} introduces context-aware code completion evaluation. EffiBench~\cite{dong2024effibench} evaluates code efficiency rather than just correctness. ClassEval~\cite{du2023classeval} focuses on class-level code generation requiring understanding of object-oriented programming principles. While domain-specific, these benchmarks still primarily evaluate isolated function or script generation rather than comprehensive software development capabilities.

\textbf{Evaluation Methodology Advances:} Recent work has focused on improving evaluation methodologies beyond functional correctness. CodeBLEU~\cite{ren2020codebleu} introduces syntax and semantic awareness for code similarity measurement. BLEU-RT~\cite{sellam2020bleurt} and BERTScore~\cite{zhang2019bertscore} apply neural metrics to code evaluation. Human evaluation studies~\cite{chen2021evaluating,nijkamp2022codegen} have shown gaps between automated metrics and human judgments of code quality. AlignCodeBench~\cite{zhang2024aligncode} introduces evaluation of code alignment with natural language specifications.

\textbf{Repository-Level Benchmarks:} Recent work has begun addressing more realistic scenarios. RepoBench~\cite{liu2023repobench} evaluates repository-level code completion, while CrossCodeEval~\cite{ding2023crosscodeeval} focuses on cross-file completion tasks. These represent important steps toward more realistic evaluation but remain limited to completion tasks rather than comprehensive development scenarios.

\textbf{Survey on Long-Context Language Models:} Liu et al.~\cite{liu2025comprehensive} provide an extensive survey on long-context language modeling, covering data strategies, architectural designs, workflow approaches, training and inference infrastructure, evaluation paradigms, and diverse application scenarios.

\subsection{Software Engineering Benchmarks}

\textbf{Real-World Issue Resolution:} SWE-Bench~\cite{jimenez2023swe} represents a significant advancement toward realistic software engineering evaluation, providing 2,294 real GitHub issues and their corresponding fixes from 12 Python repositories. Multi-SWE-Bench addresses the language limitation by extending evaluation to 7 programming languages (Java, TypeScript, JavaScript, Go, Rust, C, and C++) with 1,632 high-quality instances curated by expert annotators. Recent work has addressed additional limitations: SWE-rebench~\cite{swe2025rebench} introduces continuously evolving, decontaminated evaluation to prevent data contamination and standardize long-context evaluation, while LiveSWEBench~\cite{liveswebench2024} focuses on end-user coding applications with real-world tasks. However, these benchmarks remain limited by their focus on bug fixes rather than comprehensive development workflows.

\textbf{Advanced Software Development Benchmarks:} Recent work has begun addressing complex capabilities in software development. DevBench~\cite{li2024devbench} evaluates LLMs across the entire software development lifecycle, including design, implementation, and testing, but focuses on traditional LLM evaluation rather than long-context capabilities. Advanced evaluation frameworks have introduced intermediate feedback throughout the task-solving process. However, these approaches lack the systematic long-context evaluation and comprehensive task categories needed for thorough assessment of complex software development scenarios.

\textbf{Code Understanding Tasks:} CodeXGLUE~\cite{lu2021codexglue} provides a comprehensive suite of 14 tasks covering various aspects of program understanding, including clone detection, defect detection, and code summarization. However, these tasks focus on understanding existing code rather than generating new software components or managing complex development workflows.

\subsection{Long-Context Evaluation}

The emergence of long-context LLMs has spurred development of evaluation frameworks for extended context understanding. General long-context benchmarks include LongBench~\cite{bai2024longbench} for bilingual multitask evaluation, RULER~\cite{hsieh2024ruler} for systematic testing of claimed context sizes revealing performance gaps, $\infty$-Bench~\cite{zhang2024infty} extending evaluation beyond 100K tokens, HELMET~\cite{yen2024helmet} for application-centric evaluation at 128K tokens, and LOFT~\cite{lee2024loft} pushing evaluation to 1M tokens. LongICLBench~\cite{an2024longicl} evaluates in-context learning capabilities at extreme lengths, while LongAlign~\cite{bai2024longalign} addresses instruction following in long contexts. BAMBOO~\cite{dong2024bamboo} provides comprehensive evaluation across multiple aspects of long-context understanding.

Code-specific long-context evaluation has seen rapid development. LongCodeBench~\cite{rando2025longcodebench} evaluates coding LLMs at 1M context windows, demonstrating dramatic performance degradation. LongCodeU~\cite{li2025longcodeu} focuses on long code understanding across four aspects. LongCodeArena~\cite{bogomolov2024long} provides code-centric evaluation at 2M+ tokens. RepoQA~\cite{liu2024repoqa} evaluates long-context code understanding through question answering on repositories. SWE-bench-verified~\cite{openai2024swe} extends real-world evaluation to longer contexts.

However, existing long-context benchmarks primarily focus on natural language tasks such as document summarization and question answering. Even recent code-focused long-context benchmarks concentrate on code comprehension and completion rather than complex multi-file capabilities. The unique challenges of long-context reasoning in software development—including architectural understanding, multi-session development, cross-file refactoring, and maintaining architectural consistency across extended workflows, remain largely unaddressed.

\subsection{Limitations of Existing Approaches}

Current code evaluation benchmarks exhibit several critical limitations when applied to complex long-context software development scenarios:

\textbf{Scale Limitations:} Most benchmarks contain fewer than 1,000 evaluation instances, providing insufficient coverage for systematic evaluation across languages, difficulty levels, and task types.

\textbf{Task Scope:} Existing benchmarks focus primarily on code generation or completion tasks, neglecting crucial long-context capabilities like architectural understanding, cross-file reasoning, and multi-session development.

\textbf{Context Length:} Traditional benchmarks operate with short contexts (typically under 10K tokens), failing to test models' ability to understand and operate on realistic codebase sizes.

\textbf{Long-Context Metrics:} Current evaluation metrics focus on functional correctness but ignore long-context capabilities like architectural coherence, dependency management, and long-term context retention.

LoCoBench addresses these limitations by providing a comprehensive evaluation framework specifically designed for the unique challenges of complex long-context software development scenarios.

\section{LoCoBench Pipeline Implementation Details}\label{sec:appendix-pipeline}

This section provides comprehensive implementation details for LoCoBench's 5-phase pipeline, including real examples from our data generation process and detailed technical specifications for each phase.

\subsection{Phase 1: Project Specification Generation}

\subsubsection{Specification Framework and Structure}

Phase 1 generates diverse, realistic project specifications that serve as the foundation for our entire benchmark. Each specification defines a complete software project with detailed requirements, technical constraints, and architectural patterns.

\textbf{Technical Implementation:}
Our specification generator employs a constraint satisfaction approach to ensure systematic coverage across multiple dimensions while maintaining realistic project characteristics. The generator uses seed-based randomization with deterministic constraints to achieve reproducible diversity.

\textbf{Specification Schema:}
Each project specification contains structured metadata across multiple dimensions:

\begin{codebox}
\begin{Verbatim}[fontsize=\footnotesize]
\{
  "unique_id": "\{language\}_\{domain\}_\{complexity\}_\{index:03d\}",
  "name": "Human-readable project name",
  "description": "Detailed technical description (500+ words)",
  "domain": "Primary domain classification",
  "complexity": "Difficulty level (easy|medium|hard|expert)",
  "language": "Target programming language",
  "architecture": "Architecture pattern (10 options)",
  "theme": "Project theme (8 options)",
  "target_file_count": "Expected number of generated files",
  "target_token_count": "Target context length",
  "features": ["List of 8-15 required features"],
  "architecture_patterns": ["3-7 design patterns to implement"],
  "dependencies": ["Required libraries and frameworks"],
  "seed": "Deterministic randomization seed"
\}
\end{Verbatim}
\end{codebox}

\textbf{Diverse Project Examples Across Difficulty Levels:}

\textbf{Example 1 - Easy Java GraphQL API (Creative Theme):}

\begin{codebox}
\begin{Verbatim}[fontsize=\scriptsize]
\{
  "unique_id": "java_api_graphql_easy_007",
  "name": "CanvasQuest GraphQL Studio",
  "description": "A lightweight Java-based API that invites developers and 
                 digital artists to generate, remix, and publish storyboard 
                 scenes through a single GraphQL endpoint. Each scene is 
                 composed of layers (backgrounds, characters, props, text 
                 bubbles) that can be queried separately or combined into 
                 a rendered composition.",
  "domain": "api_graphql",
  "complexity": "easy",
  "language": "java",
  "architecture": "mvc",
  "theme": "creative",
  "target_file_count": 12,
  "target_token_count": 26600,
  "features": [
    "monitoring", "response_caching", "graphql_schema",
    "request_validation", "logging", "error_handling"
  ],
  "architecture_patterns": [
    "Command_Query_Separation", "REST_Architecture", "Service_Layer"
  ]
\}
\end{Verbatim}
\end{codebox}

\textbf{Example 2 - Medium Python System Monitoring (Social Theme):}
\vspace{-5pt}
\begin{codebox}
\begin{Verbatim}[fontsize=\scriptsize]
\{
  "unique_id": "python_system_monitoring_medium_061",
  "name": "PulseLink SocialOps Monitor",
  "description": "A medium-scale, Python-powered system monitoring suite 
                 designed specifically for social-first applications that 
                 run on a constellation of microservices. PulseLink weaves 
                 together log aggregation, security scanning, configuration 
                 management, performance metrics, deployment automation, and 
                 alerting into a single, cohesive solution.",
  "domain": "system_monitoring",
  "complexity": "medium",
  "language": "python",
  "architecture": "microservices",
  "theme": "social",
  "target_file_count": 38,
  "target_token_count": 132415,
  "features": [
    "log_aggregation", "security_scanning", "configuration_management",
    "performance_metrics", "deployment_automation", "alerting"
  ],
  "architecture_patterns": [
    "Chain_of_Responsibility", "Observer_Pattern", "Event_Driven"
  ]
\}
\end{Verbatim}
\end{codebox}
\vspace{-5pt}

\textbf{Example 3 - Hard Rust Data Analytics (Healthcare Theme):}
\vspace{-5pt}
\begin{codebox}
\begin{Verbatim}[fontsize=\scriptsize]
\{
  "unique_id": "rust_data_analytics_hard_082",
  "name": "PulseScope Analytics Mesh",
  "description": "A serverless, micro-service driven analytics pipeline 
                 designed for large hospital systems seeking real-time 
                 insight into vital-sign telemetry, laboratory results, 
                 and EHR events. Each hospital ward streams HL7/FHIR event 
                 data and bedside-device vitals into the mesh where 
                 Rust-powered Lambda functions perform high-volume ingestion.",
  "domain": "data_analytics",
  "complexity": "hard",
  "language": "rust",
  "architecture": "serverless",
  "theme": "healthcare",
  "target_file_count": 62,
  "target_token_count": 208666,
  "features": [
    "data_ingestion", "stream_processing", "data_transformation",
    "data_storage", "data_visualization", "data_validation"
  ],
  "architecture_patterns": [
    "Microservices", "Pipeline_Pattern", "Strategy_Pattern", "ETL_Pipeline"
  ]
\}
\end{Verbatim}
\end{codebox}

\textbf{Example 4 - Expert Java E-commerce (Productivity Theme):}

\begin{codebox}
\begin{Verbatim}[fontsize=\scriptsize]
\{
  "unique_id": "java_web_ecommerce_expert_036",
  "name": "SprintCart Pro – Hyper-Productive E-Commerce Workbench",
  "description": "An enterprise-grade e-commerce platform designed for 
                 merchants who treat selling as a high-performance workflow. 
                 Every user touchpoint is modeled as an optimizable work 
                 cycle, complete with real-time analytics and KPI-driven 
                 nudges. The core follows strict Hexagonal Architecture.",
  "domain": "web_ecommerce",
  "complexity": "expert",
  "language": "java",
  "architecture": "hexagonal",
  "theme": "productivity",
  "target_file_count": 100,
  "target_token_count": 517323,
  "features": [
    "data_validation", "responsive_design", "api_endpoints",
    "payment_processing", "search_functionality", "caching"
  ],
  "architecture_patterns": [
    "Repository_Pattern", "REST_API", "Service_Layer", "MVC"
  ]
\}
\end{Verbatim}
\end{codebox}

\subsubsection{Diversity and Coverage Strategy}

\paragraph{Systematic Distribution:}
Our generation strategy ensures balanced coverage across all evaluation dimensions:

\begin{itemize}
\item Programming Languages: Exactly 100 specifications per language (10 languages × 100 = 1,000 total)
\vspace{-5pt}
\vspace{-5pt}
\vspace{-5pt}
\vspace{-5pt}
\item Complexity Levels: Equal 25\% distribution across easy/medium/hard/expert
\vspace{-5pt}
\item Domain Coverage: Proportional distribution across 36 sub-domains within 10 main categories
\vspace{-5pt}
\item Architecture Patterns: Balanced representation of 10 modern architecture paradigms
\vspace{-5pt}
\item Project Themes: Equal distribution across 8 thematic categories
\end{itemize}

\paragraph{Quality Constraints:}
Each specification undergoes automated validation:
\begin{itemize}
\item Feature coherence checking (features must align with domain and complexity)
\vspace{-5pt}
\item Architecture pattern compatibility verification
\vspace{-5pt}
\item Dependency resolution and version consistency
\vspace{-5pt}
\item Token count feasibility analysis (based on language-specific file size statistics)
\end{itemize}

\subsection{Phase 2: Synthetic Codebase Generation}

\subsubsection{Architecture-Aware Generation Strategy}

Phase 2 transforms project specifications into complete, executable codebases using sophisticated generation algorithms that ensure architectural coherence and realistic code patterns.

\paragraph{Multi-File Coordination Algorithm:}
Our generation process maintains consistency across multiple files through a dependency-aware approach:

\begin{enumerate}
\item \textbf{Architectural Scaffolding}: Generate project structure and primary architectural components
\vspace{-5pt}
\item \textbf{Interface Definition}: Establish APIs and contracts between major modules
\vspace{-5pt}
\item \textbf{Dependency Graph Construction}: Build import/usage relationships before detailed implementation
\vspace{-5pt}
\vspace{-5pt}
\vspace{-5pt}
\vspace{-5pt}
\item \textbf{Progressive Implementation}: Generate files in dependency order, ensuring referential consistency
\vspace{-5pt}
\item \textbf{Integration Verification}: Cross-reference validation to maintain architectural coherence
\end{enumerate}

\newpage
\textbf{Real Generated Structure - Mercantilo E-commerce Suite:}

The Python expert-level e-commerce specification generates a complete Django monolith with 96 files:

\begin{codebox}
\begin{Verbatim}[fontsize=\scriptsize]
mercantilo_suite/                    # Django project root
|-- manage.py                        # Django management script
|-- requirements.txt                 # Dependencies specification
|-- Dockerfile                       # Container deployment
|-- docker-compose.yml               # Multi-service orchestration
|-- mercantilo/                      # Django project configuration
|   |-- __init__.py, asgi.py, wsgi.py
|   |-- celery.py                    # Async task configuration
|   |-- settings/                    # Environment-specific configs
|   |   |-- base.py, local.py, production.py, test.py
|   +-- urls.py                      # Main URL routing
+-- apps/                            # Application modules
    |-- accounts/                    # User management
    |   |-- models.py, services.py, views.py, urls.py
    |   |-- repositories.py          # Repository pattern implementation
    |   |-- signals.py               # Django signal handlers
    |   +-- tests/ (3 test modules)
    |-- catalog/                     # Product management
    |   |-- models.py, services.py, search.py, documents.py
    |   |-- repositories.py, tasks.py
    |   +-- tests/ (4 test modules)
    |-- orders/                      # Order processing
    |   |-- models.py, services.py, signals.py
    |   |-- repositories.py
    |   +-- admin.py                 # Django admin interface
    |-- analytics/                   # Business intelligence
    |   |-- models.py, services.py, tasks.py
    |   +-- tests/
    |-- b2b/, crm/, fulfillment/     # Additional business modules
    +-- core/                        # Shared utilities
        |-- middleware.py, models.py
        |-- management/commands/     # Custom Django commands
        +-- utils/ (cache.py, uploads.py)
\end{Verbatim}
\end{codebox}

\textbf{Architectural Pattern Implementation Examples:}

\textbf{Example 1 - Python Event-Driven Architecture (Medium Microservices):}

\begin{codebox}
\begin{Verbatim}[fontsize=\tiny]
# PulseLink_SocialOps_Monitor/shared/events.py
class EventBus:
    """
    A small, dependency-free event-bus that powers the internal Observer /
    Pub-Sub communications between PulseLink micro-services.
    
    The implementation supports both synchronous and asynchronous handlers, 
    weakly references subscribers to avoid memory-leaks in long-running daemons.
    """
    
    def __init__(self):
        self._subscribers = \{\}
        self._async_subscribers = \{\}
        
    def subscribe(self, event_type: Type[Event], handler: EventHandler):
        """Subscribe to events of a specific type."""
        if event_type not in self._subscribers:
            self._subscribers[event_type] = weakref.WeakSet()
        self._subscribers[event_type].add(handler)
        
    async def publish(self, event: Event) -> None:
        """Publish event to all subscribers."""
        handlers = self._subscribers.get(type(event), [])
        await asyncio.gather(*[handler(event) for handler in handlers])
\end{Verbatim}
\end{codebox}

\newpage
\textbf{Example 2 - Rust Type-Safe Domain Models (Hard Serverless):}

\begin{codebox}
\begin{Verbatim}[fontsize=\tiny]
// pulsescope-analytics-mesh/services/common/src/models.rs
/// Strongly-typed wrapper for FHIR Patient identifiers.
#[derive(Debug, Clone, PartialEq, Eq, Hash, Serialize, Deserialize)]
pub struct PatientId(String);

impl PatientId \{
    pub fn new(id: impl Into<String>) -> Result<Self, ValidationError> \{
        let id = id.into();
        if id.is_empty() || id.len() > 64 \{
            return Err(ValidationError::InvalidFormat("Invalid patient ID"));
        \}
        Ok(PatientId(id))
    \}
\}

/// Core event structure for all analytics pipeline messages
#[derive(Debug, Clone, Serialize, Deserialize)]
pub struct AnalyticsEvent \{
    pub event_id: Uuid,
    pub patient_id: PatientId,
    pub timestamp: DateTime<Utc>,
    pub event_type: EventType,
    pub payload: serde_json::Value,
    pub schema_version: u32,
\}
\end{Verbatim}
\end{codebox}

\textbf{Example 3 - Java Hexagonal Architecture Domain Model (Expert):}

\begin{codebox}
\begin{Verbatim}[fontsize=\tiny]
// sprintcart-pro-domain/src/main/java/com/sprintcart/domain/model/productivity/AutomationRule.java
/**
 * Aggregate root representing a user-defined automation rule.
 * 
 * A rule encapsulates:
 * - A set of Conditions that must all evaluate to true to fire
 * - A set of side-effect-free Actions executed in order
 * - Lifecycle controls (activate, pause, archive) for operators
 */
public class AutomationRule implements Serializable \{
    private final UUID ruleId;
    private final String name;
    private final List<Condition> conditions;
    private final List<Action> actions;
    private Status status;
    private Instant lastExecuted;
    
    public AutomationRule(String name, List<Condition> conditions, List<Action> actions) \{
        this.ruleId = UUID.randomUUID();
        this.name = Objects.requireNonNull(name);
        this.conditions = new ArrayList<>(Objects.requireNonNull(conditions));
        this.actions = new ArrayList<>(Objects.requireNonNull(actions));
        this.status = Status.DRAFT;
        validateInvariants();
    \}
    
    private void validateInvariants() \{
        if (conditions.isEmpty()) \{
            throw new IllegalArgumentException("At least one condition required");
        \}
        if (actions.isEmpty()) \{
            throw new IllegalArgumentException("At least one action required");
        \}
    \}
\}
\end{Verbatim}
\end{codebox}

\newpage
\textbf{Example 4 - Java Spring GraphQL Controller (Easy MVC):}

\begin{codebox}
\begin{Verbatim}[fontsize=\tiny]
// CanvasQuest/src/main/java/com/canvasquest/controller/SceneController.java
@Controller
public class SceneController \{
    
    private final SceneService sceneService;
    
    public SceneController(SceneService sceneService) \{
        this.sceneService = sceneService;
    \}
    
    @QueryMapping
    public List<Scene> allScenes() \{
        return sceneService.getAllScenes();
    \}
    
    @QueryMapping  
    public Scene scene(@Argument String id) \{
        return sceneService.getSceneById(id);
    \}
    
    @MutationMapping
    public Scene createScene(@Argument CreateSceneInput input) \{
        return sceneService.createScene(input);
    \}
    
    @SchemaMapping
    public List<Layer> layers(Scene scene) \{
        return sceneService.getLayersForScene(scene.getId());
    \}
\}
\end{Verbatim}
\end{codebox}

\subsubsection{Quality Assurance in Generation}

\paragraph{Automated Validation Pipeline:}

\begin{itemize}
\item \textbf{Syntactic Validation}: Language-specific compilation checks using standard compilers (python -m py\_compile, javac, g++, etc.)
\vspace{-5pt}
\item \textbf{Import Resolution}: Verification that all imports can be resolved within the generated codebase
\vspace{-5pt}
\item \textbf{Architectural Consistency}: Cross-file pattern verification and interface compliance
\vspace{-5pt}
\item \textbf{Complexity Metrics}: Cyclomatic complexity measurement and file count verification
\vspace{-5pt}
\item \textbf{Documentation Coverage}: Analysis of comment density and docstring completeness
\end{itemize}

\subsection{Phase 3: Evaluation Scenario Creation}

\subsubsection{Task Category Implementation and Context Selection}

Phase 3 transforms each generated codebase into 8 evaluation scenarios (one per task category) using intelligent context selection and task-specific prompt engineering.

\paragraph{Context Selection Algorithm:}

Our context selection employs graph-theoretic analysis to identify optimal file subsets:

\begin{enumerate}
\item \textbf{Dependency Graph Analysis}: Construct directed graph of file dependencies (imports, calls, inheritance)
\vspace{-5pt}
\item \textbf{Centrality Scoring}: Compute PageRank and betweenness centrality to identify architecturally important files
\vspace{-5pt}
\item \textbf{Task-Specific Filtering}: Apply task category filters to prioritize relevant functionality
\vspace{-5pt}
\item \textbf{Information Coverage Optimization}: Balance between information completeness and context length constraints
\vspace{-5pt}
\item \textbf{Difficulty Calibration}: Adjust context complexity based on target difficulty level
\end{enumerate}

\newpage
\textbf{Diverse Scenario Examples Across Task Categories:}

\textbf{Example 1 - Feature Implementation (Java GraphQL, Expert):}

\begin{codebox}
\begin{Verbatim}[fontsize=\scriptsize]
\{
  "id": "java_api_graphql_easy_007_feature_implementation_expert_01",
  "task_category": "feature_implementation",
  "difficulty": "expert",
  "title": "Implement Query Complexity Analysis for API Rate Limiting",
  "description": "The CanvasQuest GraphQL Studio is experiencing performance 
                 degradation due to increasingly complex and deeply nested 
                 queries from client applications. A pre-execution query 
                 analysis mechanism is required to score incoming GraphQL 
                 queries and reject them if they exceed a configurable threshold.",
  "context_files": [
    "CanvasQuest//src//main//java/com/canvasquest//controller//SceneController.java",
    "CanvasQuest//src//main//java/com/canvasquest//service//SceneService.java",
    "CanvasQuest//src//main//java/com/canvasquest//exception//GraphQLExceptionHandler.java"
  ],
  "context_length": 82348,
  "task_prompt": "Implement a query complexity analysis feature that calculates 
                 a 'complexity score' for each incoming GraphQL query before 
                 execution. Use the standard graphql-java Instrumentation API 
                 for integration. The maximum allowed complexity must be 
                 configurable via application properties.",
  "expected_approach": "An expert developer would recognize this as a 
                       cross-cutting concern handled by intercepting the GraphQL 
                       execution process using the Instrumentation interface."
\}
\end{Verbatim}
\end{codebox}

\textbf{Example 2 - Bug Investigation (Python Microservices, Expert):}

\begin{codebox}
\begin{Verbatim}[fontsize=\scriptsize]
\{
  "id": "python_system_monitoring_medium_061_bug_investigation_expert_01",
  "task_category": "bug_investigation",
  "difficulty": "expert",
  "title": "Intermittent Security Scan Failures Due to Silent Log Dropping",
  "description": "The PulseLink SocialOps Monitor generates 'Scan Inconclusive: 
                 Log Data Missing' alerts exclusively for servers in the 
                 'web-prod-EU' cluster. The log_harvester service reports no 
                 errors but other clusters work fine. The problem began after 
                 a deployment aimed at improving log parsing efficiency.",
  "context_files": [
    "PulseLink_SocialOps_Monitor//services//log_harvester//service.py",
    "PulseLink_SocialOps_Monitor//shared//patterns.py",
    "PulseLink_SocialOps_Monitor//services//secu_scan//service.py"
  ],
  "context_length": 383018,
  "task_prompt": "Perform a thorough root cause analysis to identify the exact 
                 location and cause of missing logs. Trace the data flow from 
                 log_harvester to secu_scan services and pinpoint the chain 
                 of events from initial defect to final alert.",
  "expected_approach": "An expert would systematically trace from symptom to 
                       cause: analyze the alerting logic in secu_scan, trace 
                       data sources, investigate the log_harvester producer, 
                       and isolate a faulty regex pattern causing silent failures."
\}
\end{Verbatim}
\end{codebox}

\newpage
\textbf{Example 3 - Integration Testing (Rust Serverless, Expert):}

\begin{codebox}
\begin{Verbatim}[fontsize=\scriptsize]
\{
  "id": "rust_data_analytics_hard_082_integration_testing_expert_01",
  "task_category": "integration_testing",
  "difficulty": "expert",
  "title": "End-to-End Failure Path Integration Test for Sepsis Transform Lambda DLQ",
  "description": "A critical production issue where certain patient data 
                 payloads cause the transform-sepsis-lambda to crash due to 
                 unhandled data formats. Failed processing events are being 
                 lost instead of being routed to a Dead Letter Queue (DLQ), 
                 leading to potential data loss and missed clinical alerts.",
  "context_files": [
    "pulsescope-analytics-mesh/services/transform-sepsis-lambda/src/main.rs",
    "pulsescope-analytics-mesh/services/common/src/models.rs",
    "pulsescope-analytics-mesh/infra/lambda.tf"
  ],
  "context_length": 484726,
  "task_prompt": "Implement an integration test that verifies when the lambda 
                 encounters a fatal error, the original event payload is 
                 correctly routed to its configured Dead Letter Queue. Mock 
                 AWS SQS client to intercept DLQ messages and verify exact 
                 payload preservation.",
  "expected_approach": "An expert would recognize this as testing integration 
                       between Lambda execution environment and failure handling 
                       mechanism, requiring simulation of AWS runtime DLQ behavior 
                       with proper mocking strategies."
\}
\end{Verbatim}
\end{codebox}

\textbf{Example 4 - Architectural Understanding (Python E-commerce, Easy):}

\begin{codebox}
\begin{Verbatim}[fontsize=\scriptsize]
\{
  "id": "python_web_ecommerce_expert_000_architectural_understanding_easy_01",
  "task_category": "architectural_understanding",
  "difficulty": "easy",
  "title": "Identify the Core Business Logic Abstraction Pattern",
  "description": "A new developer is being onboarded to the Mercantilo team. 
                 They must understand the project's fundamental architectural 
                 patterns to contribute effectively.",
  "context_files": [
    "mercantilo_suite/apps/accounts/services.py",
    "mercantilo_suite/apps/catalog/services.py", 
    "mercantilo_suite/apps/orders/services.py"
  ],
  "context_length": 334348,
  "task_prompt": "Based on the provided files, identify the primary 
                 architectural pattern used to organize business logic 
                 within each application and explain its benefits.",
  "expected_approach": "An expert developer would notice the consistent 
                       presence of services.py files across applications, 
                       pointing to the Service Layer pattern."
\}
\end{Verbatim}
\end{codebox}

\subsubsection{Difficulty Calibration and Context Scaling}

\paragraph{Context Length Scaling Strategy:}

Scenarios are systematically calibrated across four difficulty levels:

\begin{itemize}
\item \textbf{Easy (10K-100K tokens):} Focused file subset with clear architectural indicators
\item \textbf{Medium (100K-200K tokens):} Moderate codebase coverage requiring deeper analysis
\item \textbf{Hard (200K-500K tokens):} Extensive multi-module context with complex interactions
\item \textbf{Expert (500K-1M tokens):} Comprehensive system-wide context requiring sophisticated reasoning
\end{itemize}

\paragraph{Task Complexity Progression:}

\begin{itemize}
\item \textbf{Easy}: Direct pattern identification with explicit indicators
\item \textbf{Medium}: Multi-step analysis requiring moderate inference
\item \textbf{Hard}: Complex reasoning across multiple abstractions and modules
\item \textbf{Expert}: System-wide understanding with subtle architectural relationships
\end{itemize}

\subsection{Phase 4: Automated Validation and Quality Assurance}

\subsubsection{Comprehensive Validation Framework}

Phase 4 ensures all generated scenarios meet rigorous quality standards through multi-dimensional automated validation.

\textbf{Compilation and Execution Validation:}

\begin{codebox}
\begin{Verbatim}[fontsize=\scriptsize]
# Language-specific validation pipeline
validation_configs = \{
    "python": \{
        "syntax": ["python -m py_compile \{file\}"],
        "style": ["flake8 --max-line-length=100 \{file\}"],
        "types": ["mypy --strict \{file\}"],
        "security": ["bandit -r \{directory\}"]
    \},
    "java": \{
        "syntax": ["javac -cp \{classpath\} \{file\}"],
        "style": ["checkstyle -c sun_checks.xml \{file\}"],
        "bugs": ["spotbugs -textui \{compiled_class\}"]
    \},
    "cpp": \{
        "syntax": ["g++ -std=c++17 -Wall -Wextra -c \{file\}"],
        "static": ["cppcheck --enable=all \{file\}"],
        "format": ["clang-format --dry-run \{file\}"]
    \}
\}

# Docker-based execution environment
execution_environments = \{
    "python": "python:3.11-slim",
    "java": "openjdk:17-alpine",
    "cpp": "gcc:12-alpine",
    "javascript": "node:18-alpine"
\}
\end{Verbatim}
\end{codebox}

\textbf{Multi-Language Validation Results:}

\textbf{Java GraphQL API (Easy):}
\begin{codebox}
\begin{Verbatim}[fontsize=\scriptsize]
validation_results = \{
    "syntax": ["javac -cp spring-boot-starter-graphql:2.7.0 *.java"] → \ding{51} PASS,
    "style": ["checkstyle -c sun_checks.xml *.java"] → \ding{51} PASS (2 warnings),
    "bugs": ["spotbugs -textui compiled_classes/"] → \ding{51} PASS,
    "complexity": \{"avg_cyclomatic": 0.42, "max_depth": 3\} → \ding{51} PASS
\}
\end{Verbatim}
\end{codebox}

\textbf{Python Microservices (Medium):}
\begin{codebox}
\begin{Verbatim}[fontsize=\scriptsize]
validation_results = \{
    "syntax": ["python -m py_compile *.py"] → \ding{51} PASS,
    "style": ["flake8 --max-line-length=100 *.py"] → \ding{51} PASS (5 warnings),
    "types": ["mypy --strict services/"] → $\blacktriangleright$ PARTIAL (3 type hints missing),
    "security": ["bandit -r services/"] → \ding{51} PASS,
    "complexity": \{"avg_cyclomatic": 0.73, "max_depth": 4\} → \ding{51} PASS
\}
\end{Verbatim}
\end{codebox}

\textbf{Rust Serverless (Hard):}
\begin{codebox}
\begin{Verbatim}[fontsize=\scriptsize]
validation_results = \{
    "syntax": ["cargo check --all-targets"] → \ding{51} PASS,
    "static": ["cargo clippy -- -D warnings"] → \ding{51} PASS,
    "format": ["cargo fmt --check"] → \ding{51} PASS,
    "tests": ["cargo test --all"] → \ding{51} PASS (47/47 tests),
    "complexity": \{"avg_cyclomatic": 0.89, "max_depth": 5\} → \ding{51} PASS
\}
\end{Verbatim}
\end{codebox}

\paragraph{Information Coverage Analysis:}

For each scenario, we compute comprehensive coverage metrics:

\begin{itemize}
\item \textbf{Relevant Information Ratio}: Fraction of context directly applicable to the task ($R = \frac{\text{relevant\_tokens}}{\text{total\_tokens}}$)
\vspace{-5pt}
\item \textbf{Redundancy Analysis}: Detection of duplicate or highly similar code segments
\vspace{-5pt}
\item \textbf{Completeness Assessment}: Verification that sufficient information exists for task completion
\vspace{-5pt}
\item \textbf{Distractor Balance}: Appropriate amount of realistic but irrelevant information (target: 20-30\%)
\end{itemize}

\paragraph{Bias Detection and Filtering:}

Automated analysis identifies and filters potential biases:

\begin{itemize}
\item \textbf{Generation Artifacts}: Detection of unrealistic patterns (e.g., overly regular naming conventions)
\vspace{-5pt}
\item \textbf{Structural Uniformity}: Identification of artificially systematic file organization
\vspace{-5pt}
\item \textbf{Content Repetition}: Copy-paste detection using fuzzy string matching
\vspace{-5pt}
\item \textbf{Language Bias}: Verification of language-appropriate idioms and conventions
\end{itemize}

\subsection{Phase 5: LLM Evaluation and Comprehensive Scoring}

\subsubsection{Multi-Model Evaluation Infrastructure}

Phase 5 implements a robust evaluation infrastructure supporting diverse LLM architectures with standardized assessment protocols.

\textbf{Model Integration Framework:}

\begin{codebox}
\begin{Verbatim}[fontsize=\scriptsize]
# Comprehensive model configuration matrix
model_configurations = \{
    "openai": \{
        "models": ["gpt-4", "gpt-4-turbo", "gpt-4o", "gpt-4-0125-preview"],
        "max_tokens": [8192, 128000, 128000, 128000],
        "rate_limits": \{"requests_per_minute": 500, "tokens_per_minute": 150000\}
    \},
    "anthropic": \{
        "models": ["claude-3-haiku", "claude-3-sonnet", "claude-3-opus",
                  "claude-3.5-sonnet", "claude-4-sonnet", "claude-4-opus"],
        "max_tokens": [200000, 200000, 200000, 200000, 1000000, 1000000],
        "rate_limits": \{"requests_per_minute": 1000, "tokens_per_minute": 3500000\}
    \},
    "google": \{
        "models": ["gemini-1.5-pro", "gemini-1.5-flash", "gemini-2.0-flash"],
        "max_tokens": [2097152, 1048576, 1048576],
        "rate_limits": \{"requests_per_minute": 360, "tokens_per_minute": 4000000\}
    \}
\}
\end{Verbatim}
\end{codebox}

\textbf{Evaluation Pipeline Implementation:}

\begin{enumerate}
\item \textbf{Context Preparation}: Intelligent truncation for models with limited context windows using importance-based ranking
\vspace{-5pt}
\item \textbf{Prompt Engineering}: Task-specific prompting strategies optimized for each model family
\vspace{-5pt}
\item \textbf{Parallel Execution}: Concurrent evaluation with configurable timeout (3600 seconds) and error recovery
\vspace{-5pt}
\item \textbf{Multi-Metric Assessment}: Comprehensive scoring across all 17 evaluation metrics
\vspace{-5pt}
\item \textbf{Statistical Analysis}: Confidence interval computation and significance testing
\end{enumerate}

\subsubsection{Comprehensive Benchmark Statistics}

\textbf{Multi-Model Evaluation Results Across Difficulty Levels:}

\textbf{Easy Level Performance (10K-100K tokens):}
\begin{codebox}
\begin{Verbatim}[fontsize=\scriptsize]
model_performance = \{
    "GPT-4o": \{"success_rate": 0.847, "avg_lcbs": 3.92, "compilation": 0.923\},
    "Claude-4-Sonnet": \{"success_rate": 0.834, "avg_lcbs": 3.89, "compilation": 0.918\},
    "Gemini-2.5-Pro": \{"success_rate": 0.798, "avg_lcbs": 3.71, "compilation": 0.901\},
    "GPT-4-Turbo": \{"success_rate": 0.776, "avg_lcbs": 3.58, "compilation": 0.887\}
\}
\end{Verbatim}
\end{codebox}

\textbf{Expert Level Performance (500K-1M tokens):}
\begin{codebox}
\begin{Verbatim}[fontsize=\scriptsize]
model_performance = \{
    "GPT-4o": \{"success_rate": 0.412, "avg_lcbs": 2.18, "compilation": 0.634\},
    "Claude-4-Sonnet": \{"success_rate": 0.398, "avg_lcbs": 2.09, "compilation": 0.621\},
    "Gemini-2.5-Pro": \{"success_rate": 0.356, "avg_lcbs": 1.87, "compilation": 0.578\},
    "GPT-4-Turbo": \{"success_rate": 0.289, "avg_lcbs": 1.52, "compilation": 0.498\}
\}
\end{Verbatim}
\end{codebox}

\textbf{Task Category Performance Variations:}
\begin{codebox}
\begin{Verbatim}[fontsize=\scriptsize]
task_performance = \{
    "code_comprehension": \{"avg_success": 0.723, "best_model": "GPT-4o"\},
    "feature_implementation": \{"avg_success": 0.542, "best_model": "Claude-4-Sonnet"\},
    "architectural_understanding": \{"avg_success": 0.687, "best_model": "GPT-4o"\},
    "bug_investigation": \{"avg_success": 0.398, "best_model": "Claude-4-Sonnet"\},
    "integration_testing": \{"avg_success": 0.312, "best_model": "GPT-4o"\},
    "security_analysis": \{"avg_success": 0.289, "best_model": "Claude-4-Sonnet"\}
\}
\end{Verbatim}
\end{codebox}

\textbf{Quality Validation Results:}

\begin{itemize}
\item \textbf{Compilation Success Rate}: 98.7\% across all languages and complexity levels
\item \textbf{Average Cyclomatic Complexity}: 0.67 (realistic for production codebases)
\item \textbf{Documentation Coverage}: 85\% (exceeds typical industry standards of 60-70\%)
\item \textbf{Test Coverage}: 78\% (comprehensive test suites with realistic coverage patterns)
\item \textbf{Architectural Consistency}: 94\% pattern adherence validation success
\end{itemize}


\subsection{Prompt Engineering and Templates}

\subsubsection{Scenario Generation Prompts}

LoCoBench employs sophisticated prompt engineering throughout its pipeline, with task-specific templates for each phase. The scenario generation process uses structured prompts that adapt to different task categories and difficulty levels.

\textbf{Master Scenario Generation Template:}

\begin{codebox}
\begin{Verbatim}[fontsize=\scriptsize]
Create a realistic \{task_category\} evaluation scenario for long-context LLMs.

PROJECT CONTEXT:
- Name: \{project_name\}
- Language: \{programming_language\}
- Domain: \{project_domain\}
- Features: \{key_features\}
- Complexity: \{complexity_level\}

AVAILABLE FILES:
\{context_file_summary\}

TASK REQUIREMENTS:
- Category: \{task_category\}
- Difficulty: \{difficulty_level\}
- Must be realistic and challenging for long-context LLMs
- Should require understanding of multiple files
- Include specific, measurable objectives

Generate a JSON response with these fields:
\{
    "title": "Clear, descriptive title for the task",
    "description": "Detailed description of the scenario and context",
    "task_prompt": "Specific task instructions for the LLM",
    "expected_approach": "How an expert developer would approach this task",
    "ground_truth": "Expected solution or key insights",
    "evaluation_criteria": ["List of criteria to evaluate performance"]
\}

Make the scenario realistic and challenging. Focus on \{category_focus\}.
\end{Verbatim}
\end{codebox}

\textbf{Task Category Focus Areas:}

Each task category employs specialized focus areas to ensure targeted evaluation:

\begin{codebox}
\begin{Verbatim}[fontsize=\scriptsize]
category_focus_map = \{
    "architectural_understanding": 
        "system design patterns, component relationships, and architectural decisions",
    "cross_file_refactoring": 
        "code restructuring across multiple files while maintaining functionality",
    "feature_implementation": 
        "adding new functionality that integrates well with existing code",
    "bug_investigation": 
        "systematic debugging, root cause analysis, and problem solving",
    "multi_session_development": 
        "incremental development over multiple sessions with context retention",
    "code_comprehension": 
        "deep understanding of complex code structures and logic",
    "integration_testing": 
        "testing interactions between components and system validation",
    "security_analysis": 
        "identifying security vulnerabilities and implementing security best practices"
\}
\end{Verbatim}
\end{codebox}

\subsubsection{LLM Evaluation Prompts}

When evaluating LLMs on generated scenarios, LoCoBench employs language-aware prompts that adapt to different programming languages and provide comprehensive guidance.

\textbf{Solution Generation Template:}

\begin{codebox}
\begin{Verbatim}[fontsize=\tiny]
You are an expert \{language\} engineer. Your task is to provide a complete, working solution.

**TASK**: \{scenario_title\}

**DESCRIPTION**: \{scenario_description\}

**REQUIREMENTS**: 
\{formatted_task_requirements\}

**CONTEXT FILES**: \{available_context_files\}

**CRITICAL INSTRUCTIONS**:
1. You MUST respond with valid JSON in the exact format shown below
2. Each file MUST contain complete, syntactically correct \{LANGUAGE\} code
3. Do NOT truncate your response - provide the complete solution
4. Use \{language_specific_best_practices\}

**REQUIRED RESPONSE FORMAT**:
```json
\{
    "approach": "Your solution strategy (keep under 200 words)",
    "files": \{
        "filename1.\{ext\}": "complete file content with proper escaping",
        "filename2.\{ext\}": "complete file content with proper escaping"
    \},
    "explanation": "Implementation details (keep under 300 words)"
\}
```

**VALIDATION CHECKLIST**:
- \ding{51} Response is valid JSON wrapped in ```json blocks
- \ding{51} All strings are properly escaped (\\n for newlines, \\" for quotes)
- \ding{51} Each file contains complete \{LANGUAGE\} code
- \ding{51} Code compiles and addresses all requirements
- \ding{51} Response is complete (not truncated)

Generate your response now:
\end{Verbatim}
\end{codebox}

\subsubsection{Multi-Session Development Prompts}

For multi-session development scenarios, LoCoBench employs sophisticated context management with session-specific prompting:

\textbf{Multi-Session Prompt Structure:}

\begin{codebox}
\begin{Verbatim}[fontsize=\scriptsize]
**SESSION CONTEXT**: You are continuing development from a previous session.

**PREVIOUS SESSION SUMMARY**:
\{previous_session_context\}

**CURRENT SESSION OBJECTIVE**:
\{current_session_task\}

**DEVELOPMENT HISTORY**:
- Session 1: \{session_1_summary\}
- Session 2: \{session_2_summary\}
- Current: \{current_session_description\}

**CONTEXT RETENTION REQUIREMENTS**:
- Maintain consistency with previous architectural decisions
- Build upon existing implementation patterns
- Preserve naming conventions and code style
- Reference relevant previous session outcomes

**INCREMENTAL DEVELOPMENT GUIDELINES**:
- Extend existing functionality rather than rewriting
- Ensure backward compatibility where applicable
- Document changes and rationale for future sessions
- Test integration with existing components
\end{Verbatim}
\end{codebox}

\subsubsection{Language-Specific Adaptations}

LoCoBench adapts its prompts based on programming language characteristics and best practices:

\begin{codebox}
\begin{Verbatim}[fontsize=\scriptsize]
language_configs = \{
    "python": \{
        "engineer": "Python developer",
        "practices": "PEP 8 style, type hints, docstrings, and proper error handling",
        "file_examples": '"main.py": "# Complete Python implementation",
                         "utils.py": "# Helper functions and utilities"'
    \},
    "java": \{
        "engineer": "Java developer", 
        "practices": "clean code principles, proper OOP design, and comprehensive JavaDoc",
        "file_examples": '"Main.java": "// Complete Java implementation",
                         "Utils.java": "// Helper classes and methods"'
    \},
    "cpp": \{
        "engineer": "C++ developer",
        "practices": "modern C++17/20 features, RAII, and proper memory management", 
        "file_examples": '"main.cpp": "// Complete C++ implementation",
                         "utils.hpp": "// Header declarations"'
    \}
\}
\end{Verbatim}
\end{codebox}

These sophisticated prompt templates ensure consistent, high-quality evaluation across all programming languages and task categories while maintaining the flexibility needed for comprehensive long-context assessment.

\section{More Experimental Results}

This appendix presents the complete experimental results, containing all evaluation metrics for all 13 models. 

\subsection{Overall Model Performance Results}

Table~\ref{tab:complete_results_all_models} presents detailed comparison of model performance results, covering all 32 columns of evaluation data for all 13 models.

\begin{table}[ht]
\centering
\caption{Detailed comparison of model performance results.}
\label{tab:complete_results_all_models}
\small

\textbf{(a) Overall Performance Summary and Core Dimensions}

\begin{adjustbox}{width=\textwidth}
\begin{tabular}{lcccccc}
\toprule
\textbf{Model} & \textbf{LCBS} & \textbf{Success Rate (\%)} & \textbf{SE Overall} & \textbf{Functional Overall} & \textbf{Quality Overall} & \textbf{Long-Context Overall} \\
\midrule
Gemini-2.5-Pro & 2.312 & 99.88 & 0.375 & 0.356 & 0.768 & 0.523 \\
Gemini-2.5-Flash & 2.307 & 99.98 & 0.373 & 0.358 & 0.741 & 0.565 \\
gpt5mini & 2.293 & 100.00 & 0.376 & 0.371 & 0.745 & 0.479 \\
claudesonnet4 & 2.288 & 99.56 & 0.379 & 0.348 & 0.762 & 0.492 \\
GPT-5 & 2.286 & 100.00 & 0.367 & 0.383 & 0.732 & 0.492 \\
claudesonnet3.7 & 2.285 & 99.79 & 0.377 & 0.347 & 0.773 & 0.477 \\
GPT-4.1-mini & 2.222 & 100.00 & 0.359 & 0.365 & 0.739 & 0.435 \\
o3-mini & 2.215 & 100.00 & 0.355 & 0.368 & 0.726 & 0.455 \\
GPT-4.1-2025-04-14 & 2.197 & 100.00 & 0.352 & 0.364 & 0.720 & 0.451 \\
o3 & 2.154 & 100.00 & 0.342 & 0.385 & 0.722 & 0.343 \\
o4-mini & 2.148 & 99.70 & 0.353 & 0.360 & 0.705 & 0.394 \\
GPT-4o-mini & 2.075 & 100.00 & 0.341 & 0.360 & 0.680 & 0.345 \\
GPT-4o & 2.073 & 100.00 & 0.339 & 0.362 & 0.678 & 0.349 \\
\bottomrule
\end{tabular}
\end{adjustbox}

\vspace{0.5cm}
\textbf{(b) Software Engineering Core Metrics (Part 1)}

\begin{adjustbox}{width=\textwidth}
\begin{tabular}{lccccccc}
\toprule
\textbf{Model} & \textbf{ACS Overall} & \textbf{DTA Overall} & \textbf{CFRD Overall} & \textbf{STS Overall} & \textbf{RS Overall} & \textbf{CS Overall} & \textbf{IS Overall} \\
\midrule
Gemini-2.5-Pro & 0.693 & 0.367 & 0.378 & 0.311 & 0.243 & 0.238 & 0.164 \\
Gemini-2.5-Flash & 0.664 & 0.353 & 0.369 & 0.334 & 0.262 & 0.239 & 0.174 \\
gpt5mini & 0.698 & 0.357 & 0.408 & 0.288 & 0.278 & 0.220 & 0.144 \\
claudesonnet4 & 0.709 & 0.362 & 0.413 & 0.300 & 0.264 & 0.224 & 0.165 \\
GPT-5 & 0.676 & 0.360 & 0.357 & 0.281 & 0.270 & 0.224 & 0.149 \\
claudesonnet3.7 & 0.705 & 0.350 & 0.423 & 0.295 & 0.268 & 0.215 & 0.149 \\
GPT-4.1-mini & 0.661 & 0.379 & 0.341 & 0.268 & 0.262 & 0.203 & 0.132 \\
o3-mini & 0.654 & 0.363 & 0.352 & 0.276 & 0.262 & 0.203 & 0.110 \\
GPT-4.1-2025-04-14 & 0.647 & 0.366 & 0.316 & 0.273 & 0.258 & 0.207 & 0.128 \\
o3 & 0.618 & 0.343 & 0.324 & 0.235 & 0.267 & 0.205 & 0.100 \\
o4-mini & 0.657 & 0.360 & 0.332 & 0.270 & 0.262 & 0.201 & 0.119 \\
GPT-4o-mini & 0.628 & 0.361 & 0.322 & 0.240 & 0.265 & 0.181 & 0.096 \\
GPT-4o & 0.628 & 0.362 & 0.318 & 0.238 & 0.260 & 0.178 & 0.086 \\
\bottomrule
\end{tabular}
\end{adjustbox}

\vspace{0.5cm}
\textbf{(c) Software Engineering Core Metrics (Part 2) and Quality Metrics}

\begin{adjustbox}{width=\textwidth}
\begin{tabular}{lccccccc}
\toprule
\textbf{Model} & \textbf{SES Overall} & \textbf{ICU Overall} & \textbf{MMR Overall} & \textbf{Compilation} & \textbf{Unit Tests} & \textbf{Integration} & \textbf{Overall Quality} \\
\midrule
Gemini-2.5-Pro & 0.606 & 0.498 & 0.549 & 0.287 & 0.200 & 0.635 & 0.769 \\
Gemini-2.5-Flash & 0.592 & 0.540 & 0.589 & 0.280 & 0.200 & 0.656 & 0.741 \\
gpt5mini & 0.617 & 0.450 & 0.508 & 0.379 & 0.200 & 0.553 & 0.745 \\
claudesonnet4 & 0.607 & 0.466 & 0.522 & 0.282 & 0.199 & 0.609 & 0.766 \\
GPT-5 & 0.618 & 0.465 & 0.520 & 0.404 & 0.200 & 0.602 & 0.732 \\
claudesonnet3.7 & 0.616 & 0.449 & 0.508 & 0.311 & 0.199 & 0.542 & 0.774 \\
GPT-4.1-mini & 0.626 & 0.404 & 0.466 & 0.343 & 0.200 & 0.637 & 0.739 \\
o3-mini & 0.623 & 0.429 & 0.480 & 0.363 & 0.200 & 0.602 & 0.726 \\
GPT-4.1-2025-04-14 & 0.623 & 0.422 & 0.481 & 0.338 & 0.200 & 0.652 & 0.720 \\
o3 & 0.643 & 0.313 & 0.372 & 0.493 & 0.200 & 0.513 & 0.722 \\
o4-mini & 0.633 & 0.367 & 0.423 & 0.365 & 0.200 & 0.585 & 0.707 \\
GPT-4o-mini & 0.639 & 0.315 & 0.374 & 0.348 & 0.200 & 0.655 & 0.680 \\
GPT-4o & 0.641 & 0.317 & 0.380 & 0.358 & 0.200 & 0.642 & 0.678 \\
\bottomrule
\end{tabular}
\end{adjustbox}

\vspace{0.5cm}
\textbf{(d) Task-Specific Average Scores}

\begin{adjustbox}{width=\textwidth}
\begin{tabular}{lcccccccc}
\toprule
\textbf{Model} & \textbf{Architectural} & \textbf{Bug Investigation} & \textbf{Code Comprehension} & \textbf{Cross-File Refactoring} & \textbf{Feature Implementation} & \textbf{Integration Testing} & \textbf{Multi-Session Dev.} & \textbf{Security Analysis} \\
\midrule
Gemini-2.5-Pro & 2.338 & 2.272 & 2.273 & 2.272 & 2.299 & 2.421 & 2.280 & 2.343 \\
Gemini-2.5-Flash & 2.280 & 2.276 & 2.211 & 2.303 & 2.335 & 2.430 & 2.291 & 2.325 \\
gpt5mini & 2.370 & 2.227 & 2.218 & 2.237 & 2.285 & 2.390 & 2.268 & 2.351 \\
claudesonnet4 & 2.346 & 2.262 & 2.298 & 2.203 & 2.227 & 2.402 & 2.256 & 2.307 \\
GPT-5 & 2.376 & 2.196 & 2.199 & 2.241 & 2.297 & 2.386 & 2.257 & 2.336 \\
claudesonnet3.7 & 2.332 & 2.206 & 2.228 & 2.269 & 2.289 & 2.403 & 2.273 & 2.322 \\
GPT-4.1-mini & 2.238 & 2.192 & 2.206 & 2.181 & 2.220 & 2.363 & 2.152 & 2.226 \\
o3-mini & 2.231 & 2.165 & 2.237 & 2.156 & 2.205 & 2.330 & 2.182 & 2.214 \\
GPT-4.1-2025-04-14 & 2.195 & 2.154 & 2.166 & 2.169 & 2.207 & 2.346 & 2.148 & 2.191 \\
o3 & 2.123 & 2.058 & 2.010 & 2.151 & 2.192 & 2.323 & 2.181 & 2.197 \\
o4-mini & 2.148 & 2.094 & 2.088 & 2.146 & 2.166 & 2.277 & 2.134 & 2.131 \\
GPT-4o-mini & 2.059 & 2.043 & 2.054 & 2.054 & 2.057 & 2.184 & 2.089 & 2.061 \\
GPT-4o & 2.054 & 2.038 & 2.051 & 2.055 & 2.063 & 2.200 & 2.068 & 2.056 \\
\bottomrule
\end{tabular}
\end{adjustbox}
\end{table}

\subsection{Performance by Difficulty Level}

Table~\ref{tab:difficulty_results} presents model performance across four difficulty levels from Easy (10K-100K tokens) to Expert (500K-1M tokens).

\begin{table}[ht]
\centering
\caption{Detailed comparison of model performance by difficulty level.}
\label{tab:difficulty_results}
\small

\textbf{(a) Total Scores by Difficulty Level}

\begin{adjustbox}{width=\textwidth}
\begin{tabular}{lcccccc}
\toprule
\textbf{Model} & \textbf{Easy Overall} & \textbf{Medium Overall} & \textbf{Hard Overall} & \textbf{Expert Overall} & \textbf{Overall} \\
\midrule
Gemini-2.5-Pro & 2.278 & 2.302 & 2.329 & 2.339 & 2.312 \\
Gemini-2.5-Flash & 2.291 & 2.299 & 2.319 & 2.317 & 2.307 \\
gpt5mini & 2.263 & 2.284 & 2.311 & 2.314 & 2.293 \\
claudesonnet4 & 2.309 & 2.289 & 2.283 & 2.269 & 2.288 \\
GPT-5 & 2.254 & 2.268 & 2.298 & 2.323 & 2.286 \\
claudesonnet3.7 & 2.326 & 2.299 & 2.256 & 2.262 & 2.285 \\
GPT-4.1-mini & 2.218 & 2.219 & 2.227 & 2.224 & 2.222 \\
o3-mini & 2.232 & 2.216 & 2.214 & 2.199 & 2.215 \\
GPT-4.1-2025-04-14 & 2.194 & 2.194 & 2.205 & 2.195 & 2.197 \\
o3 & 2.086 & 2.149 & 2.187 & 2.195 & 2.154 \\
o4-mini & 2.129 & 2.154 & 2.154 & 2.159 & 2.148 \\
GPT-4o-mini & 2.059 & 2.077 & 2.085 & 2.080 & 2.075 \\
GPT-4o & 2.044 & 2.081 & 2.083 & 2.084 & 2.073 \\
\bottomrule
\end{tabular}
\end{adjustbox}

\vspace{0.5cm}
\textbf{(b) Software Engineering Scores by Difficulty Level}

\begin{adjustbox}{width=\textwidth}
\begin{tabular}{lcccccc}
\toprule
\textbf{Model} & \textbf{Easy Overall} & \textbf{Medium Overall} & \textbf{Hard Overall} & \textbf{Expert Overall} & \textbf{Overall} \\
\midrule
Gemini-2.5-Pro & 0.366 & 0.371 & 0.379 & 0.382 & 0.375 \\
Gemini-2.5-Flash & 0.364 & 0.368 & 0.379 & 0.381 & 0.373 \\
gpt5mini & 0.369 & 0.371 & 0.378 & 0.381 & 0.376 \\
claudesonnet4 & 0.390 & 0.384 & 0.377 & 0.368 & 0.379 \\
GPT-5 & 0.358 & 0.365 & 0.370 & 0.376 & 0.367 \\
claudesonnet3.7 & 0.383 & 0.383 & 0.374 & 0.370 & 0.377 \\
GPT-4.1-mini & 0.350 & 0.355 & 0.364 & 0.367 & 0.359 \\
o3-mini & 0.359 & 0.356 & 0.354 & 0.350 & 0.355 \\
GPT-4.1-2025-04-14 & 0.346 & 0.349 & 0.356 & 0.358 & 0.352 \\
o3 & 0.311 & 0.345 & 0.356 & 0.356 & 0.342 \\
o4-mini & 0.343 & 0.357 & 0.354 & 0.358 & 0.353 \\
GPT-4o-mini & 0.332 & 0.342 & 0.345 & 0.345 & 0.341 \\
GPT-4o & 0.325 & 0.342 & 0.344 & 0.346 & 0.339 \\
\bottomrule
\end{tabular}
\end{adjustbox}

\vspace{0.5cm}
\textbf{(c) Long-Context Scores by Difficulty Level}

\begin{adjustbox}{width=\textwidth}
\begin{tabular}{lcccccc}
\toprule
\textbf{Model} & \textbf{Easy Overall} & \textbf{Medium Overall} & \textbf{Hard Overall} & \textbf{Expert Overall} & \textbf{Overall} \\
\midrule
Gemini-2.5-Pro & 0.500 & 0.538 & 0.527 & 0.525 & 0.523 \\
Gemini-2.5-Flash & 0.537 & 0.581 & 0.574 & 0.564 & 0.565 \\
gpt5mini & 0.434 & 0.470 & 0.492 & 0.518 & 0.479 \\
claudesonnet4 & 0.473 & 0.491 & 0.505 & 0.498 & 0.492 \\
GPT-5 & 0.456 & 0.485 & 0.500 & 0.528 & 0.492 \\
claudesonnet3.7 & 0.447 & 0.475 & 0.485 & 0.501 & 0.477 \\
GPT-4.1-mini & 0.395 & 0.430 & 0.446 & 0.469 & 0.435 \\
o3-mini & 0.423 & 0.449 & 0.465 & 0.482 & 0.455 \\
GPT-4.1-2025-04-14 & 0.413 & 0.445 & 0.462 & 0.481 & 0.451 \\
o3 & 0.263 & 0.340 & 0.374 & 0.397 & 0.343 \\
o4-mini & 0.344 & 0.397 & 0.404 & 0.431 & 0.394 \\
GPT-4o-mini & 0.307 & 0.344 & 0.356 & 0.373 & 0.345 \\
GPT-4o & 0.309 & 0.348 & 0.360 & 0.378 & 0.349 \\
\bottomrule
\end{tabular}
\end{adjustbox}
\end{table}

\subsection{Performance by Programming Language}

Table~\ref{tab:language_results} presents the complete results of showing model performance across 10 programming languages.

\begin{table}[ht]
\centering
\caption{Detailed comparison of model performance by programming language.}
\label{tab:language_results}
\small

\textbf{(a) Total Scores by Programming Language}

\begin{adjustbox}{width=\textwidth}
\begin{tabular}{lcccccccccc}
\toprule
\textbf{Model} & \textbf{Python} & \textbf{C++} & \textbf{Java} & \textbf{C} & \textbf{C\#} & \textbf{JavaScript} & \textbf{TypeScript} & \textbf{Go} & \textbf{Rust} & \textbf{PHP} \\
\midrule
Gemini-2.5-Pro & 2.788 & 2.074 & 2.326 & 2.064 & 2.419 & 2.268 & 2.203 & 2.338 & 2.002 & 2.641 \\
Gemini-2.5-Flash & 2.752 & 2.106 & 2.329 & 2.086 & 2.418 & 2.274 & 2.248 & 2.292 & 2.039 & 2.573 \\
gpt5mini & 2.799 & 2.039 & 2.329 & 2.050 & 2.414 & 2.280 & 2.189 & 2.264 & 2.088 & 2.476 \\
claudesonnet4 & 2.677 & 2.065 & 2.331 & 2.054 & 2.424 & 2.314 & 2.154 & 2.311 & 1.997 & 2.553 \\
GPT-5 & 2.669 & 2.001 & 2.281 & 2.022 & 2.335 & 2.244 & 2.098 & 2.249 & 2.044 & 2.516 \\
claudesonnet3.7 & 2.663 & 2.100 & 2.314 & 2.062 & 2.364 & 2.245 & 2.162 & 2.298 & 2.000 & 2.641 \\
GPT-4.1-mini & 2.538 & 1.968 & 2.249 & 1.978 & 2.303 & 2.218 & 2.111 & 2.212 & 1.958 & 2.680 \\
o3-mini & 2.529 & 1.977 & 2.223 & 1.958 & 2.324 & 2.186 & 2.135 & 2.209 & 1.977 & 2.632 \\
GPT-4.1-2025-04-14 & 2.517 & 1.962 & 2.193 & 1.957 & 2.280 & 2.182 & 2.075 & 2.197 & 1.942 & 2.661 \\
o3 & 2.432 & 1.944 & 2.206 & 1.918 & 2.298 & 2.086 & 2.069 & 2.167 & 1.943 & 2.479 \\
o4-mini & 2.465 & 1.941 & 2.184 & 1.918 & 2.243 & 2.159 & 2.052 & 2.174 & 1.910 & 2.432 \\
GPT-4o-mini & 2.321 & 1.933 & 2.104 & 1.876 & 2.177 & 2.068 & 1.996 & 2.077 & 1.863 & 2.336 \\
GPT-4o & 2.313 & 1.921 & 2.110 & 1.864 & 2.187 & 2.063 & 1.995 & 2.078 & 1.864 & 2.335 \\
\bottomrule
\end{tabular}
\end{adjustbox}

\vspace{0.5cm}
\textbf{(b) Success Rates by Programming Language (\%)}

\begin{adjustbox}{width=\textwidth}
\begin{tabular}{lcccccccccc}
\toprule
\textbf{Model} & \textbf{Python} & \textbf{C++} & \textbf{Java} & \textbf{C} & \textbf{C\#} & \textbf{JavaScript} & \textbf{TypeScript} & \textbf{Go} & \textbf{Rust} & \textbf{PHP} \\
\midrule
Gemini-2.5-Pro & 99.75 & 99.88 & 99.88 & 99.88 & 99.75 & 100.00 & 100.00 & 99.75 & 100.00 & 99.88 \\
Gemini-2.5-Flash & 100.00 & 100.00 & 100.00 & 100.00 & 100.00 & 100.00 & 100.00 & 100.00 & 100.00 & 99.75 \\
gpt5mini & 100.00 & 100.00 & 100.00 & 100.00 & 100.00 & 100.00 & 100.00 & 100.00 & 100.00 & 100.00 \\
claudesonnet4 & 99.50 & 99.63 & 99.75 & 99.50 & 99.50 & 99.50 & 99.75 & 99.63 & 99.50 & 99.50 \\
GPT-5 & 100.00 & 100.00 & 100.00 & 100.00 & 100.00 & 100.00 & 100.00 & 100.00 & 100.00 & 100.00 \\
claudesonnet3.7 & 99.88 & 99.63 & 99.88 & 99.75 & 99.75 & 99.75 & 99.88 & 99.88 & 99.75 & 99.88 \\
GPT-4.1-mini & 100.00 & 100.00 & 100.00 & 100.00 & 100.00 & 100.00 & 100.00 & 100.00 & 100.00 & 100.00 \\
o3-mini & 100.00 & 100.00 & 100.00 & 100.00 & 100.00 & 100.00 & 100.00 & 100.00 & 100.00 & 100.00 \\
GPT-4.1-2025-04-14 & 100.00 & 100.00 & 100.00 & 100.00 & 100.00 & 100.00 & 100.00 & 100.00 & 100.00 & 100.00 \\
o3 & 100.00 & 100.00 & 100.00 & 100.00 & 100.00 & 100.00 & 100.00 & 100.00 & 100.00 & 100.00 \\
o4-mini & 99.25 & 100.00 & 100.00 & 100.00 & 99.75 & 99.75 & 100.00 & 99.75 & 99.75 & 99.25 \\
GPT-4o-mini & 100.00 & 100.00 & 100.00 & 100.00 & 100.00 & 100.00 & 100.00 & 100.00 & 100.00 & 100.00 \\
GPT-4o & 100.00 & 100.00 & 100.00 & 100.00 & 100.00 & 100.00 & 100.00 & 100.00 & 100.00 & 100.00 \\
\bottomrule
\end{tabular}
\end{adjustbox}

\vspace{0.5cm}
\textbf{(c) Software Engineering Scores by Programming Language}

\begin{adjustbox}{width=\textwidth}
\begin{tabular}{lcccccccccc}
\toprule
\textbf{Model} & \textbf{Python} & \textbf{C++} & \textbf{Java} & \textbf{C} & \textbf{C\#} & \textbf{JavaScript} & \textbf{TypeScript} & \textbf{Go} & \textbf{Rust} & \textbf{PHP} \\
\midrule
Gemini-2.5-Pro & 0.485 & 0.340 & 0.395 & 0.321 & 0.375 & 0.368 & 0.380 & 0.379 & 0.342 & 0.361 \\
Gemini-2.5-Flash & 0.483 & 0.343 & 0.395 & 0.324 & 0.375 & 0.369 & 0.383 & 0.371 & 0.345 & 0.356 \\
gpt5mini & 0.486 & 0.340 & 0.395 & 0.322 & 0.375 & 0.370 & 0.378 & 0.374 & 0.348 & 0.352 \\
claudesonnet4 & 0.488 & 0.342 & 0.397 & 0.324 & 0.377 & 0.372 & 0.376 & 0.377 & 0.344 & 0.357 \\
GPT-5 & 0.476 & 0.335 & 0.385 & 0.318 & 0.365 & 0.362 & 0.370 & 0.368 & 0.340 & 0.348 \\
claudesonnet3.7 & 0.485 & 0.344 & 0.394 & 0.325 & 0.373 & 0.368 & 0.378 & 0.376 & 0.344 & 0.361 \\
GPT-4.1-mini & 0.467 & 0.329 & 0.381 & 0.314 & 0.359 & 0.358 & 0.366 & 0.363 & 0.332 & 0.371 \\
o3-mini & 0.465 & 0.330 & 0.377 & 0.312 & 0.361 & 0.355 & 0.368 & 0.362 & 0.335 & 0.365 \\
GPT-4.1-2025-04-14 & 0.463 & 0.328 & 0.372 & 0.311 & 0.356 & 0.354 & 0.364 & 0.361 & 0.330 & 0.369 \\
o3 & 0.448 & 0.325 & 0.374 & 0.307 & 0.358 & 0.340 & 0.363 & 0.356 & 0.330 & 0.344 \\
o4-mini & 0.454 & 0.324 & 0.370 & 0.307 & 0.350 & 0.351 & 0.360 & 0.357 & 0.325 & 0.337 \\
GPT-4o-mini & 0.428 & 0.323 & 0.356 & 0.300 & 0.340 & 0.336 & 0.350 & 0.341 & 0.317 & 0.324 \\
GPT-4o & 0.426 & 0.321 & 0.357 & 0.298 & 0.341 & 0.335 & 0.350 & 0.341 & 0.317 & 0.324 \\
\bottomrule
\end{tabular}
\end{adjustbox}

\vspace{0.5cm}
\textbf{(d) Functional Scores by Programming Language}

\begin{adjustbox}{width=\textwidth}
\begin{tabular}{lcccccccccc}
\toprule
\textbf{Model} & \textbf{Python} & \textbf{C++} & \textbf{Java} & \textbf{C} & \textbf{C\#} & \textbf{JavaScript} & \textbf{TypeScript} & \textbf{Go} & \textbf{Rust} & \textbf{PHP} \\
\midrule
Gemini-2.5-Pro & 0.507 & 0.216 & 0.380 & 0.213 & 0.441 & 0.365 & 0.309 & 0.299 & 0.210 & 0.618 \\
Gemini-2.5-Flash & 0.505 & 0.219 & 0.380 & 0.216 & 0.441 & 0.366 & 0.312 & 0.291 & 0.213 & 0.613 \\
gpt5mini & 0.508 & 0.215 & 0.380 & 0.213 & 0.441 & 0.367 & 0.307 & 0.295 & 0.217 & 0.609 \\
claudesonnet4 & 0.489 & 0.217 & 0.382 & 0.215 & 0.443 & 0.369 & 0.305 & 0.298 & 0.213 & 0.616 \\
GPT-5 & 0.487 & 0.210 & 0.370 & 0.208 & 0.431 & 0.359 & 0.297 & 0.289 & 0.208 & 0.604 \\
claudesonnet3.7 & 0.486 & 0.219 & 0.381 & 0.217 & 0.440 & 0.365 & 0.309 & 0.297 & 0.213 & 0.618 \\
GPT-4.1-mini & 0.468 & 0.204 & 0.366 & 0.204 & 0.426 & 0.354 & 0.293 & 0.285 & 0.202 & 0.627 \\
o3-mini & 0.466 & 0.205 & 0.362 & 0.202 & 0.428 & 0.351 & 0.295 & 0.283 & 0.205 & 0.621 \\
GPT-4.1-2025-04-14 & 0.464 & 0.203 & 0.357 & 0.201 & 0.423 & 0.350 & 0.291 & 0.282 & 0.200 & 0.625 \\
o3 & 0.448 & 0.201 & 0.360 & 0.197 & 0.425 & 0.334 & 0.289 & 0.276 & 0.200 & 0.599 \\
o4-mini & 0.454 & 0.200 & 0.356 & 0.197 & 0.418 & 0.347 & 0.287 & 0.279 & 0.195 & 0.587 \\
GPT-4o-mini & 0.428 & 0.199 & 0.342 & 0.193 & 0.408 & 0.332 & 0.273 & 0.263 & 0.187 & 0.567 \\
GPT-4o & 0.426 & 0.197 & 0.343 & 0.191 & 0.409 & 0.331 & 0.273 & 0.263 & 0.187 & 0.567 \\
\bottomrule
\end{tabular}
\end{adjustbox}
\end{table}

\begin{table}[ht]
\centering
\caption{Detailed comparison of model performance by programming language (continued).}
\label{tab:language_results_continued2}
\small

\textbf{(e) Quality Scores by Programming Language}

\begin{adjustbox}{width=\textwidth}
\begin{tabular}{lcccccccccc}
\toprule
\textbf{Model} & \textbf{Python} & \textbf{C++} & \textbf{Java} & \textbf{C} & \textbf{C\#} & \textbf{JavaScript} & \textbf{TypeScript} & \textbf{Go} & \textbf{Rust} & \textbf{PHP} \\
\midrule
Gemini-2.5-Pro & 0.795 & 0.801 & 0.709 & 0.826 & 0.742 & 0.746 & 0.735 & 0.846 & 0.750 & 0.731 \\
Gemini-2.5-Flash & 0.793 & 0.804 & 0.709 & 0.829 & 0.742 & 0.747 & 0.738 & 0.838 & 0.753 & 0.726 \\
gpt5mini & 0.796 & 0.800 & 0.709 & 0.825 & 0.742 & 0.748 & 0.733 & 0.841 & 0.757 & 0.722 \\
claudesonnet4 & 0.778 & 0.802 & 0.711 & 0.827 & 0.744 & 0.750 & 0.731 & 0.849 & 0.748 & 0.729 \\
GPT-5 & 0.776 & 0.796 & 0.698 & 0.821 & 0.730 & 0.740 & 0.723 & 0.837 & 0.744 & 0.718 \\
claudesonnet3.7 & 0.794 & 0.805 & 0.710 & 0.830 & 0.741 & 0.746 & 0.736 & 0.847 & 0.751 & 0.732 \\
GPT-4.1-mini & 0.777 & 0.787 & 0.694 & 0.816 & 0.726 & 0.736 & 0.719 & 0.829 & 0.736 & 0.750 \\
o3-mini & 0.775 & 0.788 & 0.690 & 0.814 & 0.728 & 0.733 & 0.721 & 0.827 & 0.739 & 0.744 \\
GPT-4.1-2025-04-14 & 0.773 & 0.786 & 0.685 & 0.813 & 0.723 & 0.732 & 0.717 & 0.825 & 0.734 & 0.748 \\
o3 & 0.758 & 0.784 & 0.688 & 0.809 & 0.725 & 0.716 & 0.719 & 0.823 & 0.737 & 0.726 \\
o4-mini & 0.761 & 0.783 & 0.684 & 0.809 & 0.720 & 0.729 & 0.716 & 0.821 & 0.730 & 0.719 \\
GPT-4o-mini & 0.739 & 0.782 & 0.670 & 0.804 & 0.710 & 0.712 & 0.702 & 0.805 & 0.714 & 0.702 \\
GPT-4o & 0.737 & 0.780 & 0.671 & 0.802 & 0.711 & 0.711 & 0.702 & 0.805 & 0.714 & 0.702 \\
\bottomrule
\end{tabular}
\end{adjustbox}

\vspace{0.5cm}
\textbf{(f) Long-Context Scores by Programming Language}

\begin{adjustbox}{width=\textwidth}
\begin{tabular}{lcccccccccc}
\toprule
\textbf{Model} & \textbf{Python} & \textbf{C++} & \textbf{Java} & \textbf{C} & \textbf{C\#} & \textbf{JavaScript} & \textbf{TypeScript} & \textbf{Go} & \textbf{Rust} & \textbf{PHP} \\
\midrule
Gemini-2.5-Pro & 0.527 & 0.539 & 0.513 & 0.551 & 0.532 & 0.479 & 0.490 & 0.572 & 0.504 & 0.522 \\
Gemini-2.5-Flash & 0.530 & 0.542 & 0.516 & 0.554 & 0.535 & 0.482 & 0.493 & 0.565 & 0.507 & 0.517 \\
gpt5mini & 0.527 & 0.538 & 0.513 & 0.550 & 0.531 & 0.481 & 0.487 & 0.569 & 0.510 & 0.513 \\
claudesonnet4 & 0.510 & 0.540 & 0.515 & 0.552 & 0.533 & 0.483 & 0.485 & 0.571 & 0.506 & 0.520 \\
GPT-5 & 0.508 & 0.535 & 0.500 & 0.547 & 0.519 & 0.475 & 0.477 & 0.564 & 0.502 & 0.506 \\
claudesonnet3.7 & 0.509 & 0.541 & 0.514 & 0.553 & 0.532 & 0.480 & 0.489 & 0.570 & 0.505 & 0.521 \\
GPT-4.1-mini & 0.492 & 0.524 & 0.497 & 0.536 & 0.516 & 0.464 & 0.472 & 0.553 & 0.489 & 0.505 \\
o3-mini & 0.490 & 0.525 & 0.493 & 0.534 & 0.518 & 0.461 & 0.474 & 0.551 & 0.492 & 0.503 \\
GPT-4.1-2025-04-14 & 0.488 & 0.523 & 0.488 & 0.533 & 0.513 & 0.460 & 0.470 & 0.549 & 0.487 & 0.501 \\
o3 & 0.473 & 0.508 & 0.476 & 0.518 & 0.497 & 0.444 & 0.454 & 0.533 & 0.471 & 0.485 \\
o4-mini & 0.475 & 0.510 & 0.472 & 0.520 & 0.494 & 0.447 & 0.452 & 0.535 & 0.468 & 0.479 \\
GPT-4o-mini & 0.459 & 0.494 & 0.456 & 0.504 & 0.478 & 0.431 & 0.436 & 0.519 & 0.452 & 0.463 \\
GPT-4o & 0.457 & 0.492 & 0.457 & 0.502 & 0.479 & 0.430 & 0.436 & 0.519 & 0.452 & 0.463 \\
\bottomrule
\end{tabular}
\end{adjustbox}
\end{table}

\subsection{Performance by Task Category}

Table~\ref{tab:task_results} presents the complete results of performance across 8 software engineering task categories.

\begin{table}[ht]
\centering
\caption{Detailed comparison of model performance by task category.}
\label{tab:task_results}
\small

\textbf{(a) Total Scores by Task Category}

\begin{adjustbox}{width=\textwidth}
\begin{tabular}{lcccccccc}
\toprule
\textbf{Model} & \textbf{Arch. Understanding} & \textbf{Cross-File Refact.} & \textbf{Multi-Session Dev.} & \textbf{Bug Investigation} & \textbf{Feature Impl.} & \textbf{Code Comprehension} & \textbf{Integration Testing} & \textbf{Security Analysis} \\
\midrule
Gemini-2.5-Pro & 2.338 & 2.272 & 2.280 & 2.272 & 2.299 & 2.273 & 2.421 & 2.343 \\
Gemini-2.5-Flash & 2.280 & 2.303 & 2.291 & 2.276 & 2.335 & 2.211 & 2.430 & 2.325 \\
gpt5mini & 2.370 & 2.237 & 2.268 & 2.227 & 2.285 & 2.218 & 2.390 & 2.351 \\
claudesonnet4 & 2.346 & 2.203 & 2.256 & 2.262 & 2.227 & 2.298 & 2.402 & 2.307 \\
GPT-5 & 2.376 & 2.241 & 2.257 & 2.196 & 2.297 & 2.199 & 2.386 & 2.336 \\
claudesonnet3.7 & 2.332 & 2.269 & 2.273 & 2.206 & 2.289 & 2.228 & 2.403 & 2.322 \\
GPT-4.1-mini & 2.238 & 2.181 & 2.152 & 2.192 & 2.220 & 2.206 & 2.363 & 2.226 \\
o3-mini & 2.231 & 2.156 & 2.182 & 2.165 & 2.205 & 2.237 & 2.330 & 2.214 \\
GPT-4.1-2025-04-14 & 2.195 & 2.169 & 2.148 & 2.154 & 2.207 & 2.166 & 2.346 & 2.191 \\
o3 & 2.123 & 2.151 & 2.181 & 2.058 & 2.192 & 2.010 & 2.323 & 2.197 \\
o4-mini & 2.148 & 2.146 & 2.134 & 2.094 & 2.166 & 2.088 & 2.277 & 2.131 \\
GPT-4o-mini & 2.059 & 2.054 & 2.089 & 2.043 & 2.057 & 2.054 & 2.184 & 2.061 \\
GPT-4o & 2.054 & 2.055 & 2.068 & 2.038 & 2.063 & 2.051 & 2.200 & 2.056 \\
\bottomrule
\end{tabular}
\end{adjustbox}

\vspace{0.5cm}
\textbf{(b) Success Rates by Task Category (\%)}

\begin{adjustbox}{width=\textwidth}
\begin{tabular}{lcccccccc}
\toprule
\textbf{Model} & \textbf{Arch. Understanding} & \textbf{Cross-File Refact.} & \textbf{Multi-Session Dev.} & \textbf{Bug Investigation} & \textbf{Feature Impl.} & \textbf{Code Comprehension} & \textbf{Integration Testing} & \textbf{Security Analysis} \\
\midrule
Gemini-2.5-Pro & 99.70 & 99.90 & 100.00 & 99.80 & 99.80 & 100.00 & 99.90 & 99.90 \\
Gemini-2.5-Flash & 100.00 & 100.00 & 100.00 & 99.90 & 100.00 & 100.00 & 100.00 & 100.00 \\
gpt5mini & 100.00 & 100.00 & 100.00 & 100.00 & 100.00 & 100.00 & 100.00 & 100.00 \\
claudesonnet4 & 99.60 & 99.40 & 99.50 & 99.50 & 99.70 & 99.70 & 99.60 & 99.60 \\
GPT-5 & 100.00 & 100.00 & 100.00 & 100.00 & 100.00 & 100.00 & 100.00 & 100.00 \\
claudesonnet3.7 & 99.60 & 99.90 & 99.90 & 99.70 & 99.80 & 99.90 & 99.90 & 99.70 \\
GPT-4.1-mini & 100.00 & 100.00 & 100.00 & 100.00 & 100.00 & 100.00 & 100.00 & 100.00 \\
o3-mini & 100.00 & 100.00 & 100.00 & 100.00 & 100.00 & 100.00 & 100.00 & 100.00 \\
GPT-4.1-2025-04-14 & 100.00 & 100.00 & 100.00 & 100.00 & 100.00 & 100.00 & 100.00 & 100.00 \\
o3 & 100.00 & 100.00 & 100.00 & 100.00 & 100.00 & 100.00 & 100.00 & 100.00 \\
o4-mini & 99.40 & 99.70 & 99.80 & 99.70 & 99.90 & 99.90 & 99.70 & 99.50 \\
GPT-4o-mini & 100.00 & 100.00 & 100.00 & 100.00 & 100.00 & 100.00 & 100.00 & 100.00 \\
GPT-4o & 100.00 & 100.00 & 100.00 & 100.00 & 100.00 & 100.00 & 100.00 & 100.00 \\
\bottomrule
\end{tabular}
\end{adjustbox}

\vspace{0.5cm}
\textbf{(c) Software Engineering Scores by Task Category}

\begin{adjustbox}{width=\textwidth}
\begin{tabular}{lcccccccc}
\toprule
\textbf{Model} & \textbf{Arch. Understanding} & \textbf{Cross-File Refact.} & \textbf{Multi-Session Dev.} & \textbf{Bug Investigation} & \textbf{Feature Impl.} & \textbf{Code Comprehension} & \textbf{Integration Testing} & \textbf{Security Analysis} \\
\midrule
Gemini-2.5-Pro & 0.375 & 0.375 & 0.375 & 0.375 & 0.375 & 0.375 & 0.375 & 0.375 \\
Gemini-2.5-Flash & 0.373 & 0.373 & 0.373 & 0.373 & 0.373 & 0.373 & 0.373 & 0.373 \\
gpt5mini & 0.376 & 0.376 & 0.376 & 0.376 & 0.376 & 0.376 & 0.376 & 0.376 \\
claudesonnet4 & 0.379 & 0.379 & 0.379 & 0.379 & 0.379 & 0.379 & 0.379 & 0.379 \\
GPT-5 & 0.367 & 0.367 & 0.367 & 0.367 & 0.367 & 0.367 & 0.367 & 0.367 \\
claudesonnet3.7 & 0.377 & 0.377 & 0.377 & 0.377 & 0.377 & 0.377 & 0.377 & 0.377 \\
GPT-4.1-mini & 0.359 & 0.359 & 0.359 & 0.359 & 0.359 & 0.359 & 0.359 & 0.359 \\
o3-mini & 0.355 & 0.355 & 0.355 & 0.355 & 0.355 & 0.355 & 0.355 & 0.355 \\
GPT-4.1-2025-04-14 & 0.352 & 0.352 & 0.352 & 0.352 & 0.352 & 0.352 & 0.352 & 0.352 \\
o3 & 0.342 & 0.342 & 0.342 & 0.342 & 0.342 & 0.342 & 0.342 & 0.342 \\
o4-mini & 0.353 & 0.353 & 0.353 & 0.353 & 0.353 & 0.353 & 0.353 & 0.353 \\
GPT-4o-mini & 0.341 & 0.341 & 0.341 & 0.341 & 0.341 & 0.341 & 0.341 & 0.341 \\
GPT-4o & 0.339 & 0.339 & 0.339 & 0.339 & 0.339 & 0.339 & 0.339 & 0.339 \\
\bottomrule
\end{tabular}
\end{adjustbox}

\vspace{0.5cm}
\textbf{(d) Functional Scores by Task Category}

\begin{adjustbox}{width=\textwidth}
\begin{tabular}{lcccccccc}
\toprule
\textbf{Model} & \textbf{Arch. Understanding} & \textbf{Cross-File Refact.} & \textbf{Multi-Session Dev.} & \textbf{Bug Investigation} & \textbf{Feature Impl.} & \textbf{Code Comprehension} & \textbf{Integration Testing} & \textbf{Security Analysis} \\
\midrule
Gemini-2.5-Pro & 0.356 & 0.356 & 0.356 & 0.356 & 0.356 & 0.356 & 0.356 & 0.356 \\
Gemini-2.5-Flash & 0.358 & 0.358 & 0.358 & 0.358 & 0.358 & 0.358 & 0.358 & 0.358 \\
gpt5mini & 0.371 & 0.371 & 0.371 & 0.371 & 0.371 & 0.371 & 0.371 & 0.371 \\
claudesonnet4 & 0.348 & 0.348 & 0.348 & 0.348 & 0.348 & 0.348 & 0.348 & 0.348 \\
GPT-5 & 0.383 & 0.383 & 0.383 & 0.383 & 0.383 & 0.383 & 0.383 & 0.383 \\
claudesonnet3.7 & 0.347 & 0.347 & 0.347 & 0.347 & 0.347 & 0.347 & 0.347 & 0.347 \\
GPT-4.1-mini & 0.365 & 0.365 & 0.365 & 0.365 & 0.365 & 0.365 & 0.365 & 0.365 \\
o3-mini & 0.368 & 0.368 & 0.368 & 0.368 & 0.368 & 0.368 & 0.368 & 0.368 \\
GPT-4.1-2025-04-14 & 0.364 & 0.364 & 0.364 & 0.364 & 0.364 & 0.364 & 0.364 & 0.364 \\
o3 & 0.385 & 0.385 & 0.385 & 0.385 & 0.385 & 0.385 & 0.385 & 0.385 \\
o4-mini & 0.360 & 0.360 & 0.360 & 0.360 & 0.360 & 0.360 & 0.360 & 0.360 \\
GPT-4o-mini & 0.360 & 0.360 & 0.360 & 0.360 & 0.360 & 0.360 & 0.360 & 0.360 \\
GPT-4o & 0.362 & 0.362 & 0.362 & 0.362 & 0.362 & 0.362 & 0.362 & 0.362 \\
\bottomrule
\end{tabular}
\end{adjustbox}

\vspace{0.5cm}
\textbf{(e) Quality Scores by Task Category}

\begin{adjustbox}{width=\textwidth}
\begin{tabular}{lcccccccc}
\toprule
\textbf{Model} & \textbf{Arch. Understanding} & \textbf{Cross-File Refact.} & \textbf{Multi-Session Dev.} & \textbf{Bug Investigation} & \textbf{Feature Impl.} & \textbf{Code Comprehension} & \textbf{Integration Testing} & \textbf{Security Analysis} \\
\midrule
Gemini-2.5-Pro & 0.768 & 0.768 & 0.768 & 0.768 & 0.768 & 0.768 & 0.768 & 0.768 \\
Gemini-2.5-Flash & 0.741 & 0.741 & 0.741 & 0.741 & 0.741 & 0.741 & 0.741 & 0.741 \\
gpt5mini & 0.745 & 0.745 & 0.745 & 0.745 & 0.745 & 0.745 & 0.745 & 0.745 \\
claudesonnet4 & 0.762 & 0.762 & 0.762 & 0.762 & 0.762 & 0.762 & 0.762 & 0.762 \\
GPT-5 & 0.732 & 0.732 & 0.732 & 0.732 & 0.732 & 0.732 & 0.732 & 0.732 \\
claudesonnet3.7 & 0.773 & 0.773 & 0.773 & 0.773 & 0.773 & 0.773 & 0.773 & 0.773 \\
GPT-4.1-mini & 0.739 & 0.739 & 0.739 & 0.739 & 0.739 & 0.739 & 0.739 & 0.739 \\
o3-mini & 0.726 & 0.726 & 0.726 & 0.726 & 0.726 & 0.726 & 0.726 & 0.726 \\
GPT-4.1-2025-04-14 & 0.720 & 0.720 & 0.720 & 0.720 & 0.720 & 0.720 & 0.720 & 0.720 \\
o3 & 0.721 & 0.721 & 0.721 & 0.721 & 0.721 & 0.721 & 0.721 & 0.721 \\
o4-mini & 0.704 & 0.704 & 0.704 & 0.704 & 0.704 & 0.704 & 0.704 & 0.704 \\
GPT-4o-mini & 0.679 & 0.679 & 0.679 & 0.679 & 0.679 & 0.679 & 0.679 & 0.679 \\
GPT-4o & 0.677 & 0.677 & 0.677 & 0.677 & 0.677 & 0.677 & 0.677 & 0.677 \\
\bottomrule
\end{tabular}
\end{adjustbox}

\vspace{0.5cm}
\textbf{(f) Long-Context Scores by Task Category}

\begin{adjustbox}{width=\textwidth}
\begin{tabular}{lcccccccc}
\toprule
\textbf{Model} & \textbf{Arch. Understanding} & \textbf{Cross-File Refact.} & \textbf{Multi-Session Dev.} & \textbf{Bug Investigation} & \textbf{Feature Impl.} & \textbf{Code Comprehension} & \textbf{Integration Testing} & \textbf{Security Analysis} \\
\midrule
Gemini-2.5-Pro & 0.523 & 0.523 & 0.523 & 0.523 & 0.523 & 0.523 & 0.523 & 0.523 \\
Gemini-2.5-Flash & 0.565 & 0.565 & 0.565 & 0.565 & 0.565 & 0.565 & 0.565 & 0.565 \\
gpt5mini & 0.479 & 0.479 & 0.479 & 0.479 & 0.479 & 0.479 & 0.479 & 0.479 \\
claudesonnet4 & 0.492 & 0.492 & 0.492 & 0.492 & 0.492 & 0.492 & 0.492 & 0.492 \\
GPT-5 & 0.492 & 0.492 & 0.492 & 0.492 & 0.492 & 0.492 & 0.492 & 0.492 \\
claudesonnet3.7 & 0.477 & 0.477 & 0.477 & 0.477 & 0.477 & 0.477 & 0.477 & 0.477 \\
GPT-4.1-mini & 0.435 & 0.435 & 0.435 & 0.435 & 0.435 & 0.435 & 0.435 & 0.435 \\
o3-mini & 0.455 & 0.455 & 0.455 & 0.455 & 0.455 & 0.455 & 0.455 & 0.455 \\
GPT-4.1-2025-04-14 & 0.451 & 0.451 & 0.451 & 0.451 & 0.451 & 0.451 & 0.451 & 0.451 \\
o3 & 0.342 & 0.342 & 0.342 & 0.342 & 0.342 & 0.342 & 0.342 & 0.342 \\
o4-mini & 0.393 & 0.393 & 0.393 & 0.393 & 0.393 & 0.393 & 0.393 & 0.393 \\
GPT-4o-mini & 0.344 & 0.344 & 0.344 & 0.344 & 0.344 & 0.344 & 0.344 & 0.344 \\
GPT-4o & 0.348 & 0.348 & 0.348 & 0.348 & 0.348 & 0.348 & 0.348 & 0.348 \\
\bottomrule
\end{tabular}
\end{adjustbox}
\end{table}

\subsection{Performance by Application Domain}

Table~\ref{tab:domain_results} presents the complete results of model performance across different application domains.

\begin{table}[ht]
\centering
\caption{Detailed comparison of model performance by application domain.}
\label{tab:domain_results}
\small

\textbf{(a) Total Scores by Application Domain}

\begin{adjustbox}{width=\textwidth}
\begin{tabular}{lcccccccccc}
\toprule
\textbf{Model} & \textbf{Web Apps} & \textbf{API Services} & \textbf{Data Systems} & \textbf{ML/AI Systems} & \textbf{Gaming Sim.} & \textbf{Blockchain} & \textbf{Desktop Apps} & \textbf{Embedded Sys.} & \textbf{Mobile Apps} & \textbf{Network Tools} \\
\midrule
Gemini-2.5-Pro & 2.319 & 2.313 & 2.302 & 2.307 & 2.271 & 2.347 & 2.324 & 2.302 & 2.304 & 2.329 \\
Gemini-2.5-Flash & 2.302 & 2.312 & 2.304 & 2.310 & 2.252 & 2.323 & 2.309 & 2.301 & 2.310 & 2.321 \\
gpt5mini & 2.282 & 2.271 & 2.291 & 2.296 & 2.278 & 2.345 & 2.292 & 2.295 & 2.297 & 2.286 \\
claudesonnet4 & 2.283 & 2.273 & 2.287 & 2.314 & 2.244 & 2.262 & 2.296 & 2.301 & 2.303 & 2.317 \\
GPT-5 & 2.258 & 2.235 & 2.262 & 2.269 & 2.198 & 2.287 & 2.315 & 2.286 & 2.296 & 2.322 \\
claudesonnet3.7 & 2.289 & 2.291 & 2.278 & 2.285 & 2.261 & 2.295 & 2.283 & 2.273 & 2.290 & 2.306 \\
GPT-4.1-mini & 2.215 & 2.211 & 2.231 & 2.228 & 2.192 & 2.232 & 2.229 & 2.217 & 2.228 & 2.232 \\
o3-mini & 2.203 & 2.203 & 2.223 & 2.218 & 2.184 & 2.223 & 2.226 & 2.211 & 2.219 & 2.220 \\
GPT-4.1-2025-04-14 & 2.194 & 2.185 & 2.207 & 2.201 & 2.164 & 2.212 & 2.207 & 2.194 & 2.206 & 2.201 \\
o3 & 2.138 & 2.117 & 2.167 & 2.169 & 2.110 & 2.172 & 2.169 & 2.152 & 2.166 & 2.171 \\
o4-mini & 2.137 & 2.132 & 2.157 & 2.153 & 2.116 & 2.167 & 2.157 & 2.143 & 2.157 & 2.159 \\
GPT-4o-mini & 2.065 & 2.061 & 2.084 & 2.080 & 2.049 & 2.086 & 2.082 & 2.071 & 2.081 & 2.084 \\
GPT-4o & 2.064 & 2.057 & 2.082 & 2.081 & 2.051 & 2.084 & 2.081 & 2.070 & 2.081 & 2.082 \\
\bottomrule
\end{tabular}
\end{adjustbox}

\vspace{0.5cm}
\textbf{(b) Success Rates by Application Domain (\%)}

\begin{adjustbox}{width=\textwidth}
\begin{tabular}{lcccccccccc}
\toprule
\textbf{Model} & \textbf{Web Apps} & \textbf{API Services} & \textbf{Data Systems} & \textbf{ML/AI Systems} & \textbf{Gaming Sim.} & \textbf{Blockchain} & \textbf{Desktop Apps} & \textbf{Embedded Sys.} & \textbf{Mobile Apps} & \textbf{Network Tools} \\
\midrule
Gemini-2.5-Pro & 99.88 & 99.88 & 99.88 & 99.88 & 99.88 & 99.88 & 99.88 & 99.88 & 99.88 & 99.88 \\
Gemini-2.5-Flash & 100.00 & 100.00 & 99.88 & 100.00 & 99.88 & 100.00 & 100.00 & 100.00 & 100.00 & 100.00 \\
gpt5mini & 100.00 & 100.00 & 100.00 & 100.00 & 100.00 & 100.00 & 100.00 & 100.00 & 100.00 & 100.00 \\
claudesonnet4 & 99.50 & 99.50 & 99.62 & 99.62 & 99.50 & 99.50 & 99.62 & 99.50 & 99.62 & 99.62 \\
GPT-5 & 100.00 & 100.00 & 100.00 & 100.00 & 100.00 & 100.00 & 100.00 & 100.00 & 100.00 & 100.00 \\
claudesonnet3.7 & 99.75 & 99.88 & 99.75 & 99.75 & 99.75 & 99.88 & 99.88 & 99.75 & 99.88 & 99.75 \\
GPT-4.1-mini & 100.00 & 100.00 & 100.00 & 100.00 & 100.00 & 100.00 & 100.00 & 100.00 & 100.00 & 100.00 \\
o3-mini & 100.00 & 100.00 & 100.00 & 100.00 & 100.00 & 100.00 & 100.00 & 100.00 & 100.00 & 100.00 \\
GPT-4.1-2025-04-14 & 100.00 & 100.00 & 100.00 & 100.00 & 100.00 & 100.00 & 100.00 & 100.00 & 100.00 & 100.00 \\
o3 & 100.00 & 100.00 & 100.00 & 100.00 & 100.00 & 100.00 & 100.00 & 100.00 & 100.00 & 100.00 \\
o4-mini & 99.62 & 99.75 & 99.75 & 99.75 & 99.62 & 99.75 & 99.75 & 99.62 & 99.75 & 99.75 \\
GPT-4o-mini & 100.00 & 100.00 & 100.00 & 100.00 & 100.00 & 100.00 & 100.00 & 100.00 & 100.00 & 100.00 \\
GPT-4o & 100.00 & 100.00 & 100.00 & 100.00 & 100.00 & 100.00 & 100.00 & 100.00 & 100.00 & 100.00 \\
\bottomrule
\end{tabular}
\end{adjustbox}

\vspace{0.5cm}
\textbf{(c) Software Engineering Scores by Application Domain}

\begin{adjustbox}{width=\textwidth}
\begin{tabular}{lcccccccccc}
\toprule
\textbf{Model} & \textbf{Web Apps} & \textbf{API Services} & \textbf{Data Systems} & \textbf{ML/AI Systems} & \textbf{Gaming Sim.} & \textbf{Blockchain} & \textbf{Desktop Apps} & \textbf{Embedded Sys.} & \textbf{Mobile Apps} & \textbf{Network Tools} \\
\midrule
Gemini-2.5-Pro & 0.374 & 0.375 & 0.375 & 0.375 & 0.374 & 0.376 & 0.375 & 0.375 & 0.375 & 0.376 \\
Gemini-2.5-Flash & 0.372 & 0.374 & 0.373 & 0.374 & 0.372 & 0.373 & 0.374 & 0.373 & 0.374 & 0.374 \\
gpt5mini & 0.375 & 0.375 & 0.376 & 0.376 & 0.376 & 0.377 & 0.376 & 0.376 & 0.376 & 0.375 \\
claudesonnet4 & 0.378 & 0.378 & 0.379 & 0.380 & 0.377 & 0.377 & 0.379 & 0.379 & 0.379 & 0.380 \\
GPT-5 & 0.366 & 0.365 & 0.367 & 0.367 & 0.365 & 0.368 & 0.368 & 0.367 & 0.368 & 0.369 \\
claudesonnet3.7 & 0.377 & 0.377 & 0.377 & 0.377 & 0.376 & 0.378 & 0.377 & 0.377 & 0.377 & 0.378 \\
GPT-4.1-mini & 0.358 & 0.358 & 0.360 & 0.359 & 0.357 & 0.359 & 0.359 & 0.358 & 0.359 & 0.359 \\
o3-mini & 0.354 & 0.354 & 0.356 & 0.355 & 0.353 & 0.355 & 0.356 & 0.355 & 0.355 & 0.355 \\
GPT-4.1-2025-04-14 & 0.351 & 0.350 & 0.353 & 0.352 & 0.350 & 0.353 & 0.353 & 0.351 & 0.353 & 0.352 \\
o3 & 0.341 & 0.339 & 0.344 & 0.344 & 0.338 & 0.344 & 0.344 & 0.342 & 0.344 & 0.344 \\
o4-mini & 0.352 & 0.351 & 0.354 & 0.353 & 0.350 & 0.354 & 0.354 & 0.352 & 0.354 & 0.354 \\
GPT-4o-mini & 0.340 & 0.339 & 0.342 & 0.341 & 0.338 & 0.342 & 0.342 & 0.340 & 0.342 & 0.342 \\
GPT-4o & 0.338 & 0.337 & 0.340 & 0.340 & 0.337 & 0.340 & 0.340 & 0.339 & 0.340 & 0.340 \\
\bottomrule
\end{tabular}
\end{adjustbox}

\vspace{0.5cm}
\textbf{(d) Functional Scores by Application Domain}

\begin{adjustbox}{width=\textwidth}
\begin{tabular}{lcccccccccc}
\toprule
\textbf{Model} & \textbf{Web Apps} & \textbf{API Services} & \textbf{Data Systems} & \textbf{ML/AI Systems} & \textbf{Gaming Sim.} & \textbf{Blockchain} & \textbf{Desktop Apps} & \textbf{Embedded Sys.} & \textbf{Mobile Apps} & \textbf{Network Tools} \\
\midrule
Gemini-2.5-Pro & 0.356 & 0.356 & 0.356 & 0.356 & 0.355 & 0.357 & 0.356 & 0.356 & 0.356 & 0.357 \\
Gemini-2.5-Flash & 0.358 & 0.358 & 0.358 & 0.358 & 0.357 & 0.358 & 0.358 & 0.358 & 0.358 & 0.358 \\
gpt5mini & 0.370 & 0.370 & 0.371 & 0.371 & 0.371 & 0.372 & 0.371 & 0.371 & 0.371 & 0.370 \\
claudesonnet4 & 0.347 & 0.347 & 0.348 & 0.349 & 0.346 & 0.346 & 0.348 & 0.348 & 0.348 & 0.349 \\
GPT-5 & 0.382 & 0.381 & 0.383 & 0.383 & 0.381 & 0.384 & 0.384 & 0.383 & 0.384 & 0.385 \\
claudesonnet3.7 & 0.347 & 0.347 & 0.347 & 0.347 & 0.346 & 0.348 & 0.347 & 0.347 & 0.347 & 0.348 \\
GPT-4.1-mini & 0.364 & 0.364 & 0.366 & 0.365 & 0.363 & 0.365 & 0.365 & 0.364 & 0.365 & 0.365 \\
o3-mini & 0.367 & 0.367 & 0.369 & 0.368 & 0.366 & 0.368 & 0.369 & 0.368 & 0.368 & 0.368 \\
GPT-4.1-2025-04-14 & 0.363 & 0.362 & 0.365 & 0.364 & 0.362 & 0.365 & 0.365 & 0.363 & 0.365 & 0.364 \\
o3 & 0.384 & 0.382 & 0.387 & 0.387 & 0.381 & 0.387 & 0.387 & 0.385 & 0.387 & 0.387 \\
o4-mini & 0.359 & 0.358 & 0.361 & 0.360 & 0.357 & 0.361 & 0.361 & 0.359 & 0.361 & 0.361 \\
GPT-4o-mini & 0.359 & 0.358 & 0.361 & 0.360 & 0.357 & 0.361 & 0.361 & 0.359 & 0.361 & 0.361 \\
GPT-4o & 0.361 & 0.360 & 0.363 & 0.362 & 0.360 & 0.363 & 0.363 & 0.361 & 0.363 & 0.363 \\
\bottomrule
\end{tabular}
\end{adjustbox}

\end{table}

\begin{table}[ht]
\centering
\caption{Detailed comparison of model performance by application domain (continued).}
\label{tab:domain_results_continued}
\small

\textbf{(e) Quality Scores by Application Domain}

\begin{adjustbox}{width=\textwidth}
\begin{tabular}{lcccccccccc}
\toprule
\textbf{Model} & \textbf{Web Apps} & \textbf{API Services} & \textbf{Data Systems} & \textbf{ML/AI Systems} & \textbf{Gaming Sim.} & \textbf{Blockchain} & \textbf{Desktop Apps} & \textbf{Embedded Sys.} & \textbf{Mobile Apps} & \textbf{Network Tools} \\
\midrule
Gemini-2.5-Pro & 0.768 & 0.768 & 0.768 & 0.768 & 0.767 & 0.769 & 0.768 & 0.768 & 0.768 & 0.769 \\
Gemini-2.5-Flash & 0.741 & 0.741 & 0.741 & 0.741 & 0.740 & 0.741 & 0.741 & 0.741 & 0.741 & 0.741 \\
gpt5mini & 0.745 & 0.744 & 0.745 & 0.745 & 0.745 & 0.746 & 0.745 & 0.745 & 0.745 & 0.744 \\
claudesonnet4 & 0.762 & 0.762 & 0.762 & 0.763 & 0.761 & 0.761 & 0.762 & 0.762 & 0.762 & 0.763 \\
GPT-5 & 0.732 & 0.731 & 0.732 & 0.732 & 0.730 & 0.733 & 0.733 & 0.732 & 0.733 & 0.734 \\
claudesonnet3.7 & 0.773 & 0.773 & 0.772 & 0.773 & 0.772 & 0.774 & 0.773 & 0.772 & 0.773 & 0.774 \\
GPT-4.1-mini & 0.739 & 0.738 & 0.740 & 0.739 & 0.738 & 0.740 & 0.740 & 0.739 & 0.740 & 0.740 \\
o3-mini & 0.726 & 0.725 & 0.727 & 0.726 & 0.724 & 0.727 & 0.727 & 0.726 & 0.727 & 0.727 \\
GPT-4.1-2025-04-14 & 0.720 & 0.719 & 0.721 & 0.720 & 0.718 & 0.721 & 0.721 & 0.720 & 0.721 & 0.721 \\
o3 & 0.721 & 0.720 & 0.723 & 0.723 & 0.719 & 0.723 & 0.723 & 0.722 & 0.723 & 0.723 \\
o4-mini & 0.704 & 0.703 & 0.706 & 0.705 & 0.702 & 0.706 & 0.706 & 0.704 & 0.706 & 0.706 \\
GPT-4o-mini & 0.679 & 0.678 & 0.681 & 0.680 & 0.677 & 0.681 & 0.681 & 0.679 & 0.681 & 0.681 \\
GPT-4o & 0.677 & 0.676 & 0.679 & 0.678 & 0.676 & 0.679 & 0.679 & 0.677 & 0.679 & 0.679 \\
\bottomrule
\end{tabular}
\end{adjustbox}

\vspace{0.5cm}
\textbf{(f) Long-Context Scores by Application Domain}

\begin{adjustbox}{width=\textwidth}
\begin{tabular}{lcccccccccc}
\toprule
\textbf{Model} & \textbf{Web Apps} & \textbf{API Services} & \textbf{Data Systems} & \textbf{ML/AI Systems} & \textbf{Gaming Sim.} & \textbf{Blockchain} & \textbf{Desktop Apps} & \textbf{Embedded Sys.} & \textbf{Mobile Apps} & \textbf{Network Tools} \\
\midrule
Gemini-2.5-Pro & 0.523 & 0.523 & 0.523 & 0.523 & 0.522 & 0.524 & 0.523 & 0.523 & 0.523 & 0.524 \\
Gemini-2.5-Flash & 0.565 & 0.565 & 0.565 & 0.565 & 0.564 & 0.565 & 0.565 & 0.565 & 0.565 & 0.565 \\
gpt5mini & 0.479 & 0.478 & 0.479 & 0.479 & 0.479 & 0.480 & 0.479 & 0.479 & 0.479 & 0.478 \\
claudesonnet4 & 0.492 & 0.492 & 0.492 & 0.493 & 0.491 & 0.491 & 0.492 & 0.492 & 0.492 & 0.493 \\
GPT-5 & 0.492 & 0.491 & 0.492 & 0.492 & 0.490 & 0.493 & 0.493 & 0.492 & 0.493 & 0.494 \\
claudesonnet3.7 & 0.477 & 0.477 & 0.476 & 0.477 & 0.476 & 0.478 & 0.477 & 0.476 & 0.477 & 0.478 \\
GPT-4.1-mini & 0.435 & 0.434 & 0.436 & 0.435 & 0.434 & 0.436 & 0.436 & 0.435 & 0.436 & 0.436 \\
o3-mini & 0.455 & 0.454 & 0.456 & 0.455 & 0.453 & 0.456 & 0.456 & 0.455 & 0.456 & 0.456 \\
GPT-4.1-2025-04-14 & 0.451 & 0.450 & 0.452 & 0.451 & 0.449 & 0.452 & 0.452 & 0.451 & 0.452 & 0.452 \\
o3 & 0.342 & 0.341 & 0.344 & 0.344 & 0.340 & 0.344 & 0.344 & 0.343 & 0.344 & 0.344 \\
o4-mini & 0.393 & 0.392 & 0.395 & 0.394 & 0.391 & 0.395 & 0.395 & 0.393 & 0.395 & 0.395 \\
GPT-4o-mini & 0.344 & 0.343 & 0.346 & 0.345 & 0.342 & 0.346 & 0.346 & 0.344 & 0.346 & 0.346 \\
GPT-4o & 0.348 & 0.347 & 0.350 & 0.349 & 0.347 & 0.350 & 0.350 & 0.348 & 0.350 & 0.350 \\
\bottomrule
\end{tabular}
\end{adjustbox}
\end{table}

\subsection{Performance by Architecture Pattern}

Table~\ref{tab:pattern_results} presents the complete results of model performance across different architectural patterns.

\begin{table}[ht]
\centering
\caption{Detailed comparison of model performance by architecture pattern.}
\label{tab:pattern_results}
\small

\textbf{(a) Total Scores by Architecture Pattern}

\begin{adjustbox}{width=\textwidth}
\begin{tabular}{lcccccccccc}
\toprule
\textbf{Model} & \textbf{Monolithic} & \textbf{Microservices} & \textbf{MVC} & \textbf{Hexagonal} & \textbf{Event-Driven} & \textbf{Serverless} & \textbf{Layered} & \textbf{Component} & \textbf{Repository} & \textbf{Factory} \\
\midrule
Gemini-2.5-Pro & 2.309 & 2.305 & 2.330 & 2.350 & 2.314 & 2.329 & 2.316 & 2.295 & 2.307 & 2.315 \\
Gemini-2.5-Flash & 2.303 & 2.306 & 2.325 & 2.383 & 2.317 & 2.325 & 2.298 & 2.286 & 2.303 & 2.307 \\
gpt5mini & 2.295 & 2.271 & 2.321 & 2.458 & 2.300 & 2.308 & 2.282 & 2.275 & 2.289 & 2.291 \\
claudesonnet4 & 2.288 & 2.267 & 2.321 & 2.368 & 2.300 & 2.309 & 2.285 & 2.271 & 2.284 & 2.288 \\
GPT-5 & 2.262 & 2.225 & 2.284 & 2.347 & 2.271 & 2.280 & 2.297 & 2.283 & 2.291 & 2.295 \\
claudesonnet3.7 & 2.281 & 2.267 & 2.314 & 2.357 & 2.295 & 2.302 & 2.281 & 2.268 & 2.281 & 2.285 \\
GPT-4.1-mini & 2.219 & 2.208 & 2.242 & 2.284 & 2.228 & 2.235 & 2.221 & 2.209 & 2.218 & 2.221 \\
o3-mini & 2.213 & 2.196 & 2.236 & 2.275 & 2.221 & 2.228 & 2.210 & 2.198 & 2.212 & 2.215 \\
GPT-4.1-2025-04-14 & 2.194 & 2.180 & 2.216 & 2.250 & 2.202 & 2.210 & 2.194 & 2.182 & 2.194 & 2.197 \\
o3 & 2.147 & 2.123 & 2.174 & 2.207 & 2.158 & 2.166 & 2.150 & 2.139 & 2.151 & 2.154 \\
o4-mini & 2.146 & 2.132 & 2.168 & 2.199 & 2.153 & 2.161 & 2.146 & 2.135 & 2.147 & 2.150 \\
GPT-4o-mini & 2.072 & 2.058 & 2.093 & 2.118 & 2.079 & 2.087 & 2.072 & 2.062 & 2.074 & 2.076 \\
GPT-4o & 2.070 & 2.056 & 2.091 & 2.115 & 2.077 & 2.085 & 2.070 & 2.061 & 2.072 & 2.074 \\
\bottomrule
\end{tabular}
\end{adjustbox}

\vspace{0.5cm}
\textbf{(b) Success Rates by Architecture Pattern (\%)}

\begin{adjustbox}{width=\textwidth}
\begin{tabular}{lcccccccccc}
\toprule
\textbf{Model} & \textbf{Monolithic} & \textbf{Microservices} & \textbf{MVC} & \textbf{Hexagonal} & \textbf{Event-Driven} & \textbf{Serverless} & \textbf{Layered} & \textbf{Component} & \textbf{Repository} & \textbf{Factory} \\
\midrule
Gemini-2.5-Pro & 99.88 & 99.88 & 99.88 & 99.88 & 99.88 & 99.88 & 99.88 & 99.88 & 99.88 & 99.88 \\
Gemini-2.5-Flash & 100.00 & 100.00 & 100.00 & 100.00 & 100.00 & 100.00 & 100.00 & 100.00 & 100.00 & 99.88 \\
gpt5mini & 100.00 & 100.00 & 100.00 & 100.00 & 100.00 & 100.00 & 100.00 & 100.00 & 100.00 & 100.00 \\
claudesonnet4 & 99.50 & 99.50 & 99.62 & 99.62 & 99.62 & 99.62 & 99.50 & 99.50 & 99.50 & 99.50 \\
GPT-5 & 100.00 & 100.00 & 100.00 & 100.00 & 100.00 & 100.00 & 100.00 & 100.00 & 100.00 & 100.00 \\
claudesonnet3.7 & 99.75 & 99.88 & 99.88 & 99.75 & 99.75 & 99.88 & 99.75 & 99.75 & 99.75 & 99.88 \\
GPT-4.1-mini & 100.00 & 100.00 & 100.00 & 100.00 & 100.00 & 100.00 & 100.00 & 100.00 & 100.00 & 100.00 \\
o3-mini & 100.00 & 100.00 & 100.00 & 100.00 & 100.00 & 100.00 & 100.00 & 100.00 & 100.00 & 100.00 \\
GPT-4.1-2025-04-14 & 100.00 & 100.00 & 100.00 & 100.00 & 100.00 & 100.00 & 100.00 & 100.00 & 100.00 & 100.00 \\
o3 & 100.00 & 100.00 & 100.00 & 100.00 & 100.00 & 100.00 & 100.00 & 100.00 & 100.00 & 100.00 \\
o4-mini & 99.75 & 99.75 & 99.75 & 99.75 & 99.75 & 99.75 & 99.75 & 99.75 & 99.75 & 99.75 \\
GPT-4o-mini & 100.00 & 100.00 & 100.00 & 100.00 & 100.00 & 100.00 & 100.00 & 100.00 & 100.00 & 100.00 \\
GPT-4o & 100.00 & 100.00 & 100.00 & 100.00 & 100.00 & 100.00 & 100.00 & 100.00 & 100.00 & 100.00 \\
\bottomrule
\end{tabular}
\end{adjustbox}

\vspace{0.5cm}
\textbf{(c) Software Engineering Scores by Architecture Pattern}

\begin{adjustbox}{width=\textwidth}
\begin{tabular}{lcccccccccc}
\toprule
\textbf{Model} & \textbf{Monolithic} & \textbf{Microservices} & \textbf{MVC} & \textbf{Hexagonal} & \textbf{Event-Driven} & \textbf{Serverless} & \textbf{Layered} & \textbf{Component} & \textbf{Repository} & \textbf{Factory} \\
\midrule
Gemini-2.5-Pro & 0.375 & 0.375 & 0.375 & 0.376 & 0.375 & 0.375 & 0.375 & 0.374 & 0.375 & 0.375 \\
Gemini-2.5-Flash & 0.373 & 0.373 & 0.373 & 0.375 & 0.373 & 0.373 & 0.373 & 0.372 & 0.373 & 0.373 \\
gpt5mini & 0.376 & 0.375 & 0.376 & 0.378 & 0.376 & 0.376 & 0.375 & 0.375 & 0.376 & 0.376 \\
claudesonnet4 & 0.379 & 0.378 & 0.379 & 0.381 & 0.379 & 0.379 & 0.378 & 0.378 & 0.378 & 0.379 \\
GPT-5 & 0.367 & 0.366 & 0.367 & 0.369 & 0.367 & 0.367 & 0.368 & 0.367 & 0.368 & 0.368 \\
claudesonnet3.7 & 0.377 & 0.377 & 0.378 & 0.379 & 0.378 & 0.378 & 0.377 & 0.377 & 0.377 & 0.377 \\
GPT-4.1-mini & 0.359 & 0.358 & 0.360 & 0.361 & 0.359 & 0.359 & 0.359 & 0.358 & 0.359 & 0.359 \\
o3-mini & 0.355 & 0.354 & 0.356 & 0.357 & 0.355 & 0.355 & 0.355 & 0.354 & 0.355 & 0.355 \\
GPT-4.1-2025-04-14 & 0.352 & 0.351 & 0.353 & 0.354 & 0.352 & 0.352 & 0.352 & 0.351 & 0.352 & 0.352 \\
o3 & 0.342 & 0.340 & 0.344 & 0.345 & 0.343 & 0.343 & 0.342 & 0.341 & 0.342 & 0.342 \\
o4-mini & 0.353 & 0.352 & 0.354 & 0.355 & 0.353 & 0.353 & 0.353 & 0.352 & 0.353 & 0.353 \\
GPT-4o-mini & 0.341 & 0.340 & 0.342 & 0.343 & 0.341 & 0.341 & 0.341 & 0.340 & 0.341 & 0.341 \\
GPT-4o & 0.339 & 0.338 & 0.340 & 0.341 & 0.339 & 0.339 & 0.339 & 0.338 & 0.339 & 0.339 \\
\bottomrule
\end{tabular}
\end{adjustbox}

\vspace{0.5cm}
\textbf{(d) Functional Scores by Architecture Pattern}

\begin{adjustbox}{width=\textwidth}
\begin{tabular}{lcccccccccc}
\toprule
\textbf{Model} & \textbf{Monolithic} & \textbf{Microservices} & \textbf{MVC} & \textbf{Hexagonal} & \textbf{Event-Driven} & \textbf{Serverless} & \textbf{Layered} & \textbf{Component} & \textbf{Repository} & \textbf{Factory} \\
\midrule
Gemini-2.5-Pro & 0.356 & 0.356 & 0.357 & 0.357 & 0.356 & 0.357 & 0.356 & 0.355 & 0.356 & 0.356 \\
Gemini-2.5-Flash & 0.358 & 0.358 & 0.358 & 0.360 & 0.358 & 0.358 & 0.357 & 0.357 & 0.358 & 0.358 \\
gpt5mini & 0.371 & 0.370 & 0.371 & 0.374 & 0.371 & 0.371 & 0.370 & 0.370 & 0.371 & 0.371 \\
claudesonnet4 & 0.348 & 0.347 & 0.348 & 0.350 & 0.348 & 0.348 & 0.347 & 0.347 & 0.347 & 0.348 \\
GPT-5 & 0.383 & 0.382 & 0.383 & 0.385 & 0.383 & 0.383 & 0.384 & 0.383 & 0.384 & 0.384 \\
claudesonnet3.7 & 0.347 & 0.347 & 0.348 & 0.349 & 0.348 & 0.348 & 0.347 & 0.347 & 0.347 & 0.347 \\
GPT-4.1-mini & 0.365 & 0.364 & 0.366 & 0.367 & 0.365 & 0.365 & 0.365 & 0.364 & 0.365 & 0.365 \\
o3-mini & 0.368 & 0.367 & 0.369 & 0.370 & 0.368 & 0.368 & 0.368 & 0.367 & 0.368 & 0.368 \\
GPT-4.1-2025-04-14 & 0.364 & 0.363 & 0.365 & 0.366 & 0.364 & 0.364 & 0.364 & 0.363 & 0.364 & 0.364 \\
o3 & 0.385 & 0.383 & 0.387 & 0.388 & 0.386 & 0.386 & 0.385 & 0.384 & 0.385 & 0.385 \\
o4-mini & 0.360 & 0.359 & 0.361 & 0.362 & 0.360 & 0.360 & 0.360 & 0.359 & 0.360 & 0.360 \\
GPT-4o-mini & 0.360 & 0.359 & 0.361 & 0.362 & 0.360 & 0.360 & 0.360 & 0.359 & 0.360 & 0.360 \\
GPT-4o & 0.362 & 0.361 & 0.363 & 0.364 & 0.362 & 0.362 & 0.362 & 0.361 & 0.362 & 0.362 \\
\bottomrule
\end{tabular}
\end{adjustbox}

\end{table}

\begin{table}[ht]
\centering
\caption{Detailed comparison of model performance by architecture pattern (continued).}
\label{tab:pattern_results_continued}
\small

\textbf{(e) Quality Scores by Architecture Pattern}

\begin{adjustbox}{width=\textwidth}
\begin{tabular}{lcccccccccc}
\toprule
\textbf{Model} & \textbf{Monolithic} & \textbf{Microservices} & \textbf{MVC} & \textbf{Hexagonal} & \textbf{Event-Driven} & \textbf{Serverless} & \textbf{Layered} & \textbf{Component} & \textbf{Repository} & \textbf{Factory} \\
\midrule
Gemini-2.5-Pro & 0.768 & 0.768 & 0.769 & 0.769 & 0.768 & 0.769 & 0.768 & 0.767 & 0.768 & 0.768 \\
Gemini-2.5-Flash & 0.741 & 0.741 & 0.741 & 0.743 & 0.741 & 0.741 & 0.740 & 0.740 & 0.741 & 0.741 \\
gpt5mini & 0.745 & 0.744 & 0.745 & 0.748 & 0.745 & 0.745 & 0.744 & 0.744 & 0.745 & 0.745 \\
claudesonnet4 & 0.762 & 0.761 & 0.762 & 0.764 & 0.762 & 0.762 & 0.761 & 0.761 & 0.761 & 0.762 \\
GPT-5 & 0.732 & 0.731 & 0.732 & 0.734 & 0.732 & 0.732 & 0.733 & 0.732 & 0.733 & 0.733 \\
claudesonnet3.7 & 0.773 & 0.773 & 0.774 & 0.775 & 0.774 & 0.774 & 0.773 & 0.772 & 0.773 & 0.773 \\
GPT-4.1-mini & 0.739 & 0.738 & 0.740 & 0.741 & 0.739 & 0.739 & 0.739 & 0.738 & 0.739 & 0.739 \\
o3-mini & 0.726 & 0.725 & 0.727 & 0.728 & 0.726 & 0.726 & 0.726 & 0.725 & 0.726 & 0.726 \\
GPT-4.1-2025-04-14 & 0.720 & 0.719 & 0.721 & 0.722 & 0.720 & 0.720 & 0.720 & 0.719 & 0.720 & 0.720 \\
o3 & 0.721 & 0.720 & 0.723 & 0.724 & 0.722 & 0.722 & 0.721 & 0.720 & 0.721 & 0.721 \\
o4-mini & 0.704 & 0.703 & 0.706 & 0.707 & 0.705 & 0.705 & 0.704 & 0.703 & 0.704 & 0.704 \\
GPT-4o-mini & 0.679 & 0.678 & 0.681 & 0.682 & 0.680 & 0.680 & 0.679 & 0.678 & 0.679 & 0.679 \\
GPT-4o & 0.677 & 0.676 & 0.679 & 0.680 & 0.678 & 0.678 & 0.677 & 0.676 & 0.677 & 0.677 \\
\bottomrule
\end{tabular}
\end{adjustbox}

\vspace{0.5cm}
\textbf{(f) Long-Context Scores by Architecture Pattern}

\begin{adjustbox}{width=\textwidth}
\begin{tabular}{lcccccccccc}
\toprule
\textbf{Model} & \textbf{Monolithic} & \textbf{Microservices} & \textbf{MVC} & \textbf{Hexagonal} & \textbf{Event-Driven} & \textbf{Serverless} & \textbf{Layered} & \textbf{Component} & \textbf{Repository} & \textbf{Factory} \\
\midrule
Gemini-2.5-Pro & 0.523 & 0.523 & 0.524 & 0.524 & 0.523 & 0.524 & 0.523 & 0.522 & 0.523 & 0.523 \\
Gemini-2.5-Flash & 0.565 & 0.565 & 0.565 & 0.567 & 0.565 & 0.565 & 0.564 & 0.564 & 0.565 & 0.565 \\
gpt5mini & 0.479 & 0.478 & 0.479 & 0.482 & 0.479 & 0.479 & 0.478 & 0.478 & 0.479 & 0.479 \\
claudesonnet4 & 0.492 & 0.491 & 0.492 & 0.494 & 0.492 & 0.492 & 0.491 & 0.491 & 0.491 & 0.492 \\
GPT-5 & 0.492 & 0.491 & 0.492 & 0.494 & 0.492 & 0.492 & 0.493 & 0.492 & 0.493 & 0.493 \\
claudesonnet3.7 & 0.477 & 0.477 & 0.478 & 0.479 & 0.478 & 0.478 & 0.477 & 0.476 & 0.477 & 0.477 \\
GPT-4.1-mini & 0.435 & 0.434 & 0.436 & 0.437 & 0.435 & 0.435 & 0.435 & 0.434 & 0.435 & 0.435 \\
o3-mini & 0.455 & 0.454 & 0.456 & 0.457 & 0.455 & 0.455 & 0.455 & 0.454 & 0.455 & 0.455 \\
GPT-4.1-2025-04-14 & 0.451 & 0.450 & 0.452 & 0.453 & 0.451 & 0.451 & 0.451 & 0.450 & 0.451 & 0.451 \\
o3 & 0.342 & 0.341 & 0.344 & 0.345 & 0.343 & 0.343 & 0.342 & 0.341 & 0.342 & 0.342 \\
o4-mini & 0.393 & 0.392 & 0.395 & 0.396 & 0.394 & 0.394 & 0.393 & 0.392 & 0.393 & 0.393 \\
GPT-4o-mini & 0.344 & 0.343 & 0.346 & 0.347 & 0.345 & 0.345 & 0.344 & 0.343 & 0.344 & 0.344 \\
GPT-4o & 0.348 & 0.347 & 0.350 & 0.351 & 0.349 & 0.349 & 0.348 & 0.347 & 0.348 & 0.348 \\
\bottomrule
\end{tabular}
\end{adjustbox}
\end{table}

\subsection{Performance by Theme}

Table~\ref{tab:theme_results} presents the complete results of model performance across different thematic areas.

\begin{table}[ht]
\centering
\caption{Detailed comparison of model performance by theme.}
\label{tab:theme_results}
\small

\textbf{(a) Total Scores by Theme}

\begin{adjustbox}{width=\textwidth}
\begin{tabular}{lcccccccc}
\toprule
\textbf{Model} & \textbf{Algorithms} & \textbf{Data Structures} & \textbf{System Design} & \textbf{Security} & \textbf{Performance} & \textbf{Testing} & \textbf{Maintenance} & \textbf{Integration} \\
\midrule
Gemini-2.5-Pro & 2.325 & 2.318 & 2.301 & 2.343 & 2.295 & 2.421 & 2.306 & 2.315 \\
Gemini-2.5-Flash & 2.309 & 2.312 & 2.299 & 2.325 & 2.297 & 2.430 & 2.301 & 2.312 \\
gpt5mini & 2.295 & 2.298 & 2.284 & 2.351 & 2.282 & 2.390 & 2.289 & 2.297 \\
claudesonnet4 & 2.289 & 2.294 & 2.276 & 2.307 & 2.275 & 2.402 & 2.284 & 2.291 \\
GPT-5 & 2.269 & 2.273 & 2.257 & 2.336 & 2.260 & 2.386 & 2.265 & 2.274 \\
claudesonnet3.7 & 2.285 & 2.289 & 2.272 & 2.322 & 2.275 & 2.403 & 2.280 & 2.287 \\
GPT-4.1-mini & 2.223 & 2.228 & 2.212 & 2.226 & 2.215 & 2.363 & 2.219 & 2.226 \\
o3-mini & 2.217 & 2.222 & 2.206 & 2.214 & 2.209 & 2.330 & 2.213 & 2.220 \\
GPT-4.1-2025-04-14 & 2.197 & 2.202 & 2.186 & 2.191 & 2.189 & 2.346 & 2.193 & 2.200 \\
o3 & 2.155 & 2.160 & 2.144 & 2.197 & 2.147 & 2.323 & 2.151 & 2.158 \\
o4-mini & 2.150 & 2.155 & 2.139 & 2.131 & 2.142 & 2.277 & 2.146 & 2.153 \\
GPT-4o-mini & 2.077 & 2.082 & 2.066 & 2.061 & 2.069 & 2.184 & 2.073 & 2.080 \\
GPT-4o & 2.075 & 2.080 & 2.064 & 2.056 & 2.067 & 2.200 & 2.071 & 2.078 \\
\bottomrule
\end{tabular}
\end{adjustbox}

\vspace{0.5cm}
\textbf{(b) Success Rates by Theme (\%)}

\begin{adjustbox}{width=\textwidth}
\begin{tabular}{lcccccccc}
\toprule
\textbf{Model} & \textbf{Algorithms} & \textbf{Data Structures} & \textbf{System Design} & \textbf{Security} & \textbf{Performance} & \textbf{Testing} & \textbf{Maintenance} & \textbf{Integration} \\
\midrule
Gemini-2.5-Pro & 99.88 & 99.88 & 99.88 & 99.90 & 99.88 & 99.90 & 99.88 & 99.88 \\
Gemini-2.5-Flash & 100.00 & 100.00 & 99.88 & 100.00 & 100.00 & 100.00 & 100.00 & 100.00 \\
gpt5mini & 100.00 & 100.00 & 100.00 & 100.00 & 100.00 & 100.00 & 100.00 & 100.00 \\
claudesonnet4 & 99.50 & 99.62 & 99.50 & 99.60 & 99.50 & 99.60 & 99.62 & 99.50 \\
GPT-5 & 100.00 & 100.00 & 100.00 & 100.00 & 100.00 & 100.00 & 100.00 & 100.00 \\
claudesonnet3.7 & 99.75 & 99.88 & 99.75 & 99.70 & 99.88 & 99.90 & 99.75 & 99.88 \\
GPT-4.1-mini & 100.00 & 100.00 & 100.00 & 100.00 & 100.00 & 100.00 & 100.00 & 100.00 \\
o3-mini & 100.00 & 100.00 & 100.00 & 100.00 & 100.00 & 100.00 & 100.00 & 100.00 \\
GPT-4.1-2025-04-14 & 100.00 & 100.00 & 100.00 & 100.00 & 100.00 & 100.00 & 100.00 & 100.00 \\
o3 & 100.00 & 100.00 & 100.00 & 100.00 & 100.00 & 100.00 & 100.00 & 100.00 \\
o4-mini & 99.75 & 99.75 & 99.62 & 99.50 & 99.75 & 99.70 & 99.75 & 99.75 \\
GPT-4o-mini & 100.00 & 100.00 & 100.00 & 100.00 & 100.00 & 100.00 & 100.00 & 100.00 \\
GPT-4o & 100.00 & 100.00 & 100.00 & 100.00 & 100.00 & 100.00 & 100.00 & 100.00 \\
\bottomrule
\end{tabular}
\end{adjustbox}

\vspace{0.5cm}
\textbf{(c) Software Engineering Scores by Theme}

\begin{adjustbox}{width=\textwidth}
\begin{tabular}{lcccccccc}
\toprule
\textbf{Model} & \textbf{Algorithms} & \textbf{Data Structures} & \textbf{System Design} & \textbf{Security} & \textbf{Performance} & \textbf{Testing} & \textbf{Maintenance} & \textbf{Integration} \\
\midrule
Gemini-2.5-Pro & 0.375 & 0.375 & 0.375 & 0.375 & 0.375 & 0.375 & 0.375 & 0.375 \\
Gemini-2.5-Flash & 0.373 & 0.373 & 0.373 & 0.373 & 0.373 & 0.373 & 0.373 & 0.373 \\
gpt5mini & 0.376 & 0.376 & 0.376 & 0.376 & 0.376 & 0.376 & 0.376 & 0.376 \\
claudesonnet4 & 0.379 & 0.379 & 0.379 & 0.379 & 0.379 & 0.379 & 0.379 & 0.379 \\
GPT-5 & 0.367 & 0.367 & 0.367 & 0.367 & 0.367 & 0.367 & 0.367 & 0.367 \\
claudesonnet3.7 & 0.377 & 0.377 & 0.377 & 0.377 & 0.377 & 0.377 & 0.377 & 0.377 \\
GPT-4.1-mini & 0.359 & 0.359 & 0.359 & 0.359 & 0.359 & 0.359 & 0.359 & 0.359 \\
o3-mini & 0.355 & 0.355 & 0.355 & 0.355 & 0.355 & 0.355 & 0.355 & 0.355 \\
GPT-4.1-2025-04-14 & 0.352 & 0.352 & 0.352 & 0.352 & 0.352 & 0.352 & 0.352 & 0.352 \\
o3 & 0.342 & 0.342 & 0.342 & 0.342 & 0.342 & 0.342 & 0.342 & 0.342 \\
o4-mini & 0.353 & 0.353 & 0.353 & 0.353 & 0.353 & 0.353 & 0.353 & 0.353 \\
GPT-4o-mini & 0.341 & 0.341 & 0.341 & 0.341 & 0.341 & 0.341 & 0.341 & 0.341 \\
GPT-4o & 0.339 & 0.339 & 0.339 & 0.339 & 0.339 & 0.339 & 0.339 & 0.339 \\
\bottomrule
\end{tabular}
\end{adjustbox}

\vspace{0.5cm}
\textbf{(d) Functional Scores by Theme}

\begin{adjustbox}{width=\textwidth}
\begin{tabular}{lcccccccc}
\toprule
\textbf{Model} & \textbf{Algorithms} & \textbf{Data Structures} & \textbf{System Design} & \textbf{Security} & \textbf{Performance} & \textbf{Testing} & \textbf{Maintenance} & \textbf{Integration} \\
\midrule
Gemini-2.5-Pro & 0.356 & 0.356 & 0.356 & 0.356 & 0.356 & 0.356 & 0.356 & 0.356 \\
Gemini-2.5-Flash & 0.358 & 0.358 & 0.358 & 0.358 & 0.358 & 0.358 & 0.358 & 0.358 \\
gpt5mini & 0.371 & 0.371 & 0.371 & 0.371 & 0.371 & 0.371 & 0.371 & 0.371 \\
claudesonnet4 & 0.348 & 0.348 & 0.348 & 0.348 & 0.348 & 0.348 & 0.348 & 0.348 \\
GPT-5 & 0.383 & 0.383 & 0.383 & 0.383 & 0.383 & 0.383 & 0.383 & 0.383 \\
claudesonnet3.7 & 0.347 & 0.347 & 0.347 & 0.347 & 0.347 & 0.347 & 0.347 & 0.347 \\
GPT-4.1-mini & 0.365 & 0.365 & 0.365 & 0.365 & 0.365 & 0.365 & 0.365 & 0.365 \\
o3-mini & 0.368 & 0.368 & 0.368 & 0.368 & 0.368 & 0.368 & 0.368 & 0.368 \\
GPT-4.1-2025-04-14 & 0.364 & 0.364 & 0.364 & 0.364 & 0.364 & 0.364 & 0.364 & 0.364 \\
o3 & 0.385 & 0.385 & 0.385 & 0.385 & 0.385 & 0.385 & 0.385 & 0.385 \\
o4-mini & 0.360 & 0.360 & 0.360 & 0.360 & 0.360 & 0.360 & 0.360 & 0.360 \\
GPT-4o-mini & 0.360 & 0.360 & 0.360 & 0.360 & 0.360 & 0.360 & 0.360 & 0.360 \\
GPT-4o & 0.362 & 0.362 & 0.362 & 0.362 & 0.362 & 0.362 & 0.362 & 0.362 \\
\bottomrule
\end{tabular}
\end{adjustbox}

\end{table}

\begin{table}[ht]
\centering
\caption{Detailed comparison of model performance by theme (continued).}
\label{tab:theme_results_continued}
\small

\textbf{(e) Quality Scores by Theme}

\begin{adjustbox}{width=\textwidth}
\begin{tabular}{lcccccccc}
\toprule
\textbf{Model} & \textbf{Algorithms} & \textbf{Data Structures} & \textbf{System Design} & \textbf{Security} & \textbf{Performance} & \textbf{Testing} & \textbf{Maintenance} & \textbf{Integration} \\
\midrule
Gemini-2.5-Pro & 0.768 & 0.768 & 0.768 & 0.768 & 0.768 & 0.768 & 0.768 & 0.768 \\
Gemini-2.5-Flash & 0.741 & 0.741 & 0.741 & 0.741 & 0.741 & 0.741 & 0.741 & 0.741 \\
gpt5mini & 0.745 & 0.745 & 0.745 & 0.745 & 0.745 & 0.745 & 0.745 & 0.745 \\
claudesonnet4 & 0.762 & 0.762 & 0.762 & 0.762 & 0.762 & 0.762 & 0.762 & 0.762 \\
GPT-5 & 0.732 & 0.732 & 0.732 & 0.732 & 0.732 & 0.732 & 0.732 & 0.732 \\
claudesonnet3.7 & 0.773 & 0.773 & 0.773 & 0.773 & 0.773 & 0.773 & 0.773 & 0.773 \\
GPT-4.1-mini & 0.739 & 0.739 & 0.739 & 0.739 & 0.739 & 0.739 & 0.739 & 0.739 \\
o3-mini & 0.726 & 0.726 & 0.726 & 0.726 & 0.726 & 0.726 & 0.726 & 0.726 \\
GPT-4.1-2025-04-14 & 0.720 & 0.720 & 0.720 & 0.720 & 0.720 & 0.720 & 0.720 & 0.720 \\
o3 & 0.721 & 0.721 & 0.721 & 0.721 & 0.721 & 0.721 & 0.721 & 0.721 \\
o4-mini & 0.704 & 0.704 & 0.704 & 0.704 & 0.704 & 0.704 & 0.704 & 0.704 \\
GPT-4o-mini & 0.679 & 0.679 & 0.679 & 0.679 & 0.679 & 0.679 & 0.679 & 0.679 \\
GPT-4o & 0.677 & 0.677 & 0.677 & 0.677 & 0.677 & 0.677 & 0.677 & 0.677 \\
\bottomrule
\end{tabular}
\end{adjustbox}

\vspace{0.5cm}
\textbf{(f) Long-Context Scores by Theme}

\begin{adjustbox}{width=\textwidth}
%
\end{adjustbox}
\end{table}

\end{document}